\documentstyle[11pt,epsf]{article}

\begin{document}

\def\been{\begin{enumerate}}
\def\enen{\end{enumerate}}
\def\beit{\begin{itemize}}
\def\enit{\end{itemize}}
\def\be{\begin{equation}}
\def\en{\end{equation}}
\def\bear{\begin{eqnarray}}
\def\enar{\end{eqnarray}}
\def\beas{\begin{eqnarray*}}
\def\enas{\end{eqnarray*}}
\def\bera{\begin{eqnarray}}
\def\enra{\end{eqnarray}}

\renewcommand{\theequation}{\thesection.\arabic{equation}}

\def\ptl{\partial}
\newcommand{\half}{\frac{1}{2}}
\def\CC{{\cal C}}
\def\buildchar#1#2#3{\null \! \mathop {\vphantom {#1}
\smash #1}\limits ^{#2}_{#3}\!\null }
\def\ut#1{\buildchar{#1}{}{^\sim}\/}
\def\dtri{\tilde{E}}
\def\CH{{\cal C_H}}
\def\CM{{\cal C_M}}
\def\CG{{\cal C_G}}
\def\mm[#1,#2,#3,#4]
 {\left(\matrix{#1 & #2 \cr #3& #4}\right)}
\def\largefbox#1#2{\begin{center}
                         \setlength{\fboxsep}{0.25cm}\fbox{\parbox[c]{#1}{#2}}
                    \end{center}}
\def\Largefbox#1#2{\begin{center}
                         \setlength{\fboxsep}{0.5cm}\fbox{\parbox[c]{#1}{#2}}
                    \end{center}}
\def\boxwidth{160mm}  
\def\boxwidthitem{150mm}  
\newenvironment{indention}[1]{\par
\addtolength{\leftskip}{#1}
\begingroup}{\endgroup\par}


\title{{\bf Re-formulating the Einstein equations \\
for stable numerical simulations}\\
-- Formulation Problem in Numerical Relativity --}

\author{Hisa-aki Shinkai  \\ {\tt hshinkai@postman.riken.go.jp}  \\
      Computational Science Division, \\
      The Institute of Physical and Chemical Research (RIKEN), \\
      Wako, Saitama, 351-0198 Japan \\
      ~\\ 
      Gen Yoneda \\  {\tt yoneda@mse.waseda.ac.jp}  \\
      Department of Mathematical Science
      Waseda University, \\
      Ookubo, Shinjuku, Tokyo, 169-8555 Japan
      }

 \date{December 1, 2002}
\maketitle

\begin{abstract}
We review recent efforts to re-formulate the Einstein equations
for fully relativistic numerical simulations.  
The so-called numerical
relativity  (computational simulations in general relativity) is 
a promising research field matching with ongoing 
astrophysical observations such as gravitational wave astronomy. 
Many trials for 
longterm stable and accurate simulations of binary compact objects 
have revealed that mathematically 
equivalent sets of evolution equations show different numerical 
stability in free evolution schemes.  
In this article, we first review 
the efforts of the community,  
categorizing them into the following three directions: 
(1) modifications of the standard Arnowitt-Deser-Misner equations 
initiated by the Kyoto group, 
(2) rewriting of the evolution equations in hyperbolic form, and 
(3) construction of an ``asymptotically constrained" system.  
We next introduce our idea
for explaining these evolution behaviors in a unified way using 
eigenvalue analysis of the 
constraint 
propagation equations.  The modifications of (or adjustments to) the evolution 
equations change  the character  of  constraint propagation, and 
several particular 
adjustments using constraints are expected to diminish the 
constraint-violating modes.  
We propose several new adjusted evolution equations, and include 
some numerical demonstrations.  We conclude by discussing some directions
for future research. 
\end{abstract}
\vfill
This article is for a part of the
book {\em Progress in Astronomy and Astrophysics} 
(Nova Science Publ., 2003?).  Also available as gr-qc/0209111.

\setcounter{tocdepth}{2}
\Largefbox{16cm}{
\tableofcontents
}
\newpage

\section{Overview}
\subsection{Numerical Relativity}
The theory of general relativity describes the nature of the 
strong gravitational field. 
The Einstein equation predicts quite unexpected phenomena such as 
gravitational collapse, gravitational waves, the expanding universe and 
so on, which are all attractive not only for researchers but also for the public. 
The Einstein equation consists of 10 partial differential equations (elliptic
and hyperbolic) for 10 metric components, and it is not easy to solve 
them for any particular situation.  
Over the decades, people have tried to study the general-relativistic 
world by finding its
exact solutions, by developing approximation methods,  or by simplifying the situations.  
Among these approaches, direct 
numerical integration  of the Einstein equations
can be said to be the most robust way to study the strong gravitational field. 
This research field is often called ``numerical relativity". 

\Largefbox{\boxwidth}{
{\bf Numerical Relativity } \hspace*{\fill} {\bf Box 1.1}\\
$\qquad =$ Necessary for unveiling the nature of strong gravity. For example: 
\beit 
\baselineskip 8pt
\item gravitational waves from colliding black holes, neutron stars, supernovae, ...
\item relativistic phenomena like cosmology, active galactic nuclei, ...
\item mathematical feedback to singularity, exact solutions, chaotic behavior, ...
\item laboratory for gravitational theories, higher-dimensional models, ...
\enit
}

Numerical relativity is now an essential field in gravity research.
The current mainstream in numerical relativity is to
analyze the final phase of compact
binary objects (black holes and/or neutron stars) 
related to gravitational wave observations
(see e.g. the conference proceedings \cite{PTPsupple}). 
Over the past decades, many groups have developed their numerical simulations
by trial and error.  Simulations require large-scale computational
facilities, and long-time stable and accurate calculations. 
So far, we have achieved certain successes in simulating the 
coalescence of binary neutron stars (see e.g. \cite{binaryNS})
and binary black holes 
(see e.g.\cite{binaryBH}). 
However, people have still been faced with unreasonable numerical
blow-ups at the end of simulations. 

Difficulties in accurate/stable long-term
evolution were supposed to be overcome by choosing proper gauge
conditions and boundary conditions.  
However, recent several
numerical experiments show
that the (standard) Arnowitt-Deser-Misner (ADM) approach
\cite{ADM,ADM-SmarrYork,ADM-York}
is not the best formulation for numerics, and
finding a better formulation
has become one of the main research topics.
A majority of workers in the field now 
believe in the existence of constraint-violating modes
in most of the formulations. 
Thus, the stability problem is 
now shedding light on the
mathematical structure of the Einstein equations.

The purpose of this article is to review the formulation problem in numerical
relativity.  
Generally speaking, there are many open issues in numerical relativity, both 
theoretical (mathematical or physical) and numerical. We list major 
topics in Box 1.2.  
More general and recent introductions to numerical
relativity are available, e.g. by 
d'Inverno (1996) \cite{reviewdInverno},
Seidel (1996/98/99) \cite{reviewSeidel}, 
Br\"ugmann (2000) \cite{reviewBruegmann}, 
Lehner (2001) \cite{reviewLehner}, 
van Putten (2001) \cite{reviewPutten}, and 
Baumgarte-Shapiro (2002) \cite{reviewBS}. 
\newlength{\listlength}
\settowidth{\listlength}{Theoretical:}
\Largefbox{\boxwidth}{
{\bf Numerical Relativity -- open issues} \hspace*{\fill} {\bf Box 1.2} 
 \baselineskip 11pt
\been 
\item[0.] How to select the foliation method of space-time \\
     $~\qquad~\qquad~\quad$Cauchy ($3+1$), characteristic ($2+2$), or combined? 
     \hspace*{\fill}
\enen
$\Rightarrow$ if the foliation is $(3+1)$, then $\cdots$  
\been 
\item How to prepare the initial data  
     \begin{list}{}{%
\setlength{\leftmargin}{\listlength}
\addtolength{\leftmargin}{\labelsep}
\setlength{\labelwidth}{\listlength}
}
     \item[Theoretical:\hfill] 
     Proper formulation for solving constraints? \\
     How to prepare realistic initial data?  \\
     Effects of background gravitational waves?  \\
     Connection to the post-Newtonian approximation?
     \item[Numerical:\hfill] 
      Techniques for solving coupled elliptic equations?\\
      Appropriate boundary conditions? 
     \end{list}
\item How to evolve the data  
     \begin{list}{}{%
\setlength{\leftmargin}{\listlength}
\addtolength{\leftmargin}{\labelsep}
\setlength{\labelwidth}{\listlength}
}
     \item[Theoretical:\hfill] 
      Free evolution or constrained evolution? \\
     \underline{Proper formulation for the evolution equations?}
     \hspace*{\fill}$\Leftarrow\Leftarrow\Leftarrow$ this review \\
      Suitable slicing conditions (gauge conditions)?
     \item[Numerical:\hfill] 
      Techniques for solving the evolution equations?\\
      Appropriate boundary treatments?  \\
      Singularity excision techniques?  \\
      Matter and shock surface treatments?\\
      Parallelization of the code?
     \end{list}
\item How to extract the physical information
     \begin{list}{}{%
\setlength{\leftmargin}{\listlength}
\addtolength{\leftmargin}{\labelsep}
\setlength{\labelwidth}{\listlength}
}
     \item[Theoretical:\hfill] 
      Gravitational wave extraction?\\
      Connection to other approximations?
     \item[Numerical:\hfill] 
      Identification of black hole horizons?  \\
      Visualization of simulations?
     \end{list}
\enen 
}

\subsection{Formulation Problem in Numerical Relativity: Overview}

There are several different approaches to simulating the Einstein equations.  
Among them the most robust way is to apply 3+1  (space + time) decomposition 
of space-time, as was first formulated by Arnowitt, Deser and 
Misner (ADM) \cite{ADM} (we call this  the ``original ADM" system). 
\footnote{
One alternative method of space-time foliation is the so-called 
characteristic approach ($2+2$ space-time
decomposition). 
See reviews e.g. by d'Inverno (1996) \cite{reviewdInverno}, 
Winicour \cite{reviewWinicour}, Lehner (2001) \cite{reviewLehner}. 
Even in the 3+1 ADM approach, we concentrate the standard
finite differential scheme to express numerical expression of space-time. 
See e.g. Brewin \cite{brewin} for a recent progress in a lattice method.}

If we divide the space-time into 3+1 dimensions, the Einstein equations 
form a constrained system: constraint equations and evolution equations. 
The system is quite similar to that of the Maxwell equations (Box 1.3), 
\largefbox{\boxwidth}{
{\bf The Maxwell equations :} \hspace*{\fill} {\bf Box 1.3} \\
\baselineskip 11pt
The evolution equations:  ($\ptl_t=\ptl / \ptl t$)
\bear
{\ptl_t {\bf E}}
&=& \mbox{rot~} {\bf B} 
-{4\pi } {\bf j}, 
\qquad \mbox{and}\qquad
{\ptl_t {\bf B} }
= - \mbox{rot~}  {\bf E} 
\enar
Constraint equations:
\bear
\mbox{div~} {\bf E} &=& 4 \pi \rho, 
\qquad \mbox{and}\qquad
\mbox{div~}{\bf B} = 0 
\enar
}
where people solve constraint equations on the initial data, 
and use evolution equations to follow the dynamical behaviors. 

In numerical relativity, this free-evolution approach is also the standard. 
This is because solving the constraints (non-linear elliptic equations) is
numerically expensive, 
and because free evolution allows us to monitor the accuracy of
numerical evolution.  In black-hole treatments, recent
``excision" techniques do not require one to impose explicit boundary 
conditions on the horizon, which is also a reason to apply free 
evolution scheme. 
As we will show in the next section, 
the standard ADM approach has two constraint
equations; the Hamiltonian (or energy) and momentum constraints. 

\begin{figure}[t]
\unitlength 1mm 
\begin{picture}(160,70)
\put(00,0){\epsfxsize=75mm \epsffile{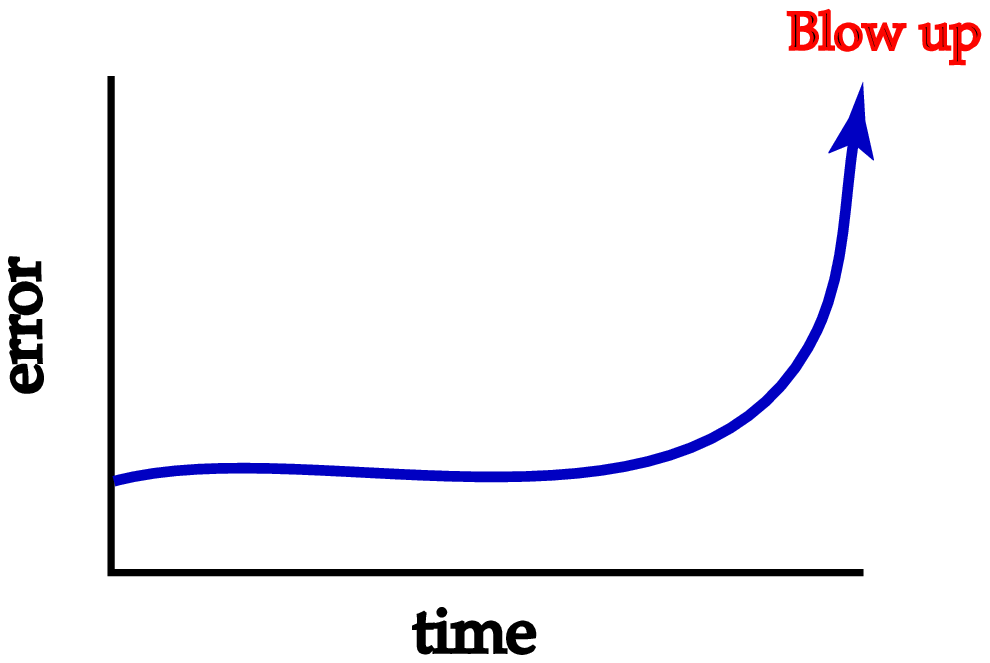} }
\put(90,0){\epsfxsize=75mm \epsffile{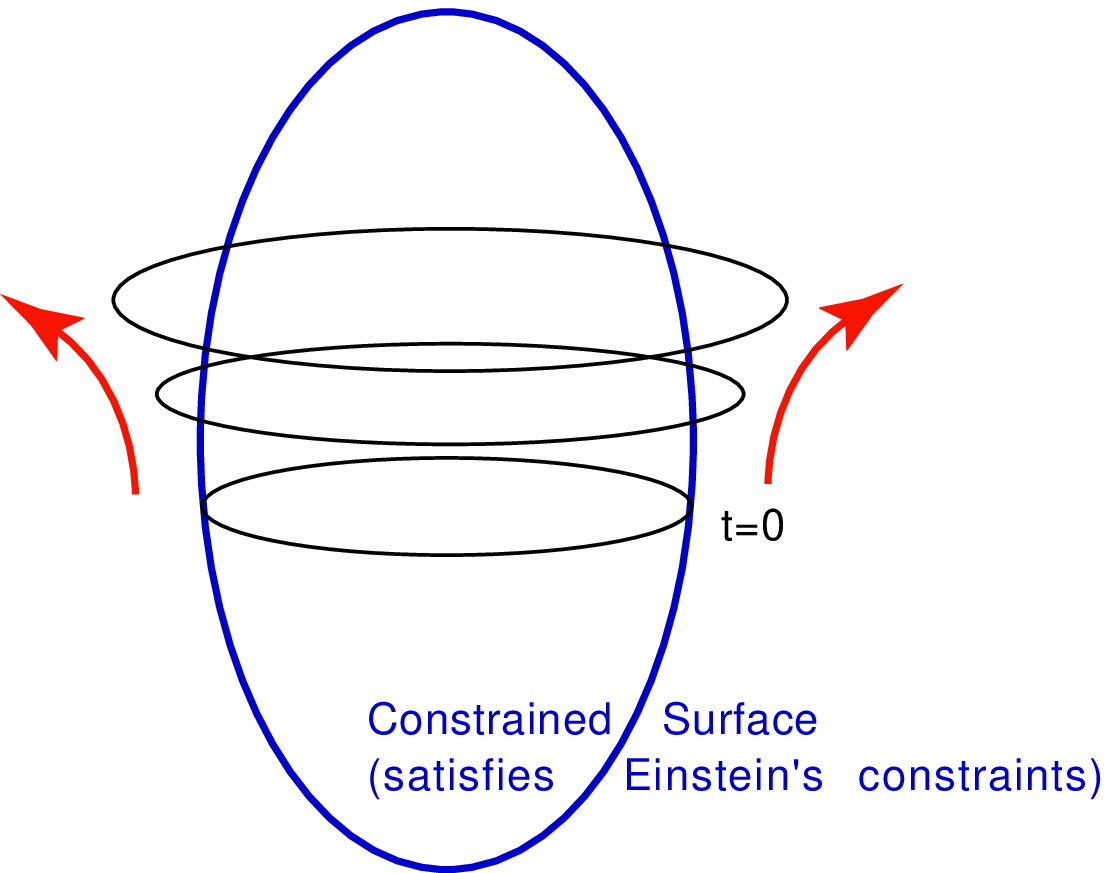} }
\end{picture}
\caption{Origin of the problem for numerical relativists: 
Numerical evolutions 
depart from the constraint surface. } \label{fig1}
\end{figure}

Up to a couple of years ago, the ``standard ADM"
decomposition  \cite{ADM-SmarrYork,ADM-York} of the Einstein
equation was taken as the standard formulation for numerical relativists.
However, numerical simulations were often interrupted by unexplained blow-ups
(Figure.\ref{fig1}).  This was thought due to the lack of resolution, or  
inappropriate  gauge choice, or the particular numerical scheme which was
applied.   However, after the accumulation of much experience,  people have 
noticed the importance of the formulation of the evolution equations,
since there are  apparent differences in numerical 
stability although the equations are mathematically equivalent 
\footnote{The word {\it stability} is used quite different ways in the community. 
\beit
\item 
We mean by {\it numerical stability} a numerical simulation which continues
without any blow-ups and in which data remains on the constrained surface.  
\item {\it Mathematical stability} is defined in terms of the well-posedness
in the theory of partial differential equations, such that 
the norm of the variables is bounded by the initial
data. See eq. (\ref{energynorm}) and around.  
\item For numerical treatments, there is also another notion of 
{\it stability}, the stability
of finite differencing schemes.  This means that
numerical errors (truncation, round-off, etc) 
are not growing by evolution, and the evaluation is obtained by   
von Neumann's analysis. 
Lax's equivalence theorem says that if a numerical scheme is consistent
(converging to  the original equations in its continuum limit) and stable (no error
growing), then the simulation represents the right (converging) solution.  
See \cite{Choptuik91} for the Einstein equations.  
\enit
}. 

At this moment, 
there are three major ways to obtain longer time evolutions. 
Of course, the ideas, procedures, and problems are mingled with each other. 
The purpose of this article is to review all three approaches and to 
introduce our idea to view them in a unified way. 
Table \ref{table:hyprefs} is a list of references. 

\begin{enumerate}
\item[(1)]
The first possibility is to use a modification of the ADM system developed
by the Kyoto group \cite{SN87,SN89} (often cited as Shibata and Nakamura
\cite{SN}) and later re-introduced by 
Baumgarte and Shapiro \cite{BS}. 
This is a combination of new variables, conformal
decomposition, rescaling of the conformal factor, 
and replacement of terms in the evolution equation using
momentum  constraints  (see \S \ref{secBSSN}). 
\item[(2)]
The second direction is to re-formulate the Einstein equations in a first-order 
hyperbolic form. 
This is motivated from the
expectation that the symmetric hyperbolic system has well-posed properties 
in its Cauchy
treatment in many systems and also that the boundary treatment can be 
improved if we
know the characteristic speed of the system.  In constructing
hyperbolic systems, the essential  procedures are to adjust equations
using constraints and to introduce new variables, normally the spatially 
derivatived metric  (see \S \ref{secHYP}). 
\item[(3)]
The third is to construct a system which is robust against 
the violation of constraints,
such that the constraint surface is an attractor.  
The idea was first proposed as a ``$\lambda$-system" 
by Brodbeck et al \cite{BFHR} 
in which they introduce artificial flow to the constraint surface 
using a new variable 
based on the symmetric hyperbolic system (see \S \ref{secASYMPT}).
\end{enumerate}
The third idea has been generalized by us as an asymptotically constrained
system.  The main procedure is to adjust the evolution equations using 
the constraint equations \cite{ronbun2,adjADM,adjADMsch}.  
The method is also applied to explain why the above approach (1) works, 
and also to 
propose alternative systems based on the ADM \cite{adjADM,adjADMsch} and 
BSSN \cite{adjBSSN} equations. 
Section \ref{secADJUSTED} is devoted to 
explain this idea with an analytical tool of the eigenvalue analysis of the
constraint propagation. 

We follow the notations of that of MTW\cite{MTW}, i.e. the signature of the
space-time is $(-+++)$, and the Riemann curvature is defined as 
\bear
 R^{\mu}_{~\nu\alpha\beta}&\equiv&
\partial_\alpha \Gamma^\mu_{\nu\beta}-
\partial_\beta \Gamma^\mu_{\nu\alpha}
+\Gamma^\mu_{~\sigma\alpha}\Gamma^\sigma_{~\nu\beta}
-\Gamma^\mu_{~\sigma\beta}\Gamma^\sigma_{~\nu\alpha} 
\\
R_{\mu\nu}&\equiv& R^{\alpha}_{~\mu\alpha\nu}
\enar
We use
$\mu,\nu=0,\cdots,3$ and
$i,j=1,\cdots,3$ as space-time indices. The unit $c=1$ is applied.  
The discussion is mostly to the vacuum space-time, but the inclusion of matter is
straightforward.   


\begin{table}
\begin{center}
\begin{tabular}{cl|l|l}
\hline
\hline
& & formulations  & numerical applications
\\
\hline
\hline
\multicolumn{4}{l}{(0) The standard ADM formulation}
\\
\hline
&ADM & 1962 Arnowitt-Deser-Misner \cite{ADM,ADM-SmarrYork} & $\Rightarrow$ many
\\
\hline
\multicolumn{4}{l}{(1) The BSSN formulation}
\\
\hline
&BSSN & 1987 Nakamura et al \cite{SN87,SN89,SN} & 
$\Rightarrow$ 1987 Nakamura et al \cite{SN87,SN89}
\\
&&& 
$\Rightarrow$ 1995 Shibata-Nakamura  \cite{SN}
\\
&&& 
$\Rightarrow$ 2002 Shibata-Uryu \cite{binaryNS} etc
\\
&& 1999 Baumgarte-Shapiro \cite{BS} & 
$\Rightarrow$ 1999 Baumgarte-Shapiro \cite{BS}
\\
&&& $\Rightarrow$ 2000 Alcubierre et al \cite{potsdam9908,potsdam0003}
\\
&&& $\Rightarrow$ 2001 Alcubierre et al \cite{binaryBH} etc
\\
&& 1999 Alcubierre et al \cite{ABMS} &
\\
&& 1999  Frittelli-Reula \cite{FR99}   & 
\\
&& 2002 Laguna-Shoemaker \cite{PabloDeirdre} &
$\Rightarrow$ 2002 Laguna-Shoemaker \cite{PabloDeirdre}
\\
\hline
\multicolumn{4}{l}{(2) The hyperbolic formulations}
\\
\hline
&BM & 1989 Bona-Mass\'o \cite{BM89,BM92,BMSS95} & 
$\Rightarrow$ 1995 Bona et al \cite{BMSS95,BMSS97,cactus1}
\\
&&&
$\Rightarrow$ 1997 Alcubierre, Mass\'o \cite{Alcubierre,AM}
\\
&&   1997 Bona et al \cite{BMSS97} &
$\Rightarrow$ 2002 Bardeen-Buchman \cite{BB}
\\
&& 1999 Arbona et al \cite{ArBona}  & 
\\
&CB-Y & 1995 Choquet-Bruhat and York \cite{CY9506071} &
$\Rightarrow$ 1997 Scheel et al \cite{SBCSThyper}
\\
&  & 1995 Abrahams et al \cite{CY9506072}  &
$\Rightarrow$ 1998 Scheel et al \cite{SBCST98}
\\
&  & 1999 Anderson-York \cite{AY}  &
$\Rightarrow$ 2002 Bardeen-Buchman \cite{BB}
\\
&FR & 1996 Frittelli-Reula \cite{FR96} & 
$\Rightarrow$ 2000 Hern \cite{HernPHD} 
\\
&& 1996 Stewart \cite{Stewart} & 
\\
&KST & 2001 Kidder-Scheel-Teukolsky \cite{KST} & 
$\Rightarrow$ 2001 Kidder-Scheel-Teukolsky \cite{KST}
\\
&& & $\Rightarrow$ 2002 Calabrese et al \cite{LSU-KST}
\\
&& & $\Rightarrow$ 2002 Lindblom-Scheel \cite{LindblomScheel} 
\\
&& 2002 Sarbach-Tiglio \cite{SarbachTiglio} & 
\\
& CFE & 1981 Friedrich\cite{FriedrichCFE}  & 
$\Rightarrow$ 1998 Frauendiener \cite{Frauendiener}
\\
&&&
$\Rightarrow$ 1999 H\"ubner \cite{Hubner}
\\
& tetrad & 1995 vanPutten-Eardley\cite{vanPutten95}  & 
$\Rightarrow$ 1997 vanPutten \cite{vanPutten97}
\\
&Ashtekar & 1986 Ashtekar \cite{Ashtekar} &
$\Rightarrow$ 2000 Shinkai-Yoneda \cite{ronbun1}
\\
&&1997 Iriondo et al \cite{Iriondo} & 
\\
&&1999 Yoneda-Shinkai \cite{ysPRL,ysIJMPD} & 
$\Rightarrow$ 2000 Shinkai-Yoneda \cite{ronbun1,ronbun2}
\\
\hline
\multicolumn{4}{l}{(3) Asymptotically constrained formulations}
\\
\hline
$\lambda$-system & to FR & 1999 Brodbeck et al \cite{BFHR} & 
$\Rightarrow$ 2001 Siebel-H\"ubner \cite{SiebelHuebner}
\\ 
   & to Ashtekar  & 1999 Shinkai-Yoneda \cite{SY-asympAsh} &
$\Rightarrow$ 2001 Yoneda-Shinkai \cite{ronbun2}
\\ 
  adjusted  & to ADM & 1987 Detweiler \cite{detweiler}  &
$\Rightarrow$ 2001 Yoneda-Shinkai \cite{adjADM}
\\ 
   & to ADM & 2001 Shinkai-Yoneda \cite{adjADM,adjADMsch} &
$\Rightarrow$ 2002 Mexico NR Workshop \cite{mexico}
\\ 
   & to BSSN & 2002 Yoneda-Shinkai \cite{adjBSSN} &
$\Rightarrow$ 2002 Mexico NR Workshop \cite{mexico}
\\ 
   & & &
$\Rightarrow$ 2002 Yo-Baumgarte-Shapiro \cite{YBS}
\\ 
\hline
\end{tabular}
\end{center}
\caption{References to recent efforts of reformulating the Einstein equations.
We list mainly those that have been applied to actual numerical comparisons.}
\label{table:hyprefs}
\end{table}

\newpage
\setcounter{equation}{0}
\section{The standard way and the three other roads}\label{sec2}
\setcounter{equation}{0}
\setcounter{subsection}{-1}
\subsection{Strategy 0: The ADM formulation}\label{secADM}

\subsubsection{The original ADM formulation}
The idea of space-time evolution was first formulated by 
Arnowitt, Deser, and Misner (ADM) \cite{ADM}.  
The formulation was first motivated by a desire to construct a canonical 
framework in general relativity, but it also gave the community to 
the fundamental idea of time evolution of space and time: such as 
foliations of 3-dimensional hypersurface (Figure \ref{fig:ADMfoliation}). 
This {\it original} ADM formulation was translated to numerical relativists
by Smarr \cite{ADM-SmarrYork} and York \cite{ADM-York} in late 70s, with slightly 
different notations.  We refer to the latter as the {\it standard}
 ADM formulation since this version is the starting point of the discussion.

The story begins by decomposing 4-dimensional space-time into 3 plus 1. 
The metric is expressed by 
\begin{equation}
ds^2= g_{\mu\nu} dx^\mu dx^\nu = - \alpha^2 dt^2+ 
\gamma_{ij}(dx^i + \beta^i dt)(dx^j + \beta^j dt), 
\end{equation}
where $\alpha$ and $\beta_j$ are defined as 
$
\alpha \equiv 1 / \sqrt{-g^{00}}$ and $\beta_j \equiv g_{0j}, 
$
and called the lapse function and shift vector, respectively. 
The projection operator or the intrinsic 3-metric $g_{ij}$ is 
defined as 
$\gamma_{\mu\nu}=g_{\mu\nu}+n_\mu n_\nu$, where
$n_\mu=(-\alpha, 0,0,0)$ [and $n^\mu=g^{\mu\nu}n_\nu=
(1/\alpha, -\beta^i/\alpha)$] is the unit normal vector of the spacelike 
hypersurface, $\Sigma$ (see Figure \ref{fig:ADMfoliation}). 
By introducing the extrinsic curvature,  
\bear
K_{ij}&=& - \half \pounds_n \gamma_{ij}, 
\enar
and using the Gauss-Codacci relation, 
the Hamiltonian density of the Einstein equations can be written as 
\bear
{\cal H}_{GR} &=& \pi^{ij}\dot{\gamma}_{ij} - {\cal L}, \quad
\mbox{where}\quad 
{\cal L}=\sqrt{-g}R=\alpha \sqrt{\gamma}[{}^{(3)\!}R - K^2 + K_{ij}K^{ij}], 
\enar
where $\pi^{ij}$ is the canonically conjugate momentum to $\gamma_{ij}$,  
\be
\pi^{ij}={\partial {\cal L} \over \partial \dot{\gamma}_{ij}}
= - \sqrt{\gamma} (K^{ij} - K \gamma^{ij}), 
\en
omitting the boundary terms. 
The variation of ${\cal H}_{GR}$ with
respect to $\alpha$ and $\beta_i$ yields the constraints, and 
the dynamical equations are given by 
$\dot{\gamma}_{ij} = {\delta {\cal H}_{GR} \over \delta \pi^{ij} } $ and 
$\dot{\pi}^{ij} = -{\delta {\cal H}_{GR} \over \delta h_{ij} }$.

\begin{figure}[b]
\unitlength 1mm 
\begin{picture}(160,50)
\end{picture}
\caption{Concept of time evolution of space-time: foliations of  
3-dimensional hypersurface. The lapse and shift functions are often denoted
$\alpha$ or $N$, and $\beta^i$ or $N^i$, respectively. (This figure is
missing in gr-qc version due to the limitation of the file size.)}
\label{fig:ADMfoliation}
\end{figure}

\subsubsection{The standard ADM formulation}
In the version of Smarr and York, $K_{ij}$ was used as a fundamental variable
instead of the conjugate momentum $\pi^{ij}$ (see also the footnote \footnote{
We remark that there is one replacement in (\ref{adm_evo2}) using (\ref{admCH})
in the process of conversion from
the original ADM to the standard ADM equations. 
This is the key issue in the later discussion, and we shall be back this point 
in \S \ref{secADJADM}.
}).

\Largefbox{\boxwidth}{
{\bf The Standard ADM formulation \cite{ADM-SmarrYork,ADM-York}:} 
\hspace*{\fill} {\bf Box 2.1}\\
The fundamental dynamical variables are $(\gamma_{ij}, K_{ij})$, 
the three-metric and extrinsic curvature.  
The three-hypersurface $\Sigma$ is foliated with gauge functions, 
($\alpha, \beta^i$), the lapse and shift vector. 
\begin{itemize}
\item 
The evolution equations:
\bera
{\partial_t} \gamma_{ij} &=& 
-2\alpha K_{ij}+D_i\beta_j+D_j\beta_i, 
 \label{adm_evo1}
 \\
{\partial_t} K_{ij} &=& 
\alpha ~^{(3)\!} R_{ij}+\alpha K K_{ij}-2\alpha K_{ik}{K^k}_j 
-D_iD_j \alpha 
\nonumber \\&& 
+(D_i \beta^k) K_{kj} +(D_j \beta^k) K_{ki} 
+\beta^k D_k K_{ij} 
\label{adm_evo2}
\enra
where $K=K^i{}_i$, and $~^{(3)\!}R_{ij}$ and $D_i$ denote
 three-dimensional Ricci curvature, 
and a covariant derivative on the three-surface, respectively.  
\item
Constraint equations:
\bera
{\cal H}^{ADM} &:=&
~^{(3)\!}R+ K^2 -K_{ij}K^{ij} \approx 0,
\label{admCH} \\
{\cal M}^{ADM}_i &:=&
D_j K^j{}_i-D_i K  
\approx 0,
\label{admCM}
\enra
where $~^{(3)\!}R=^{(3)\!}R^i{}_i$: these
are called the Hamiltonian (or energy) and momentum
constraint equations, respectively. 
\end{itemize}
}
The formulation has 12 free first-order dynamical variables 
($\gamma_{ij}, K_{ij}$), with 4 freedom of gauge choice ($\alpha, \beta_i$)
 and with 4 constraint equations,  (\ref{admCH}) and (\ref{admCM}). 
The rest freedom expresses 2 modes of gravitational waves. 

The constraint propagation equations,
which are the time evolution equations
of the Hamiltonian constraint (\ref{admCH}) and
the momentum constraints (\ref{admCM}), can be written as
\Largefbox{\boxwidth}{
{\bf The Constraint Propagations of the Standard ADM:}
\hspace*{\fill} {\bf Box 2.2}\\
\begin{eqnarray}
\partial_t {\cal H}&=&
\beta^j (\partial_j {\cal H})
+2\alpha K{\cal H}
-2\alpha \gamma^{ij}(\partial_i {\cal M}_j)
\nonumber \\ &&
+\alpha(\partial_l \gamma_{mk})
   (2\gamma^{ml}\gamma^{kj}-\gamma^{mk}\gamma^{lj}){\cal M}_j
-4\gamma^{ij} (\partial_j\alpha){\cal M}_i,
\label{CHproADM}
\\
\partial_t {\cal M}_i&=&
-(1/2)\alpha (\partial_i {\cal H})
-(\partial_i\alpha){\cal H}
+\beta^j (\partial_j {\cal M}_i)
\nonumber \\ &&
+\alpha K {\cal M}_i
-\beta^k\gamma^{jl}(\partial_i\gamma_{lk}){\cal M}_j
+(\partial_i\beta_k)\gamma^{kj}{\cal M}_j.
\label{CMproADM}
\end{eqnarray}
Further expressions of these constraint propagations are 
in Appendix \ref{appADMconpro}. 
}
From these equations, we know that  
{\it if the constraints are satisfied 
on the initial slice $\Sigma$, then 
the constraints are satisfied throughout evolution}.
The normal numerical scheme is to solve the elliptic constraints
for preparing the initial data, and to apply the free evolution (solving only 
the evolution equations).  The constraints are used to monitor the accuracy of 
simulations.

\subsubsection{Remarks}
The ADM formulation was the standard formulation for numerical relativity 
up to the middle 90s.  
Numerous successful simulations were obtained for the problems of
gravitational collapse, critical behavior, cosmology, and so on.  
However,  stability problems have arisen for the
simulations such as the
gravitational radiation from compact binary coalescence, because the
models require quite a long-term time evolution. 

The origin of the problem was that the above  statement in {\it Italics}
is true in principle, but is not always true in numerical applications. 
A long history of trial  and error  began in the early 90s. 
From the next subsection we shall 
look back of them by summarizing  ``three strategies".
We then unify these three roads as ``adjusted systems", and as its by-product 
we show in \S \ref{secADJADM} that the standard ADM equations 
{\it has} a  
constraint violating mode in its constraint propagation equations
even for a single black-hole (Schwarzschild) spacetime \cite{adjADMsch}. 

\subsection{Strategy 1: Modified ADM formulation by Nakamura et al 
(The BSSN formulation)} \label{secBSSN}

\subsubsection{Introduction}
Up to now, the most widely used formulation 
for large scale numerical simulations 
is 
a modified ADM system, which is now often cited as the 
Baumgarte-Shapiro-Shibata-Nakamura (BSSN) formulation. 
This reformulation was first introduced by Nakamura {\it et al.} 
\cite{SN87,SN89,SN}. 
The usefulness of this reformulation was re-introduced by 
Baumgarte and Shapiro \cite{BS}, then was 
confirmed by another group to show a long-term
stable numerical evolution 
\cite{potsdam9908,potsdam0003}.
The procedure is to apply conformally decomposition of the ADM variables
and to implement their dynamical equations with several replacements. 

\subsubsection{Basic variables and equations}
The widely used notation\cite{BS} introduces the variables 
($\varphi,\tilde{\gamma}_{ij}$,$K$,$\tilde{A}_{ij}$,$\tilde{\Gamma}^i$) 
instead of ($\gamma_{ij}$,$K_{ij}$), where
\bera
\varphi &=& (1/12)\log ({\rm det}\gamma_{ij}), 
\label{BSSNval1} 
\\ 
\tilde{\gamma}_{ij} & = & e^{-4\varphi}\gamma_{ij}, 
\label{BSSNval2} 
\\ 
K  &=& \gamma^{ij}K_{ij}, \label{BSSNval3} 
\\ 
\tilde{A}_{ij} &=& 
e^{-4\varphi}(K_{ij} - (1/3)\gamma_{ij}K), \label{BSSNval4} 
\\ 
\tilde{\Gamma}^i &=& 
\tilde{\Gamma}^i_{jk}\tilde{\gamma}^{jk}.
\label{BSSNval5} 
\enra
The new variable $\tilde{\Gamma}^i$ was introduced in order to 
calculate Ricci curvature more accurately.  $\tilde{\Gamma}^i$ also 
contributes to making system re-produce wave equations in its 
linear limit. 

In BSSN formulation, Ricci curvature is not calculated as 
\be 
R^{ADM}_{ij} 
= 
\partial_k \Gamma^k_{ij} 
-\partial_i\Gamma^k_{kj} 
+\Gamma^l_{ij}\Gamma^k_{lk} 
-\Gamma^l_{kj}\Gamma^k_{li}, 
\end{equation} 
but 
\bera
R^{BSSN}_{ij} 
&=& 
R^\varphi_{ij}+\tilde R_{ij}, 
\label{BSricci} 
\\ 
R^\varphi_{ij} 
&=& 
-2\tilde{D}_i\tilde{D}_j\varphi 
-2\tilde{\gamma}_{ij}\tilde{D}^k\tilde{D}_k\varphi 
+4(\tilde{D}_i\varphi)(\tilde{D}_j\varphi) 
-4\tilde{\gamma}_{ij}(\tilde{D}^k\varphi)(\tilde{D}_k\varphi), 
\\ 
\tilde{R}_{ij} 
&=& 
-(1/2)\tilde{\gamma}^{lk}\partial_{l}\partial_{k}\tilde{\gamma}_{ij} 
+\tilde{\gamma}_{k(i}\partial_{j)}\tilde{\Gamma}^k 
+\tilde{\Gamma}^k\tilde{\Gamma}_{(ij)k} 
+2\tilde{\gamma}^{lm}\tilde{\Gamma}^k_{l(i}\tilde{\Gamma}_{j)km} 
+\tilde{\gamma}^{lm}\tilde{\Gamma}^k_{im}\tilde{\Gamma}_{klj}, 
\enra
where $\tilde{D}_i$ is covariant derivative associated 
with $\tilde{\gamma}_{ij}$. 
These are approximately equivalent, but $R^{BSSN}_{ij}$ does have 
wave operator apparently in the flat background limit, so that 
we can expect more natural wave propagation behavior. 

Additionally, the BSSN requires us to impose the conformal factor as 
\begin{equation} 
\tilde{\gamma}(:={\rm det} \tilde{\gamma}_{ij})=1, \label{scalingdef} 
\end{equation} 
during evolution.  This is a kind of definition, but can also be 
treated as a constraint. 


\Largefbox{\boxwidth}{
{\bf The BSSN formulation \cite{SN87,SN89,SN,BS}:}
\hspace*{\fill} {\bf Box 2.3}\\
The fundamental dynamical variables are
($\varphi,\tilde{\gamma}_{ij}$,$K$,$\tilde{A}_{ij}$,$\tilde{\Gamma}^i$).
\\
The three-hypersurface $\Sigma$ is foliated with gauge functions, 
($\alpha, \beta^i$), the lapse and shift vector. 
\begin{itemize}
\item 
The evolution equations:
\bera
\partial_t^B \varphi 
&=& 
-(1/6)\alpha K+(1/6)\beta^i(\partial_i\varphi)+(\partial_i\beta^i), 
\label{BSSNeqmPHI} 
\\ 
\partial_t^B \tilde{\gamma}_{ij} 
&=& 
-2\alpha\tilde{A}_{ij} 
+\tilde{\gamma}_{ik}(\partial_j\beta^k) 
+\tilde{\gamma}_{jk}(\partial_i\beta^k) 
-(2/3)\tilde{\gamma}_{ij}(\partial_k\beta^k) 
+\beta^k(\partial_k\tilde{\gamma}_{ij}), 
\label{BSSNeqmtgamma} 
\\ 
\partial_t^B K 
&=& 
-D^iD_i\alpha 
+\alpha \tilde{A}_{ij}\tilde{A}^{ij} 
+(1/3) \alpha K^2 
+\beta^i (\partial_i K), 
\label{BSSNeqmK} 
\\ 
\partial_t^B \tilde{A}_{ij} 
&=& 
-e^{-4\varphi}(D_iD_j\alpha)^{TF} 
+e^{-4\varphi} \alpha (R^{BSSN}_{ij})^{TF} 
+\alpha K\tilde{A}_{ij} 
-2\alpha \tilde{A}_{ik}\tilde{A}^k{}_j 
\nonumber \\&& 
+(\partial_i\beta^k)\tilde{A}_{kj} 
+(\partial_j\beta^k)\tilde{A}_{ki} 
-(2/3)(\partial_k\beta^k)\tilde{A}_{ij} 
+\beta^k(\partial_k \tilde{A}_{ij}), 
\label{BSSNeqmTA} 
\\ 
\partial_t^B \tilde{\Gamma}^i &=& 
-2(\partial_j\alpha)\tilde{A}^{ij} 
+2\alpha 
\big(\tilde{\Gamma}^i_{jk}\tilde{A}^{kj} 
-(2/3)\tilde{\gamma}^{ij}(\partial_j K) 
+6\tilde{A}^{ij}(\partial_j\varphi) 
\big) 
\nonumber \\&& 
-\partial_j 
\big( \beta^k(\partial_k\tilde{\gamma}^{ij}) 
-\tilde{\gamma}^{kj}(\partial_k\beta^{i}) 
-\tilde{\gamma}^{ki}(\partial_k\beta^{j}) 
+(2/3)\tilde{\gamma}^{ij}(\partial_k\beta^k) 
\big). 
\label{BSSNeqmTG} 
\enra
\item
Constraint equations:
\bera
{\cal H}^{BSSN} 
&=& 
R^{BSSN}+K^2-K_{ij}K^{ij}, 
\label{BSSNconstraintH} 
\\ 
{\cal M}^{BSSN}_i 
&=& 
{\cal M}^{ADM}_i,  \label{BSSNconstraintM} 
\\
{\cal G}^i &=& \tilde{\Gamma}^i-\tilde{\gamma}^{jk} 
\tilde{\Gamma}^i_{jk}, \label{BSSNconstraintG} 
\\ 
{\cal A}&=&\tilde{A}_{ij}\tilde{\gamma}^{ij}, \label{BSSNconstraintA} 
\\ 
{\cal S} &=& 
\tilde{\gamma}-1.
\label{BSSNconstraintS} 
\enra
(\ref{BSSNconstraintH}) and (\ref{BSSNconstraintM}) are 
the Hamiltonian and momentum constraints 
(the ``kinematic" constraints), while the latter three are 
 ``algebraic" constraints due to the requirements of BSSN formulation.
\end{itemize}
}

Hereafter we will write ${\cal H}^{BSSN}$ and ${\cal M}^{BSSN}$ 
simply as ${\cal H}$ and ${\cal M}$ respectively. 

Taking careful account of these constraints, 
(\ref{BSSNconstraintH}) and (\ref{BSSNconstraintM}) can be expressed 
directly as 
\begin{eqnarray} 
{\cal H} &=& 
e^{-4\varphi}\tilde{R} 
-8e^{-4\varphi}\tilde{D}^j\tilde{D}_j\varphi 
-8e^{-4\varphi}(\tilde{D}^j\varphi)(\tilde{D}_j\varphi) 
+(2/3)K^2 
-\tilde{A}_{ij}\tilde{A}^{ij} 
-(2/3) {\cal A} K, 
\label{BSSNconstraint1} 
\\ 
{\cal M}_i 
&=& 
6\tilde{A}^j{}_{i}(\tilde{D}_j \varphi) 
-2{\cal A}(\tilde{D}_i \varphi) 
-(2/3) (\tilde{D}_i K) 
+\tilde{\gamma}^{kj}(\tilde{D}_j\tilde{A}_{ki}). 
\label{BSSNconstraint2} 
\end{eqnarray} 

In summary, the fundamental dynamical variables in BSSN are 
($\varphi,\tilde{\gamma}_{ij}$, 
$K$,$\tilde{A}_{ij}$,$\tilde{\Gamma}^i$), 
 17 in all. 
The gauge quantities are ($\alpha, \beta^i$) of which there are 4, and the 
constraints are 
$({\cal H},{\cal M}_i,{\cal G}^i,{\cal A},{\cal S})$, 
i.e. 9 components. 
As a result, 4 (2 by 2) 
components are left which correspond to 
two gravitational polarization modes.

\subsubsection{Remarks}
Why is BSSN better than the standard ADM? 
Together with numerical comparisons with the 
standard ADM case\cite{potsdam0003,LHG}, 
this question has been studied by many groups using different approaches. 
Using numerical test evolution, 
Alcubierre et al \cite{potsdam9908} found that 
the essential improvement is in 
the process of replacing terms by the momentum constraints. 
They also pointed out that 
the eigenvalues of BSSN {\it evolution equations} have fewer 
``zero eigenvalues" 
than those of ADM, and they conjectured that the instability 
might be caused by these 
``zero eigenvalues". 
Miller \cite{Miller} applied von Neumann's stability analysis 
to the plane wave propagation, and reported that BSSN has a 
wider range of 
parameters, which produces stable evolution.  
Analogical conformal decomposition of the Maxwell equations are also
reported \cite{KWB}.
An effort was made to understand the advantage of BSSN from the point
of hyperbolization of the equations in its linearized
limit \cite{potsdam9908,LSU-BSSN}.  
These studies provide some support  regarding
the advantage of BSSN, while it is also shown 
an example of an ill-posed solution in BSSN (as well in ADM) 
by Frittelli and Gomez \cite{FrittelliGomez}. 
(Inspired by the BSSN's conformal decomposition, 
several related hyperbolic formulations have 
also been proposed \cite{ArBona,FR99,ABMS}.) 

As we shall discuss in \S \ref{secADJBSSN}, the stability of the
BSSN formulation is due not only to the introductions of 
new variables, but also to the replacement of terms in the evolution 
equations using the constraints.  Further, we will show several 
additional adjustments to the BSSN equations which are expected to
give us more stable numerical simulations. 

Recently, Laguna and Shoemaker \cite{PabloDeirdre} modified the
BSSN slightly, and found a great improvement in simulating a 
Schwarzschild black-hole. 
\Largefbox{\boxwidth}{
{\bf The Laguna-Shoemaker version of BSSN \cite{PabloDeirdre}:}
\hspace*{\fill} {\bf Box 2.4}\\
Modifications to BSSN are
\beit
\item to introduce conformal scalings also to $K, A_{ij}$ and $N$ as 
$$
{}^{(LS)}\hat{K} = e^{6\,n\,\varphi}K, \qquad
{}^{(LS)}\hat{A}^i\,_j   = e^{6\,n\,\varphi}A^i\,_j, \qquad
N = e^{-6\,n\,\varphi}\alpha. 
$$
where $\hat \gamma_{ij} = e^{- 4 \varphi}\, \gamma_{ij}$ and 
$n$ is a parameter ($n=0$ recovers the BSSN variables)
\item to use a mixed indices form ${}^{(LS)}\hat{A}^i\,_j $ rather than $A_{ij}$.  
\item to use a densitized lapse $N$ rather than $\alpha$.  
\enit
}
There is no explicit explanation why these small changes work better
than before, but we expect that our method can also be applied to 
finding the reason.

\subsection{Strategy 2: Hyperbolic reformulations} 
\label{secHYP}

\subsubsection{Definitions, properties, mathematical backgrounds}

The second effort to re-formulate the Einstein equations is to make 
the evolution equations reveal a
first-order hyperbolic form explicitly. 
This is motivated by the expectation that the symmetric hyperbolic 
system has well-posed
properties in its Cauchy treatment in many systems and also that 
the boundary treatment can be
improved if we know the characteristic speed of the system.  
As a comprehensive review of the hyperbolic formulation, 
we refer to those by 
Choquet-Bruhat and York (1980) \cite{reviewChoquetYork}, 
Geroch (1996) \cite{reviewGeroch},  
Reula (1998) \cite{reviewReula}, and 
Friedrich-Rendall (2000) \cite{reviewFriedrichRendall}, 

We use the following definition:
\Largefbox{\boxwidth}{
{\bf Hyperbolic formulations} \hspace*{\fill} {\bf Box 2.5}\\
We say that the system is {\it a first-order (quasi-linear)
partial differential equation system}, 
if a certain set of
(complex-valued) variables $u_\alpha$ $(\alpha=1,\cdots, n)$
forms
\begin{equation}
\partial_t u_\alpha
= {\cal M}^{l}{}^{\beta}{}_\alpha (u) \, \partial_{l} u_\beta
+{\cal N}_\alpha(u),
\label{def}
\end{equation}
where ${\cal M}$ (the characteristic matrix) and
${\cal N}$ are functions of $u$
but do not include any derivatives of $u$.  Further we say the system
is 
\begin{itemize}
\item {\it a weakly hyperbolic system}, 
if all the eigenvalues of the characteristic matrix are real. 
\item {\it a strongly hyperbolic system}  
(or a diagonalizable / symmetrizable hyperbolic  system),
if the characteristic matrix is
diagonalizable (has a complete set of eigenvectors) and has all real eigenvalues. 
\item {\it a symmetric hyperbolic system}, 
if the characteristic matrix is a Hermitian matrix.
\end{itemize}
}
We treat ${\cal M}^{l\beta}{}_\alpha$ as a $n \times n$ matrix
when  the $l$-index is fixed.
The following properties of these matrices apply to 
every basis of $l$-index.
We say $\lambda^l$ is an
eigenvalue of ${\cal M}^{l\beta}{}_\alpha$
when the characteristic equation,
$\det ({\cal M}^{l\beta} {}_\alpha
-\lambda^l \delta^\beta {}_\alpha)=0$,
is satisfied.
The eigenvectors, $p^\alpha$, are given by solving
${\cal M}^{l}{}^{\alpha} {}_\beta \, p^{l\beta}=\lambda^l  p^{l\alpha}$.
The strong hyperbolicity
is identified,  e.g. by judging 
whether the multiplicity of its eigenvalue, $n_\lambda$, satisfies 
$\mbox{rank} ({\cal M}^{l\beta} {}_\alpha
-\lambda^l \delta^\beta {}_\alpha)=n-n_\lambda$ for all $\lambda$. 
In order to define
the symmetric hyperbolic system, 
we need to declare an inner product
$\langle u|u \rangle$
to determine whether ${\cal M}^{l\beta} {}_\alpha$ is Hermitian.
In other words, we are
required to define the way of lowering the index
$\alpha$ of $u^\alpha$.
We say ${\cal M}^{l\beta} {}_\alpha$ is Hermitian
with respect to this index rule,
when
${\cal M}^l {}_{\beta\alpha}=\bar{{\cal M}}^l {}_{\alpha\beta}$
 for every $l$,
where the overhead bar denotes complex conjugate.
To avoid this requirement of the definition of the inner product, 
people sometimes use the word {\it symmetrizable}, when
the characteristic matrix becomes Hermitian by a certain 
{\it symmetrizer} (positive definite matrix).  In our classification, 
this is only equivalent to the strongly hyperbolic system. 
\footnote{
Several groups use a slightly different definition 
for a symmetric hyperbolic system: defining when 
the principal symbol of the system  
$i{\cal M}^{l\beta}{}_\alpha k_l$  
is anti-Hermitian for arbitrary vector $k_l$. 
The two definitions are equivalent 
when the vector $k_l$ is real-valued, but different eigenvalues.
In our definition,  
all eigenvalues are real-valued, 
while in the other they are all pure imaginary.
}

Writing the system in a hyperbolic form is a
quite useful 
step in proving that the system is {\it well-posed}.  
The mathematical well-posedness of the system means
($1^\circ$) local existence (of at least one solution $u$), 
($2^\circ$) uniqueness (i.e., at most solutions),  and 
($3^\circ$) stability (or continuous dependence of
solutions $\{ u \}$ on the Cauchy data) 
of the solutions. 
The resultant statement expresses the existence of the energy inequality 
on its norm, 
\be
 || u(t) || \leq e^{\alpha \tau} || u(t=0) ||,  
\qquad \mbox{\rm where~} 0 < \tau < t, \quad
\alpha=const. 
\label{energynorm}
\en
This indicates that the norm of $u(t)$ is bounded by a certain function and
the initial norm.  Remark that this mathematical 
does not mean that the norm $u(t)$ decreases along the time evolution. 

The inclusion relation of the hyperbolicities  is,  
\begin{equation}
\mbox{\rm symmetric~hyperbolic} \subset
\mbox{\rm strongly~hyperbolic} \subset
\mbox{\rm weakly~hyperbolic}.
\end{equation}
The Cauchy problem under weak hyperbolicity is not,
  in general, $C^\infty$ well-posed.
At the strongly hyperbolic level,
we can prove the finiteness of the energy norm
if the characteristic matrix is independent of $u$
(cf \cite{Stewart}), that is one step definitely advanced over 
a weakly hyperbolic form.
Similarly, 
the well-posedness of the symmetric hyperbolic is guaranteed 
if the characteristic matrix is independent of $u$,
while if it depends on $u$ we have only limited proofs for the
well-posedness.
{}From the mathematical point of view,
proving well-posedness with less strict conditions is an 
old but still active research problem.
Therefore we have to recall that even if we construct a symmetric 
hyperbolic system in general relativity, that equation  does not necessarily
guarantee numerical stability. 


{}From the point of numerical applications, 
to hyperbolize the evolution equations 
is quite attractive, 
not only for its mathematically well-posed features. 
The expected additional advantages are the following.
\beit
\item[(a)] 
It is well known that a certain flux conservative hyperbolic
system is taken as an essential formulation in the
computational Newtonian hydrodynamics when we control shock
wave formations due to matter
(e.g. \cite{HirschBook}).
\item[(b)] 
The characteristic speed (eigenvalues of the principal matrix) 
is supposed to be the
propagation speed of the information in that system. 
Therefore it is naturally imagined that these magnitudes are
equivalent to the physical information speed of the model to 
be simulated. 
\item[(c)] 
The existence of the characteristic speed of the system 
 is expected to give us 
an improved treatment of the numerical boundary, and/or to 
give us a new well-defined 
 Cauchy problem within a finite region 
(the so-called initial boundary value problem, IBVP). 
\enit
These statements sound  reasonable, but have not yet been generally 
confirmed in actual numerical simulations. 
But we are safe in saying that the formulations are not yet 
well developed to test these issues. 
For example, IBVP studies are preliminary yet,   
and most works are based
on a particular symmetric hyperbolic and in a limited space-time symmetry
\cite{IBVP-CLT,IBVP-LSU,IBVP-FN,IBVP-IR,Stewart,IBVP-PIT1,IBVP-PIT2,IBVP-PIT3}. 
We will come back to this issue in \S \ref{sec:hypRemark} or
\S \ref{sec:outlook}, but meanwhile
let us view the hyperbolic formulations from 
the comparisons of pure evolution equations.


We note that rich
mathematical theories on partial differencial equations are
obtained in a first-order form, while there is a study on linearized ADM
equations in a second-order form \cite{KreissOrtiz}. 

\subsubsection{Hyperbolic formulations of the Einstein equations}
As was discussed by Geroch \cite{reviewGeroch},
most physical systems
can be expressed as symmetric hyperbolic systems.
In order to prove that the Einstein's theory is a well-posed system, 
to hyperbolize the Einstein equations is a long-standing research area 
in mathematical relativity.
As we mentioned in the introduction, numerical relativity shed light
on this mathematical problem.
  
The standard ADM system does not form a first
order hyperbolic system.
This can be seen immediately from the fact that the
ADM evolution equation (\ref{adm_evo2}) has Ricci curvature in RHS.
So far, several first order hyperbolic systems of the Einstein equation
have been proposed.  
In constructing
hyperbolic systems, the essential procedures are 
(1$^\circ$)
to introduce new variables, normally the spatially derivatived metric, 
(2$^\circ$)
to adjust equations using constraints.  
Occasionally, (3$^\circ$) to restrict the gauge conditions, and/or (4$^\circ$)
to rescale some variables. 
Due to process (1$^\circ$), 
the number of fundamental
dynamical variables is always larger than that of ADM. 

In the following discussion, we briefly review 
several hyperbolic systems of the Einstein
equations.  We only mention the systems which applied numerical comparisons. 
See Table. \ref{table:hyprefs} for a more comprehensive list. 


\begin{itemize}
\item The Bona-Mass\'o formulation \cite{BM89,BM92,BMSS95,BMSS97}  \\
This was the first active effort to apply 
hyperbolized equations
to numerical relativity.  
They introduced auxiliary variables to reduce the system 
first order in space:  $
A_k = \partial_k  \ln \alpha,
\quad
B_k^{~i} =  ({1}/{2}) \partial_k \beta^i,
$ and $
D_{kij} =  ({1}/{2}) \partial_k  g_{ij},
$ and construct a first order flux conservative system. 
In their latest formulation \cite{BMSS97}, the set of dynamical variables
is $(\alpha, \gamma_{ij}, K_{ij}, A_i, D_{rij}, V_i \equiv 2 D_{[ir]}{}^r)$, 
 37 functions in total, and the lapse function is restricted to a certain 
functional condition.  The system is a symmetrizable hyperbolic. 
They observed improved numerical stability compared to that of the standard 
ADM system in spherical symmetric spacetime evolution \cite{BMSS95}. 
An advantage of having characteristic speed is also applied to improve 
the treatment of the outer boundary condition \cite{cactus1}.  
However, the appearance of shock formation was also 
reported unless the lapse function is determined by solving the
elliptic or parabolic equations \cite{Alcubierre,AM}. 
\item The Einstein-Ricci system \cite{CY9506071,CY9506072} / 
Einstein-Bianchi system \cite{CY9710041} \\
Series of works by Choquet-Bruhat, York, and their colleagues developed 
hyperbolic systems in slightly different ways.  They intended to
construct wave-type equations $(\Box \phi = S)$ for $(\gamma_{ij}, K_{ij})$, 
that require the introduction of new variables  (45 or 66 variables in total) 
and the use of Bianchi identities.  The resultant system is a 
third-order system in time
for $\gamma_{ij}$ since they use the equation $\partial_t R_{ij}$, but
has only physical characteristic speed (zero or light speed). 
Scheel et al \cite{SBCSThyper,SBCST98} developed a numerial code with 
this formulation, and reported that they could evolve  
Schwarzschild black hole (with 1-dimensional code) quite long time 
($\sim 10^4M$), but futher modifications to equations were necesarrily. 
\item The Einstein-Christoffel system \cite{AY}   \\
Anderson and York \cite{AY} constructed a symmetrizable hyperbolic system 
which only has physical characteristic speed. 
The variables are fundamentally 
 $(\gamma_{ij}, K_{ij}, \Gamma_{kij})$, and they also derived a set of equations
with $(\gamma_{ij}, K_{ij}, G_{kij} \equiv \partial_k \gamma_{ij})$, 30 functions. 
Their formulation differs from the Bona-Mass\'o formulation in the way  the momentum
constraint is used to make the system hyperbolic.  Using a model of plane wave
propagation,  Bardeen and Buchman \cite{BB} numerically compared this 
formulation with ADM and Bona-Mass\'o with slight changes in variables. 
Their experiments are of plane-symmetric wave propagations, and they 
studies the boundary treatment in detail.  We will mention their work later. 
\item The Ashtekar formulation \cite{Ashtekar} \\
Ashtekar's reformulation of space-time was introduced to 
provide a new non-perturbative approach to quantum gravity. 
The new basic variables are
the densitized inverse triad, $\tilde{E}^i_a$, and the 
SO(3,C) self-dual connection, ${\cal A}^a_i$, where the indices
$i,j,\cdots$ indicate the 3-spacetime, and
$a,b,\cdots$ are for SO(3) space.  
The formulation requires additional constraints and reality conditions
in order to describe the classical Lorentzian space-time evolution, but
the evolution system itself forms a weakly hyperbolic system. 
By keeping the number of variables the same [$(\tilde{E}^i_a, {\cal A}^a_i)$=total 18 
(minimum ever)],  we can construct both strongly and symmetric hyperbolic systems by 
adjusting equations with constraints and/or restricting gauge conditions
\cite{ysPRL,ysIJMPD}.   
The authors made numerical comparisons between 
the hyperbolicity of the systems with plane symmetric gravitational wave propagation
using a periodic boundary condition \cite{ronbun1,ronbun2}. 
The outcome is that there are no drastic differences in numerical stability
between the three levels of the hyperbolic equations.  
We will describe the details in Appendix \ref{App_Ashtekar}.
\item The Frittelli-Reula formulation \cite{FR96,Stewart} \\
This is a one-parameter family of a symmetric hyperbolic system.  
The procedure is to define new variables, to adjust evolution 
equations with constraints (with one parameter), 
and to densitize the lapse function (with one parameter). 
The variables are ($\gamma_{ij}, 
M^{ij}_{~~k} \equiv  {1 \over 2} (\gamma^{ij}_{~~,k}+
a \gamma^{ij}\gamma_{rs} \gamma^{rs}_{~~,k}),  
P^{ij} \equiv K^{ij}+ b \gamma^{ij} K$) where $a, b$ are two other 
parameters, and make a total of 30. 
They define a symmetric hyperbolic system by non-diverging energy norm, 
and restricted free parameter spaces.  
A numerical comparison was also made for Gowdy spacetime \cite{HernPHD}, 
with quite similar conclusions as in the Ashtekar version. 
\item The Conformal Field equations \cite{FriedrichCFE} \\
A series of works by Friedrich \cite{FriedrichCFE} attempted to
construct a $3+1$ formulation with hyperboloidal foliations 
(i.e. asymptotically null foliations), and with comformal compactification. 
This is the ultimate plan to remove 
the outer boundary problem in numerical simulation, and to provide
a suitable foliation for gravitational radiation problem. 
However, the current equations are rather quite complicated. 
In its metric-based expression \cite{Hubner}, 
the evolution variables are 57; $\gamma_{ij}, K_{ij}$, 
the connection coefficients $\gamma^a{}_{bc}$, 
projections ${}^{(0,1)}\hat{R}_a=n^b \gamma_a{}^c \hat{R}_{bc}$ and
${}^{(1,1)}\hat{R}_{ab}=\gamma_a{}^c \gamma_b{}^d \hat{R}_{bd}$ 
 of 4-dimensional 
Ricci tensor $\hat{R}_{ab}$, the electric and magnetic components of 
the rescaled Weyl tensor $C_{abc}{}^d$, and the comformal factor $\Omega$
and its related quantities $\Omega_0\equiv n^a \nabla_a \Omega,
\nabla_a \Omega, \nabla^a \nabla_a \Omega$.  
By specifying suitable gauge functions $(\alpha, \beta^a, R)$ where $R$ is
the Ricci scalar, then the total system forms a symmetric hyperbolic system. 
Applications to numerical relativity are in progress, but 
have not yet reached the stage of applying evolution in a non-trivial metric.  
For more details, see reviews e.g. by 
Frauendiener \cite{reviewFrauendiener} or by Husa \cite{reviewHusa}. 
\item The Kidder-Scheel-Teukolsky (KST) formulation \cite{KST} \\
This set of equations is a generalized formulation of the previous ones. 
It has 12 free parameters, and includes both 
Anderson-York\cite{AY} and Frittelli-Reula\cite{FR96} formulations as a subset.
Therefore it might be useful to discuss further hyperbolization methods 
starting with this KST formulation.  We briefly summarize it in Box 2.6. 
\end{itemize}
\Largefbox{\boxwidth}{
{\bf The Kidder-Scheel-Teukolsky (KST) formulation \cite{KST}:}
\hspace*{\fill} {\bf Box 2.6}\\
\beit
\item  Starting from a set of 
$(g_{ij}, K_{ij}, d_{kij}\equiv \ptl_k g_{ij})$, 
the fundamental dynamical variables are 
defined $(g_{ij}, P_{ij}, M_{kij})$, as
\bera
P_{ij} &\equiv& K_{ij} +  \hat{z}  g_{ij} K,
\\
M_{kij} &\equiv& (1/2) [  \hat{k}  d_{kij} +   \hat{e}  d_{(ij)k}
 + g_{ij} ( \hat{a}  d_{k} +  \hat{b}  b_{k}) 
 + g_{k(i} ( \hat{c}  d_{j)} +  \hat{d}  b_{j)}) 
 ],
\enra
where $d_k=g^{ab}d_{kab}$, $b_k=g^{ab}d_{abk}$, 
and $(\hat{a}, \hat{b}, \hat{c}, \hat{d}, \hat{e}, \hat{k}, \hat{z})$
are ``kinematical" parameters. 
\item 
The 3-hypersurface $\Sigma$ is foliated with gauge functions, 
($\alpha, \beta^i$), the lapse and shift vector.  However, the densitized lapse, 
$Q=\log (\alpha g^{- \sigma} )$, (with a parameter $\sigma$) is actually used.  
\item  
The evolution equations are adjusted with constraints [in a version 
of $(g_{ij}, K_{ij}, d_{kij})$]
\bera
  \hat{\ptl}_0 g_{ij} &=& -2 \alpha K_{ij}, 
\\
  \hat{\ptl}_0 K_{ij} &=& (\cdots) +  \gamma  \alpha g_{ij} {\cal H} +  \zeta  \alpha g^{ab}
\CC_{a(ij)b}, 
\\
  \hat{\ptl}_0 d_{kij} &=& (\cdots) +  \eta  \alpha g_{k(i}{\cal M}_{j)} +  \chi
 \alpha g_{ij}
{\cal M}_{k}, 
\enra
where $\hat{\ptl}_0=\ptl_t - \pounds_\beta$ and 
$(\gamma, \zeta, \eta, \chi)$ are parameters.  The terms $(\cdots)$ are original
terms derived from the ADM equations and are available as (2.14) and (2.24) in \cite{KST}. 
\item  Constraints are 
$({\cal H}, {\cal M}_i, 
\CC_{klij})$, where
$\CC_{klij}\equiv \ptl_{[k}d_{l]ij}$. 
\enit
}
In short, the number of dynamical variables and constraints are 30 and 22
and there are  12 free parameters. 
Although there is no specific method to specify 12 free parameters
KST showed numerical examples of the 
Schwarzschild black hole (mass $M$) evolution, 
which runs quite a longer time ($t \sim 6000M$) \cite{KST,0209115}.  

We think the essential advantage of the KST system
is the introduction of  ``kinematical" parameters. 
These 6 parameters 
\begin{indention}{1cm}
\noindent
$\bullet$~do not change the eigenvalues of evolution eqs., 
\\$\bullet$~do not affect the principal part of the constraint evolution eqs.,  
\\$\bullet$~do affect the eigenvectors of the evolution system, and 
\\$\bullet$~do affect the nonlinear terms of evolution eqs/constraint evolution eqs.
\end{indention}
\noindent
As Calabrese et al \cite{LSU-KST} pointed out, KST equations at linearized level
on the flat spacetime have no non-principal terms, and these  ``kinematical" parameters
finally enable us to discuss the features of hyperbolicity in numerical experiments. 
Several variations of the KST formulation are presented in \cite{SarbachTiglio}.
Recently, Lindblom and Scheel \cite{LindblomScheel} 
tried to explain the relation between the numerical error growing rate
and the predicted error growing rate from the characteristic matrix of  
KST evolution equations.  Their trial did not succeed in matching 
the two completely, but
at least began to reveal a similar order before non-linear blow-up begins. 
\subsubsection{Remarks} \label{sec:hypRemark}
When we discuss hyperbolic systems in the context of numerical 
stability, the following questions should be considered:  

\underline{Questions to hyperbolic formulations on its applications to numerics}
\begin{itemize}
\begin{indention}{1cm} 
\item[Q(A)]
From the point of the set of evolution equations, 
does hyperbolization actually contribute to numerical accuracy
and stability?   
Under what conditions/situations will the advantages of 
hyperbolic formulation be observed?
\item[Q(B)]
If the answer to Q(A) is affirmative,
which level of hyperbolicity is practically useful for numerical
applications?  Strongly hyperbolic, symmetric hyperbolic, or other?
\item[Q(C)]
If the answer to Q(A) is negative, then can we find other 
practical advantages to hyperbolization?  
Treatment of boundary conditions, or other?
\end{indention}
\end{itemize}
 
Unfortunately, we do not have conclusive answers to these questions, but many 
experiences are being accumulated. 
Several earlier numerical comparisons reported the stability of  
hyperbolic formulations \cite{BMSS95,BMSS97,cactus1,SBCSThyper,SBCST98}.  
But we have to remember that 
this statement went against the standard ADM formulation, which has a 
constraint-violating mode for Schwarzschild spacetime as has been shown 
recently (see \S \ref{secADJADM}). 

These partial numerical successes encouraged the community to 
formulate various hyperbolic systems.  Recently, Calabrese et al 
\cite{LSU-KST_RC} reported there is a certain differences in the long-term
convergence features between weakly and strongly hyperbolic systems 
on the Minkowskii background space-time. 
However, several numerical experiments also indicate that this direction
is not a complete success. 

\underline{Objections from numerical experiments}
\beit
\begin{indention}{1cm} 
\item Above earlier numerical successes were also terminated with blow-ups.  
\item If the gauge functions are evolved according to the hyperbolic 
equations, then their finite propagation speeds may cause pathological 
shock formations in simulations \cite{Alcubierre,AM}.  
\item 
There are no
drastic differences in the evolution properties {\it between} 
hyperbolic systems (weakly, strongly and symmetric hyperbolicity)
by systematic numerical studies 
by Hern \cite{HernPHD} based on Frittelli-Reula formulation \cite{FR96}, and
by the authors \cite{ronbun1} based on Ashtekar's formulation 
\cite{Ashtekar,ysPRL,ysIJMPD}.
\item Proposed symmetric hyperbolic systems were not always the best ones 
for numerical evolution.  People are normally still required to reformulate
them for suitable evolution.  Such efforts are seen in the applications of
the Einstein-Ricci system \cite{SBCST98},  
the Einstein-Christoffel system \cite{BB},  
the conformal field equations \cite{reviewHusa}, and so on.
\item 
Bardeen and Buchmann \cite{BB} confirmed the usefulness of hyperbolicity
when they treated numerical boundary conditions
in plane-symmetric wave propagation problem, but they also mentioned that the 
same techniques can not be applied in 2 or 3-dimensional cases. 
\end{indention}
\enit
Of course, these statements only casted on a particular formulation, and therefore
we have to be careful not to over-emphasize the results.  In order to 
figure out the reasons for the above objections, 
it is worth stating the following cautions:

\underline{Remarks on hyperbolic formulations}
\beit
\begin{indention}{1cm} 
\item[(a)]
Rigorous mathematical proofs of well-posedness of PDE are
mostly for simple symmetric or strongly hyperbolic systems.  
If the matrix components or coefficients depend on dynamical variables
(as in all any versions of hyperbolized Einstein equations), 
almost nothing was proved in more general situations. 
\item[(b)]
The statement of ''stability" in the discussion of well-posedness 
refers to the bounded growth of the norm, and does not indicate
a decay of the norm in time evolution. 
\item[(c)]
The discussion of hyperbolicity only uses
the characteristic part of the evolution equations, and ignores the rest. 
\end{indention}
\enit

We think the origin of confusion  in the community results from over-expectation
on the above issues.  Mostly, point (c) is the biggest problem, 
as was already pointed out in several places \cite{ronbun2,reviewLehner,adjADMsch}.
The above numerical claims from Ashtekar and Frittelli-Reula formulations
were mostly due to the contribution (or interposition) 
of non-principal parts in evolution. 

Regarding this issue, 
the recent KST formulation finally opens the door.  
As we saw, KST's
``kinematic" parameters enable  us to reduce the non-principal part, so that
numerical experiments are hopefully expected to represent predicted
evolution features from PDE theories.  
At this moment, the agreement between numerical behavior and 
theoretical prediction is not yet perfect but close 
\cite{LindblomScheel}. 
If further studies reveal the direct correspondences between theories and numerical
results, then the direction of hyperbolization will remain as the 
essential approach in numerical relativity, and the related IBVP researches 
will become a main research subject in the future. 

Meanwhile, 
it will be useful if we have an alternative procedure
to predict stability including the effects of the 
non-principal parts of the equations, which are
neglected in the discussion of hyperbolicity. 
Our proposal of adjusted system in the next subsection may be one of them. 

\subsection{Strategy 3: Asymptotically constrained systems}  
\label{secASYMPT}

\begin{figure}[b]
\unitlength 1mm 
\begin{picture}(160,55)
\put(0, 0){\epsfxsize=70mm \epsffile{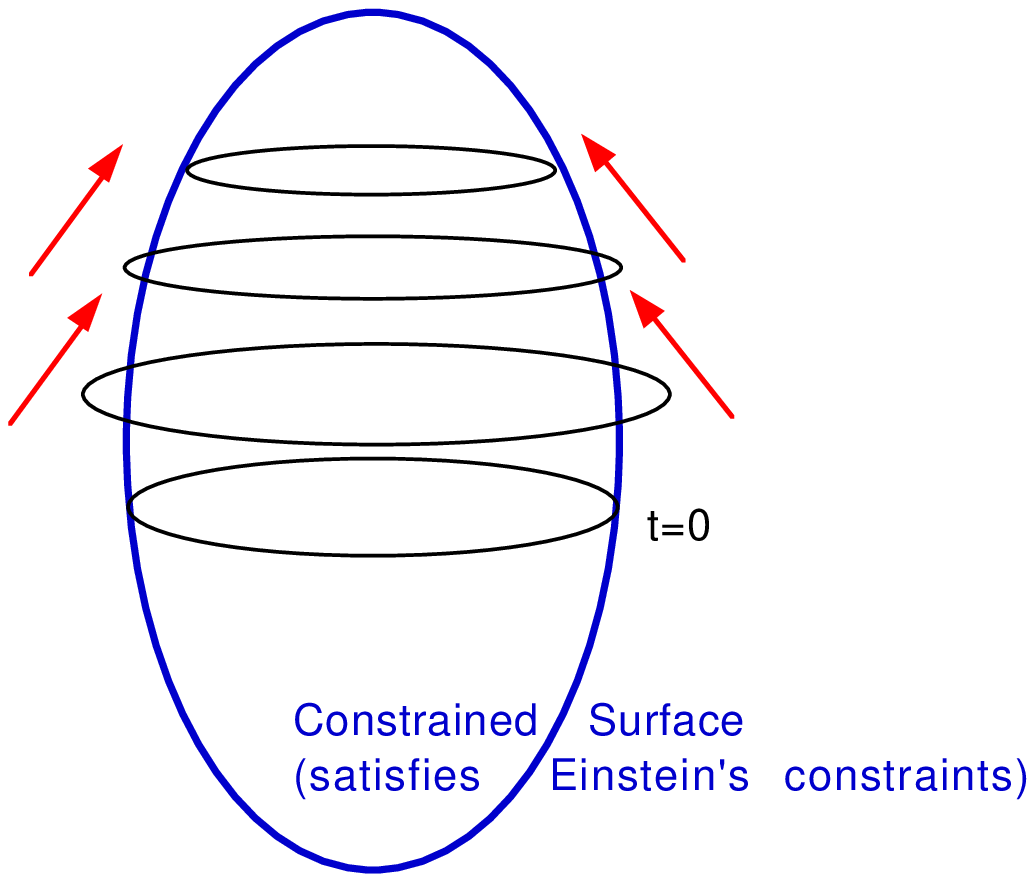} }
\put(75, 0){\epsfxsize=70mm \epsffile{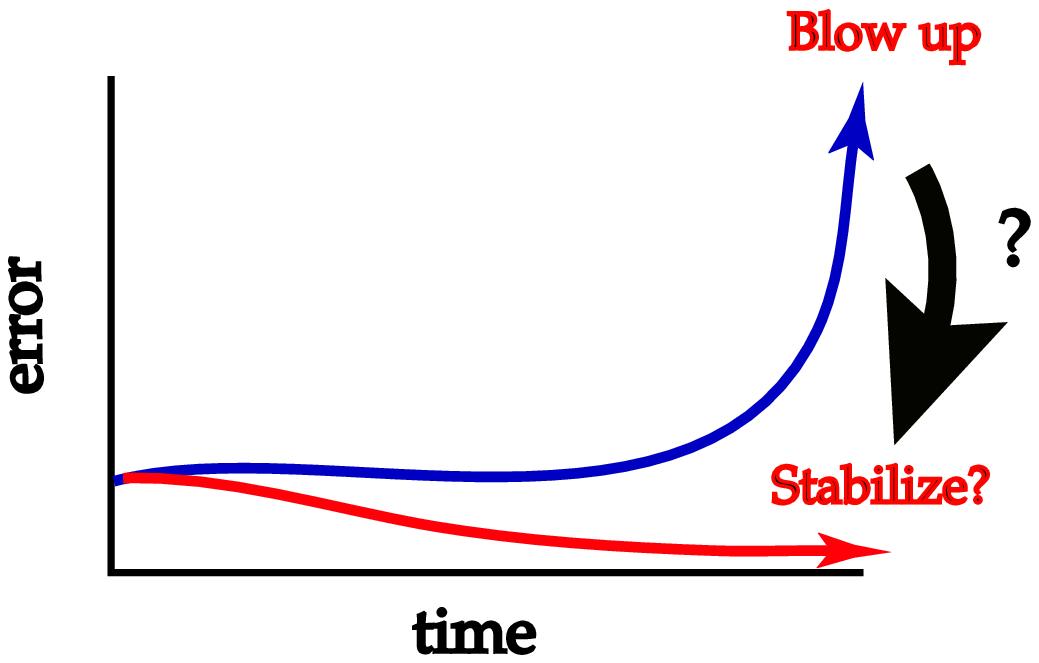} }
\end{picture}
\caption{Schematic picture of ``asymptotically constrained system". } 
\label{fig:attractor}
\end{figure}

The third strategy is to construct a robust system against 
the violation of constraints,
such that the constraint surface is an attractor (Figure \ref{fig:attractor}).  
The idea was first proposed as ``$\lambda$-system" 
by Brodbeck et al \cite{BFHR}, and then developed in more general situations 
as ``adjusted system" by the authors \cite{ronbun2}.  

\subsubsection{The ``$\lambda$-system"} 
\label{sec:lambdasystem}
Brodbeck et al \cite{BFHR} proposed a system which has additional variables
$\lambda$ that obey artificial dissipative equations.  
The variable $\lambda$s are supposed to indicate the violation of 
constraints and the target of the system is to get $\lambda=0$ as its
attractor.  
Their proposal can be summarized as in Box 2.7.

\Largefbox{\boxwidth}{
{\bf The ``$\lambda$-system"  (Brodbeck-Frittelli-H\"ubner-Reula)
\cite{BFHR}:}\hspace*{\fill}  {\bf Box 2.7}\\ 
For a symmetric hyperbolic system, 
add additional variables $\lambda$ and 
artificial force to reduce the violation of
constraints.~\\
The procedure: \\~\\
\begin{tabular}{cll}
1. & \begin{tabular}{l} {Prepare a symmetric hyperbolic evolution system}
     \end{tabular}
   & $\partial_t u=M \partial_i u+N$ 
\\[1em]
\begin{tabular}{l} 2. \\ ~~ \end{tabular} &
\begin{tabular}{l}
{Introduce $\lambda$ as an indicator of violation of }\\
{constraint which obeys dissipative eqs. of motion }
\end{tabular}
   & \begin{tabular}{l} 
$\ptl_t \lambda=\alpha C-\beta \lambda$ \\
$(\alpha \neq 0, \beta>0)$
\end{tabular}
\\[1em]
3. & \begin{tabular}{l}
     {Take a set of $(u,\lambda)$ as dynamical variables }
     \end{tabular}
   & $\ptl_t
\left(\matrix{ u \cr \lambda}\right)
\simeq
\mm[A,0,F,0]
\partial_i 
\left(\matrix{ u \cr \lambda}\right)
$
\\[1em]
\begin{tabular}{l} 4. \\ ~~ \end{tabular} &
\begin{tabular}{l}
{Modify evolution eqs so as to form  } \\
{a
symmetric hyperbolic system
} \end{tabular}
   & $
\ptl_t
\left(\matrix{ u \cr \lambda}\right)
=
\mm[A,\bar{F},F,0]
\partial_i 
\left(\matrix{ u \cr \lambda}\right)
$
\end{tabular}
}
The main ideas are
to introduce additional variables, $\lambda$s,
to impose dissipative dynamical equations for $\lambda$s, and 
to construct a symmetric hyperbolic system for both original 
variables and $\lambda$s. 
Since the total system is designed 
to have symmetric hyperbolicity,  
the evolution is supposed to be unique.  
Brodbeck et al showed analytically that such a decay of $\lambda$s can be seen
for sufficiently small $\lambda (>0)$ 
with a choice of appropriate combinations of $\alpha$s and $\beta$s.

Brodbeck et al presented a set of equations
based on Frittelli-Reula's symmetric hyperbolic formulation \cite{FR96}. 
The version of Ashtekar's variables was presented by the authors 
\cite{SY-asympAsh}
for controlling the constraints or reality conditions or both 
(see \S \ref{AshtekarLambdasystem}).  
The numerical tests of both the Maxwell-$\lambda$-system and 
the Ashtekar-$\lambda$-system were performed \cite{ronbun2}, and 
 confirmed to work as expected (see \S \ref{AshtekarLambdaNum}). 
Although it is questionable whether the
recovered solution is true evolution or 
not \cite{SiebelHuebner}, we think 
the idea is quite attractive. 
To enforce the decay of errors in its initial 
perturbative stage seems the key 
to the next improvements, which are also
 developed in the next section on ``adjusted systems". 

However, 
there is a high price to pay for constructing a $\lambda$-system.
The $\lambda$-system can not be introduced generally, because
(i) the construction of $\lambda$-system requires the original evolution 
equations to have a symmetric hyperbolic form, which is quite restrictive for
the Einstein equations, 
(ii) the final system requires many additional variables
and we also need to evaluate all the constraint equations at every time step,
which is a hard task in computation.  
Moreover, 
(iii) it is not clear that the $\lambda$-system is robust enough for 
non-linear violation of constraints, or that $\lambda$-system 
can control constraints
which do not have any spatial differential terms. 

\subsubsection{The ``adjusted system"}
\label{sec:adjsystem}

Next, we propose an alternative system
which also tries to control the violation of constraint equations
actively, which we named ``adjusted system".
We think that 
this system is more practical and robust 
than the previous $\lambda$-system.

\settowidth{\listlength}{Theoretical support:}
\Largefbox{\boxwidth}{
{\bf The Adjusted system (essentials) \cite{ronbun2}:}
\hspace*{\fill} {\bf Box 2.8}\\
\begin{list}{}{%
\setlength{\leftmargin}{\listlength}
\addtolength{\leftmargin}{\labelsep}
\setlength{\labelwidth}{\listlength}
}
\item[Purpose:\hfill]
Control the violation of constraints by reformulating the system 
so as to have a constrained surface as attractor.  
\item[Procedure:\hfill]
  Add a particular combination of constraints to the evolution equations, and
adjust its multipliers. 
\item[Theoretical support:\hfill] 
Eigenvalue analysis of the constraint propagation equations.
\item[Advantages:\hfill]
 Available even if the base system is not symmetric hyperbolic. 
\item[Advantages:\hfill]
 Keeps the number of the variables the same as in the original system.
\end{list}
}
We will describe the details in the next section, but summarize 
the procedures in advance:

\Largefbox{\boxwidth}{
{\bf The Adjusted system (procedures):}\hspace*{\fill} {\bf Box 2.9}\\~\\
\begin{tabular}{cll}
1. & \begin{tabular}{l}{Prepare a set of evolution eqs.}
     \end{tabular}
   & $\partial_t u=J \ptl_i u +K$ 
\\[1em]
2. & \begin{tabular}{l}{Add constraints in RHS }
     \end{tabular}
   & $\partial_t u=J \ptl_i u +K  \underbrace{+ \kappa C}$ 
\\[1em]
\begin{tabular}{l} 3. \\ ~~ \end{tabular} &
\begin{tabular}{l}
{Choose the coeff.  $\kappa$  so as to make the }\\
{eigenvalues of the homogenized adjusted  }\\
{$\partial_t C$ eqs negative  reals or pure imaginary.} \\
{(See Box 3.2 and 3.3)}
\end{tabular}
   & 
\begin{tabular}{l} 
   $\ptl_t C=D \ptl_iC+E C$ \\
   $\ptl_t C=D \ptl_iC+E C \underbrace{+ F \ptl_iC + G C}$
\end{tabular}
\\[1em]
\end{tabular}
\\~\\
The details are in \S \ref{secADJUSTED}.
}
The process of adjusting equations is a common technique in other re-formulating
efforts as we reviewed.  
However, we try to employ the evaluation process of constraint 
amplification factors as an alternative guideline to hyperbolization of the system. 

For the Maxwell equation and the Ashtekar version of the Einstein equations, we 
numerically found that this idea works to reduce the violation of constraints, 
and that the effects are much better  
than by constructing its symmetric hyperbolic versions \cite{ronbun2} (see \S
\ref{AshtekarAdjNum}). 
The idea  was applied to the standard ADM formulation which is
not hyperbolic and several attractive adjustments were proposed 
\cite{adjADM,adjADMsch} (see \S \ref{secADJADM}).  
This analysis  was also applied to explain  the
advantages of the BSSN formulation, 
and again several alternative adjustments to BSSN
equations were proposed 
\cite{adjBSSN} (see \S \ref{secADJBSSN}). 
We will explain these issues in the next section.


\newpage
\section{A unified treatment: Adjusted System} \label{secADJUSTED}
\setcounter{equation}{0}
This section is devoted to present our idea of ``asymptotically 
constrained system", which was briefly introduced in Box 2.8 and 
2.9 in the previous section. 
We begin with an overview of the adjusting procedure and the idea of 
background structure in \S \ref{secADJ}.  
Next, we show the applications both to ADM 
(\S \ref{secADJADM}) and BSSN (\S \ref{secADJBSSN}) formulations. 
Original references can be found in 
\cite{ronbun2,adjADM,adjADMsch,adjBSSN}.  

\subsection{Procedures : Constraint propagation equations and Proposals}
\label{secADJ}

Suppose we have a set of dynamical variables   $u^a (x^i,t)$, 
and their evolution equations, 
\begin{equation}
\partial_t u^a = f(u^a, \partial_i u^a, \cdots), \label{ueq}
\end{equation}
and the
(first class) constraints, 
\begin{equation}
C^\alpha (u^a, \partial_i u^a, \cdots) \approx 0.
\end{equation}
Note that we do not require (\ref{ueq}) forms a first order hyperbolic
form. 
We propose  to investigate
the evolution equation of $C^\alpha$ (constraint propagation),
\begin{equation}
\partial_t C^\alpha = g(C^\alpha, \partial_i C^\alpha, \cdots),
 \label{Ceq}
\end{equation}
for predicting the violation behavior of constraints in time evolution. 
We do not mean to integrate 
(\ref{Ceq}) numerically together with the original 
evolution equations (\ref{ueq}), but mean to evaluate them
analytically in advance in order to reformulate the equations (\ref{ueq}). 

There may be two major analyses of (\ref{Ceq}); 
(a) the hyperbolicity of (\ref{Ceq})
when (\ref{Ceq}) is a first order system, and 
(b) the eigenvalue analysis of the whole RHS in 
(\ref{Ceq}) after a suitable homogenization. 
\beit
\item[(a)] \underline{Hyperbolicity analysis of (\ref{Ceq})}\\
If  (\ref{Ceq}) forms a first-order system, 
the standard hyperbolic PDE analysis is applicable. 
As we viewed in \S \ref{secHYP}, 
the analysis is mainly to 
identify the level of hyperbolicity
and to calculate the characteristic speed of the system, 
from eigenvalues 
of the principal matrix.
 
For example, 
the evolution equations of the 
ADM formulations, (\ref{ueq}) 
[(\ref{adm_evo1}) and (\ref{adm_evo2})], do not form a first-order system, while
their constraint propagation equations, (\ref{Ceq}) 
[(\ref{CHproADM}) and (\ref{CMproADM})], do form. 
Therefore, we can apply the classification on the hyperbolicity (weakly,
strongly or symmetric) to the constraint propagation equations. 
However, 
if one adjusts the ADM equations with constraints, then this first-order 
characters will not be guaranteed. 
 
As we mentioned in \S \ref{sec:hypRemark}, 
another big problem in the hyperbolic analysis is that it only 
discusses the principal part of the system. 
If there is a method to characterize 
the non-principal part, then that will help to clarify 
our understanding of evolution behavior. 
\item[(b)] \underline{Eigenvalue analysis of the whole RHS in 
(\ref{Ceq}) after a suitable homogenizing procedure.}\\
This analysis may compensate for the above problems. 
\Largefbox{\boxwidthitem}{
\noindent
{\bf Amplification Factors of Constraint Propagation equations: } 
\hspace*{\fill} {\bf Box 3.1}\\
We propose  to homogenize (\ref{Ceq}) by a Fourier
transformation,  e.g. 
\begin{eqnarray}
\partial_t \hat{C}^\alpha &=& \hat{g}(\hat{C}^\alpha)
=M^\alpha{}_{\beta} \hat{C}^\beta,
\nonumber \\ &&
\mbox{~~~where~}
C(x,t)^\rho
=\displaystyle{\int} \hat{C}(k,t)^\rho\exp(ik\cdot x)d^3k,
\label{CeqF}
\end{eqnarray}
then to analyze the set of eigenvalues, say $\Lambda$s,  of the 
coefficient matrix, $M^\alpha{}_{\beta}$, in
(\ref{CeqF}).  We call $\Lambda$s the constraint 
amplification factors (CAFs)
of (\ref{Ceq}). 
}
The CAFs predict the evolution of constraint violations. 
We therefore can discuss the ``distance" to the constraint surface
using the ``norm" or ``compactness" of the constraint violations 
(although we do not have exact definitions of these ``$\cdots$" words). 

The next conjecture seems to be quite useful to predict the 
evolution feature of constraints:
\Largefbox{\boxwidthitem}{
\noindent
{\bf Conjecture on Constraint Amplification Factors (CAFs): } 
\hspace*{\fill} {\bf Box 3.2}
\begin{itemize}
\item[(A)] If CAF has 
a {\it negative real-part } (the constraints  are forced to be
diminished), then we see
more stable evolution than a system which has positive CAF.
\item[(B)] If CAF has a {\it non-zero  imaginary-part }
(the constraints are propagating away), then we see
more stable evolution than a system which has 
zero CAF.
\end{itemize}
}
We found that the system becomes more stable when
more $\Lambda$s satisfy the above criteria. 
(The first observations were in the Maxwell and Ashtekar formulations
\cite{ronbun1,ronbun2}).  
Actually, supporting mathematical proofs are available when we classify
the fate of constraint propagations as follows. 
\Largefbox{\boxwidthitem}{
\noindent
{\bf Classification of Constraint propagations: } 
\hspace*{\fill} {\bf Box 3.3}
\\
If we assume that avoiding the divergence of constraint norm is related 
to the numerical stability, the next classification would be useful: 
\beit
\item[(C1)]  {\it Asymptotically constrained : } 
All the constraints decay and converge to zero. \\ 
This case can be obtained if and only if all the real parts of CAFs are negative. 
\item[(C2)]  {\it Asymptotically bounded : } 
All the constraints are bounded at a certain value. (this includes
the above {\it asymptotically constrained} case.) \\
This case can be obtained if and only if (a) all the real parts of CAFs are not
positive and the constraint propagation matrix $M^\alpha{}_\beta$ is
diagonalizable, or  (b) all the real parts of CAFs are not positive
and the real part of the degenerated CAFs is not zero. 
\item[(C3)]  {\it Diverge : } At least one constraint will diverge. 
\enit
The details are shown in \cite{diagCP}. 
}

\begin{figure}[t]
\unitlength 1mm 
\begin{picture}(160,85)
\put(20,00){\epsfxsize=110mm \epsffile{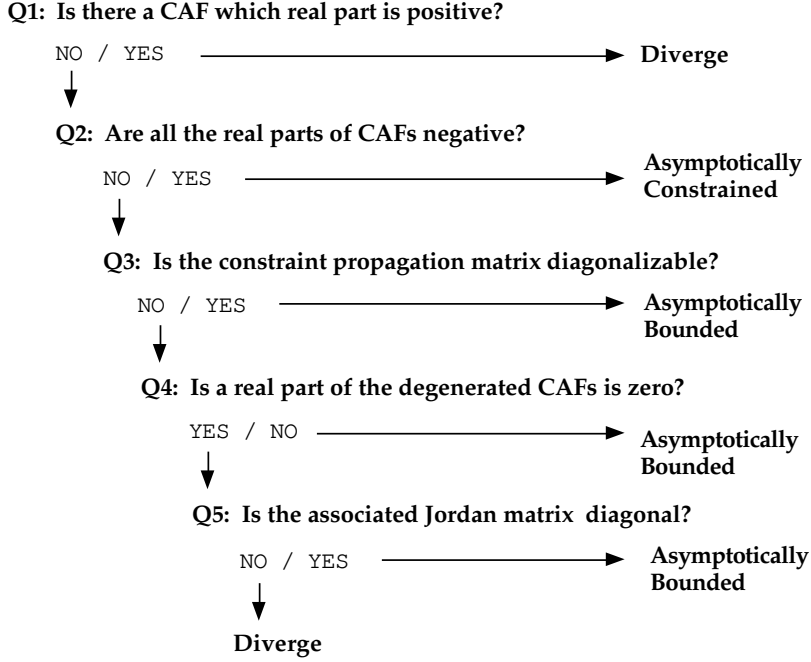} }
\end{picture}
\caption{A flowchart to classify the constraint propagations (Box 3.3). }
\label{fig:flowchart}
\end{figure}

This classification roughly indicates the heuristic statements 
(A) in Box 3.2. 
A practical procedure for this
classification is drawn in Figure 
\ref{fig:flowchart}.

We remark that this eigenvalue analysis requires the
fixing of a particular background spacetime,  
since the CAFs depend on the dynamical variables, $u^a$.

\enit

The above features of the constraint propagation, (\ref{Ceq}),
will differ when we modify the original evolution equations.
Suppose we add (adjust) the evolution equations using constraints
\begin{equation}
\partial_t u^a = f(u^a, \partial_i u^a, \cdots)
+ F(C^\alpha, \partial_i C^\alpha, \cdots), \label{DeqADJ}
\end{equation}
then (\ref{Ceq}) will also be modified as
\begin{equation}
\partial_t C^\alpha = g(C^\alpha, \partial_i C^\alpha, \cdots)
+ G(C^\alpha, \partial_i C^\alpha, \cdots). \label{CeqADJ}
\end{equation}
Therefore, the problem is how to adjust the evolution equations
so that their constraint propagations satisfy the above criteria 
as much as possible.

\subsection{Adjusted ADM formulations} \label{secADJADM}

We show an application of the above idea, to 
the standard ADM system, Box 2.1. 

\subsubsection{Adjustments to ADM equations and its effects on
constraint propagations} 
Generally, we can write the adjustment terms to
(\ref{adm_evo1}) and (\ref{adm_evo2})
using (\ref{admCH}) and (\ref{admCM}) by the following combinations
(using up to the first derivatives of constraints for simplicity):
\Largefbox{\boxwidth}{
{\bf The adjusted ADM formulation \cite{adjADMsch}:}
\hspace*{\fill} {\bf Box 3.4}\\
Modify the evolution eqs of $(\gamma_{ij}, K_{ij})$ by constraints 
${\cal H}$ and ${\cal M}_i$, 
\bera
&\mbox{adjustment terms of } 
\partial_t \gamma_{ij}:&  
+P_{ij} {\cal H}
+Q^k{}_{ij}{\cal M}_k
+p^k{}_{ij}(\nabla_k {\cal H})
+q^{kl}{}_{ij}(\nabla_k {\cal M}_l), \label{adjADM1}
\\
&\mbox{adjustment terms of } 
\partial_t K_{ij}:&  
+R_{ij} {\cal H}
+S^k{}_{ij}{\cal M}_k
+r^k{}_{ij} (\nabla_k{\cal H})
+s^{kl}{}_{ij}(\nabla_k {\cal M}_l), \label{adjADM2}
\enra
where $P, Q, R, S$ and $p, q, r, s$ are multipliers.  That is, 
\bera
{\partial_t} \gamma_{ij} &=& (\ref{adm_evo1})   +P_{ij} {\cal H}
+Q^k{}_{ij}{\cal M}_k
+p^k{}_{ij}(\nabla_k {\cal H})
+q^{kl}{}_{ij}(\nabla_k {\cal M}_l),
\\
{\partial_t} K_{ij} &=& (\ref{adm_evo2}) +R_{ij} {\cal H}
+S^k{}_{ij}{\cal M}_k
+r^k{}_{ij} (\nabla_k{\cal H})
+s^{kl}{}_{ij}(\nabla_k {\cal M}_l).
\enra
According to this adjustment, the constraint propagation equations are 
also modified as
\begin{eqnarray}
\partial_t {\cal H} &=&
\mbox{(\ref{CHproADM})}
  + H_1^{mn}  (\ref{adjADM1})
+H_2^{imn}\partial_i(\ref{adjADM1})
+H_3^{ijmn}\partial_i\partial_j (\ref{adjADM1})
+H_4^{mn}(\ref{adjADM2}), \label{Hnewhosei}
\\
\partial_t {\cal M}_i &=&
\mbox{(\ref{CMproADM})}
+ M_1{}_i{}^{mn} (\ref{adjADM1})
+M_2{}_i{}^{jmn} \partial_j(\ref{adjADM1})
+M_3{}_i{}^{mn} (\ref{adjADM2})
+M_4{}_i{}^{jmn} \partial_j(\ref{adjADM2}). \label{Mnewhosei}
\end{eqnarray}
with appropriate changes in indices.  (See 
Appendix \ref{appADMconpro}. 
 $H_1, \cdots, M_1, \cdots$ are defined there.)
}

We show two examples of adjustments here.  We will list several others later
in Table \ref{tableADMADJ}. 
\been
\item {\bf The standard ADM vs. original ADM}\\
The first comparison is to show the differences between 
the standard ADM \cite{ADM-York}
and the original ADM system \cite{ADM} (see \S \ref{secADM}).  
In
the notation of (\ref{adjADM1}) and (\ref{adjADM2}), the adjustment, 
\begin{eqnarray}
R_{ij}=\kappa_F \alpha \gamma_{ij}, \qquad
\mbox{\rm (and set the other multipliers zero)}
\label{originalADMadjust}
\end{eqnarray}
will distinguish two, 
where $\kappa_F$ is a constant. 
Here $\kappa_F=0$ corresponds to the
standard ADM (no adjustment),
and $\kappa_F=-1/4$ to the original ADM (without any
adjustment to the canonical formulation by ADM).
As one can check by (\ref{Hnewhosei}) and (\ref{Mnewhosei}) 
adding $R_{ij}$ term keeps the constraint propagation in a 
first-order form.  Frittelli \cite{Fri-con} (see also \cite{adjADM}) 
pointed out that
the  hyperbolicity of constraint propagation equations is better 
in the standard ADM system. 

\item{\bf Detweiler type}\\
Detweiler \cite{detweiler} found that with a particular combination, 
 the evolution of the
energy norm of the constraints, ${\cal H}^2+{\cal M}^2$,
can be negative definite
when we apply the maximal slicing condition, $K=0$.
His adjustment can be written in our notation in
(\ref{adjADM1}) and (\ref{adjADM2}), as
\begin{eqnarray}
P_{ij}&=&-  \kappa_L \alpha^3 \gamma_{ij}, \label{Det1}\\
R_{ij}&=& \kappa_L \alpha^3 (K_{ij}- (1 / 3) K \gamma_{ij}), \label{Det2}
\\
S^k{}_{ij}&=& \kappa_L \alpha^2 [
    {3}  (\partial_{(i} \alpha) \delta_{j)}^{k}
- (\partial_l \alpha) \gamma_{ij} \gamma^{kl}], \label{Det3}
\\
s^{kl}{}_{ij}&=&\kappa_L \alpha^3 [
  \delta^k_{(i}\delta^l_{j)}-(1/3) \gamma_{ij}\gamma^{kl}],
  \label{Det4}
\end{eqnarray}
everything else is zero, where $\kappa_L$ is a constant.
Detweiler's adjustment, (\ref{Det1})-(\ref{Det4}),
does not put constraint propagation equation
to a first order form, so we cannot discuss
hyperbolicity  or the characteristic speed of the constraints.
We confirmed numerically, using perturbation on Minkowskii
and Schwarzschild spacetime,
that Detweiler's system provides better
accuracy than the standard ADM, but only for small positive $\kappa_L$.
\enen

\subsubsection{Procedure to evaluate constraint amplification factors 
in spherically symmetric  spacetime}

Before we compare between particular adjustment examples, we describe our
procedure to evaluate CAFs in spherically symmetric spacetime. 
According to our motivation, the actual procedure to analyze the 
adjustments is to substitute the
perturbed metric to the (adjusted) evolution equations first and to 
evaluate the according
perturbative errors in the (adjusted) constraint propagation equations. 
However, for the simplicity, we apply the perturbation to the pair of
constraints directly and analyze the effects of adjustments in its 
propagation equations.  
The latter, we think, presents the feature of constraint propagation more 
clearly for our purposes. 

The discussion becomes clear if we expand the constraint 
$C_\mu := ({\cal H}, {\cal M}_i)^T$
using vector harmonics,  
\begin{equation}
C_\mu = \sum_{l,m} \left(  A^{lm} a_{lm} + B^{lm} b_{lm} + C^{lm} c_{lm} + 
D^{lm} d_{lm} \right), 
\label{vectorharmonics}
\end{equation}
where we choose the basis as 
\begin{eqnarray}
{a}_{lm} (\theta, \varphi) &=&(Y_{lm},0,0,0)^T, \label{vectorharmonics1}
\\
{b}_{lm} (\theta, \varphi) &=&(0,Y_{lm},0,0)^T,
\\
{c}_{lm} (\theta, \varphi) &=&
{r \over \sqrt{ l (l+1)}}( 0,  0, \partial_\theta Y_{lm} , \partial_\varphi 
Y_{lm})^T, 
\\
{d}_{lm} (\theta, \varphi) &=&
{r \over \sqrt{ l (l+1)}}
(0, 0,  - {1 \over \sin \theta} \partial_\varphi Y_{lm} , \sin \theta
\, \partial_\theta Y_{lm} )^T,  \label{vectorharmonics4}
\end{eqnarray}
and the  coefficients $A^{lm}, \cdots, D^{lm}$ are functions of $(t, r)$.
Here $Y_{lm}$ is the spherical harmonic function, 
\begin{equation}
Y_{lm}(\theta,\varphi) =  
 (-1)^{(m+|m|)/2} \sqrt{   {(2l+1)\over 4 
\pi}{(l-|m|)! \over  (l+|m|)!}} \, 
P^m_l (\cos \theta) e^{im \varphi}. 
\end{equation}
The basis (\ref{vectorharmonics1})-(\ref{vectorharmonics4}) are normalized 
so that they satisfy
\begin{equation}
\langle C_\mu, C_\nu \rangle = \int_0^{2 \pi} d \varphi \int_0^\pi 
\,  C^\ast_\mu C_\rho \,  \eta^{\mu \rho} 
\sin \theta d \theta,
\end{equation}
where $\eta^{\mu \rho}$ is Minkowskii metric and the asterisk denotes 
the complex conjugate. 
Therefore
\begin{equation}
A^{lm} =\langle a^{lm}_{(\nu)}, C_\nu \rangle,  \quad 
\partial_t A^{lm} =\langle a^{lm}_{(\nu)}, \partial_t C_\nu \rangle, 
\quad \mbox{etc.}
\end{equation}

\begin{figure}[t]
\setlength{\unitlength}{1cm}
\begin{picture}(15,6)
\put(1.0,0){\epsfxsize=6.0cm \epsffile{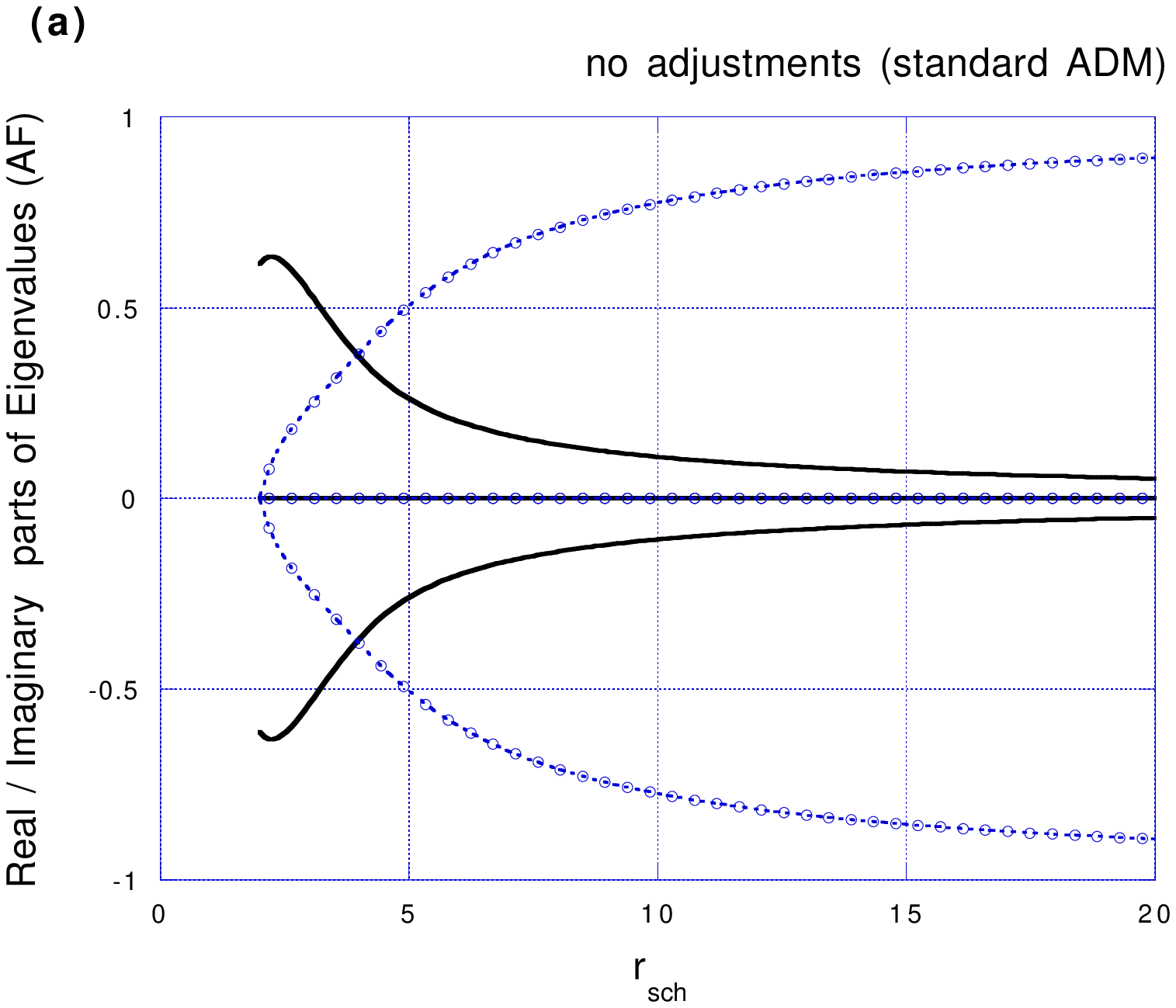} }
\put(8.5,0){\epsfxsize=6.0cm \epsffile{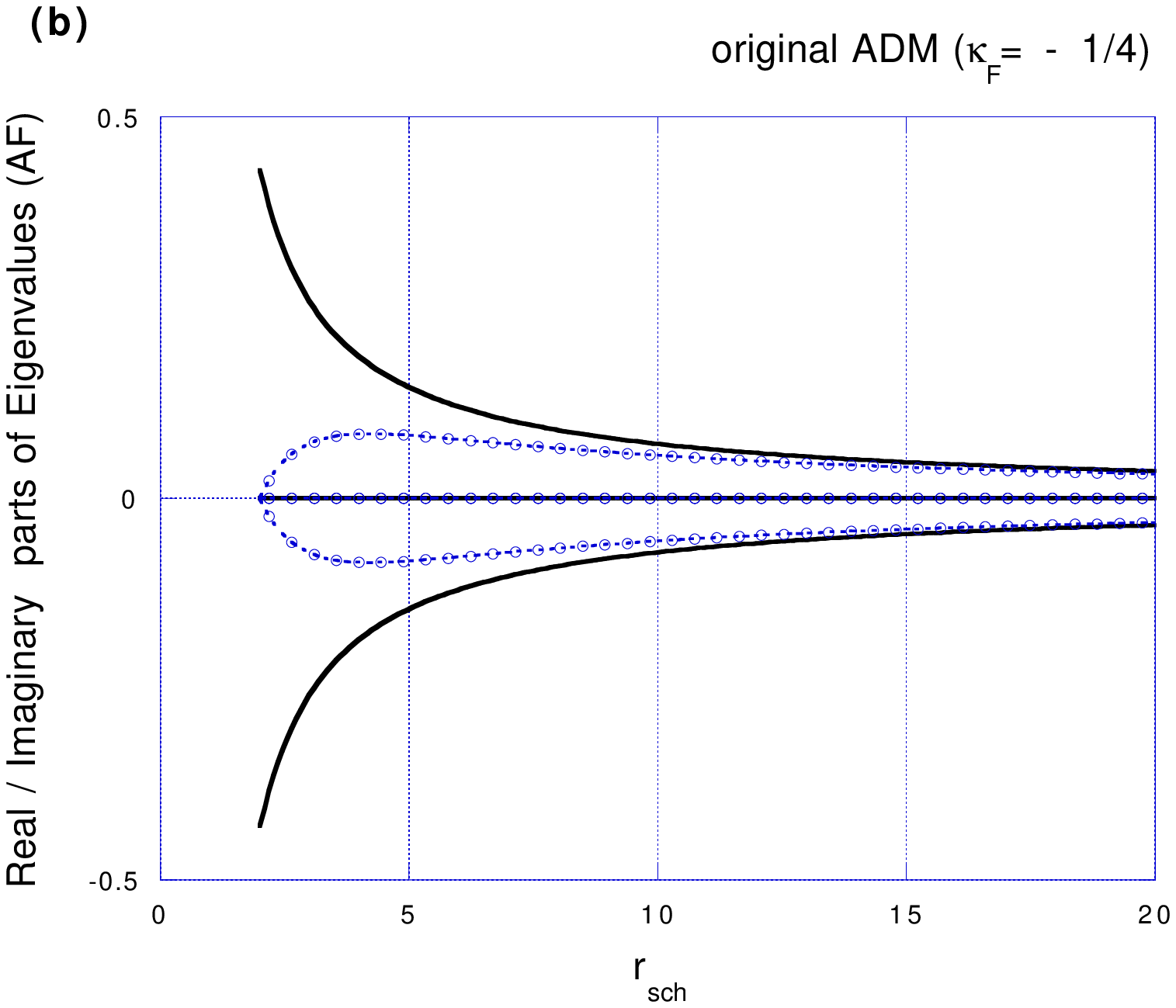} }
\end{picture}
\caption[quartic]{
Amplification factors (CAFs, eigenvalues of 
homogenized constraint propagation 
equations) 
are shown for 
the standard Schwarzschild coordinate, with 
(a) no adjustments, i.e., standard ADM, and  
(b) original ADM ($\kappa_F=-1/4$). 
[see eq. (\ref{originalADMadjust})]. 
The solid lines and the dotted lines with circles are 
real parts and imaginary parts, respectively.  
They are four lines each, but 
actually the two eigenvalues are zero for all cases.  
Plotting range is $2 < r \le 20$ using
Schwarzschild radial coordinate. We set $k=1, l=2,$ and 
$m=2$ throughout the article. 
(Reprinted from \cite{adjADMsch}, \copyright APS 2002)
}
\label{fig1schadm}
\end{figure}

In order to analyze the radial dependences, 
we also express these evolution equations using the Fourier expansion
on the radial coordinate, 
\begin{equation}
A^{lm} = \sum_k \hat A^{lm}_{(k)}(t) \, e^{ik r} \quad \mbox{etc.}
\end{equation}
So that we can obtain the RHS of the evolution equations
for $(\hat A^{lm}_{(k)}(t), \cdots, \hat D^{lm}_{(k)}(t))^T$
in a homogeneous form.

\subsubsection{Constraint amplification factors in Schwarzschild spacetime}

We present our CAF analysis 
in Schwarzschild black hole spacetime, which metric is 
\begin{equation}
ds^2=-(1-{2M \over r})dt^2+{dr^2 \over 1-{2M/ r}} +r^2 d\Omega^2,   
\qquad\mbox{(the~standard~expression)} 
\label{standardSch}
\end{equation}
where $M$ is the mass of a black hole. 
For numerical relativists, evolving a single black hole
is the essential test problem, though it is a trivial at first sight. 
The standard expression, (\ref{standardSch}), has a coordinate 
singularity at $r=2M$, so that we need to move another coordinate
for actual numerical time integrations.  

The ingoing Eddington-Finkelstein (iEF) coordinate has become
popular in numerical relativity, in order to excise black hole singularity,
since iEF penetrates the horizon without an irregular coordinate.
The expression is, 
\begin{equation}
ds^2=-(1-{2M\over r})dt_{iEF}^2+{4M \over r} dt_{iEF} dr 
+ (1+{2M\over r}) dr^2
+r^2 d\Omega^2 
\qquad\mbox{(the~iEF~expression)} 
\label{iEFSch}
\end{equation}
which is given by $t_{iEF}=t+2M \log (r - 2M)$ and 
the radial coordinate is common to
(\ref{standardSch}).

\begin{figure}[t]
\setlength{\unitlength}{1cm}
\begin{picture}(15,6)
\put(1.0,0){\epsfxsize=6.0cm \epsffile{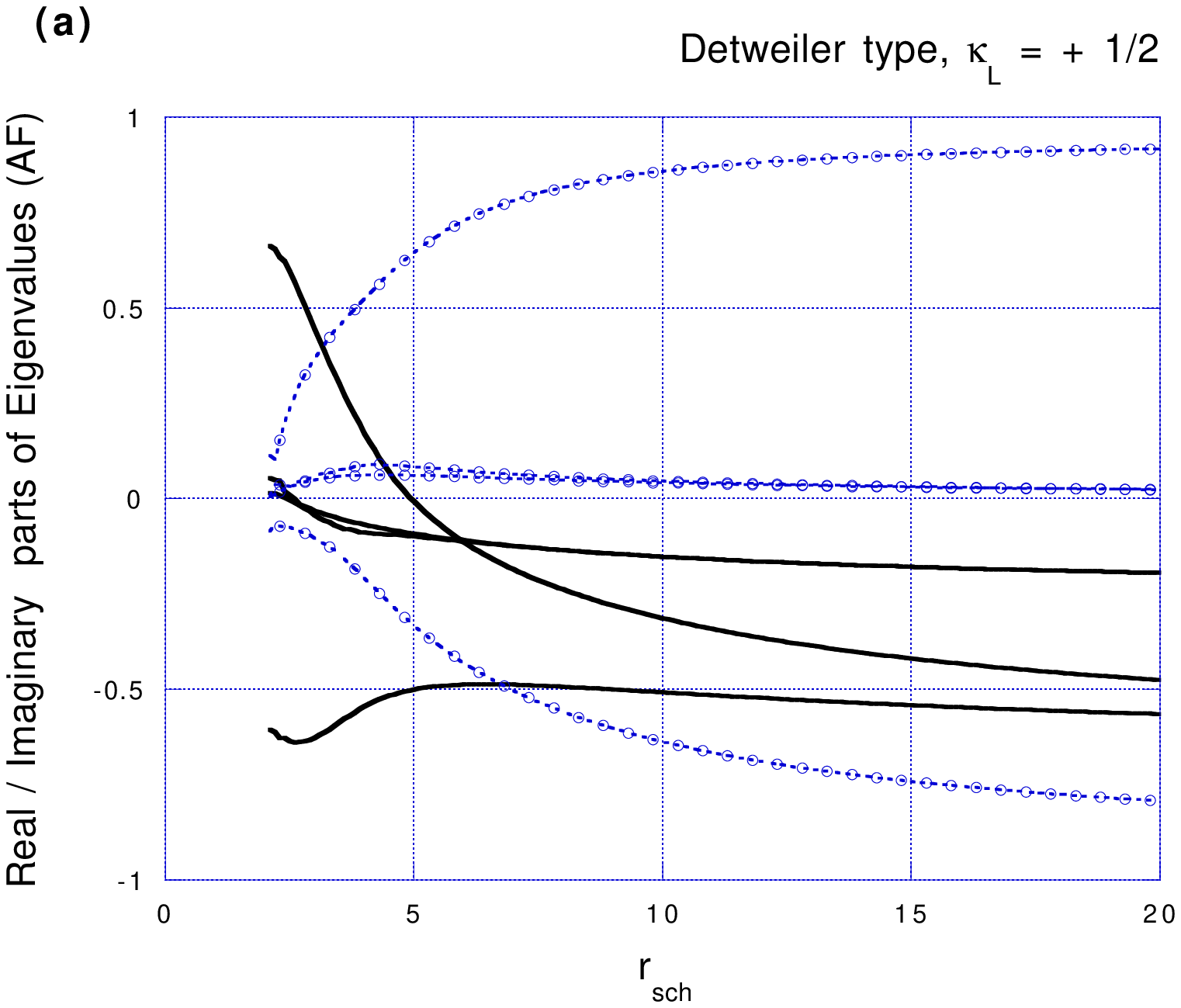} }
\put(8.5,0)  {\epsfxsize=6.0cm \epsffile{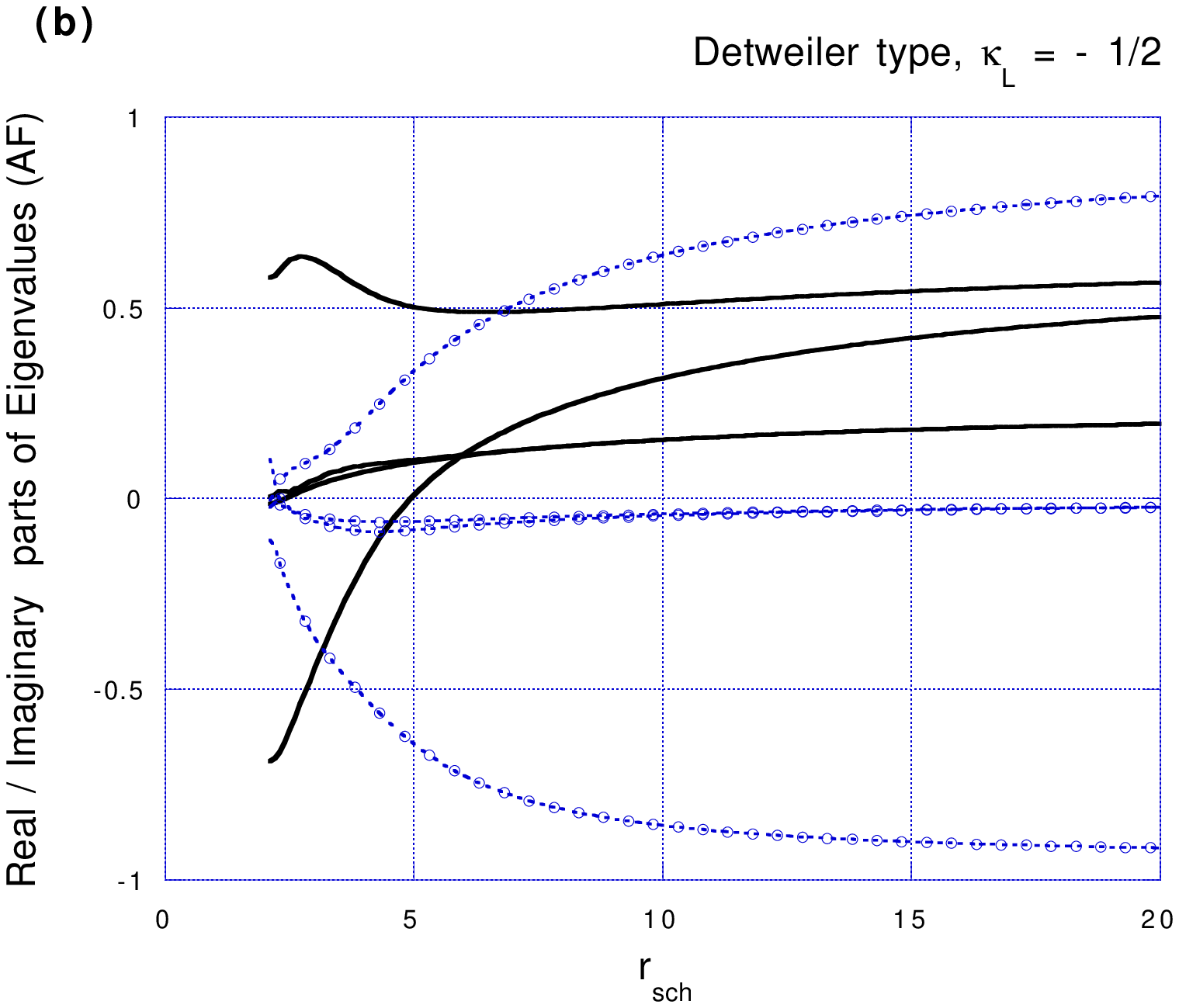} }
\end{picture}
\caption[quartic]{
Amplification factors of the standard Schwarzschild coordinate, with 
Detweiler type adjustments, (\ref{Det1})-(\ref{Det4}). 
Multipliers used in the plot are 
(a) $\kappa_L=+1/2$, and 
(b) $\kappa_L=-1/2$.
Plotting details are the same as Fig.\ref{fig1schadm}. 
(Reprinted from \cite{adjADMsch}, \copyright APS 2002)
}
\label{fig2schdet}
\end{figure}
\begin{figure}[t]
\setlength{\unitlength}{1cm}
\begin{picture}(15,6)
\put(1.0,0){\epsfxsize=6.0cm \epsffile{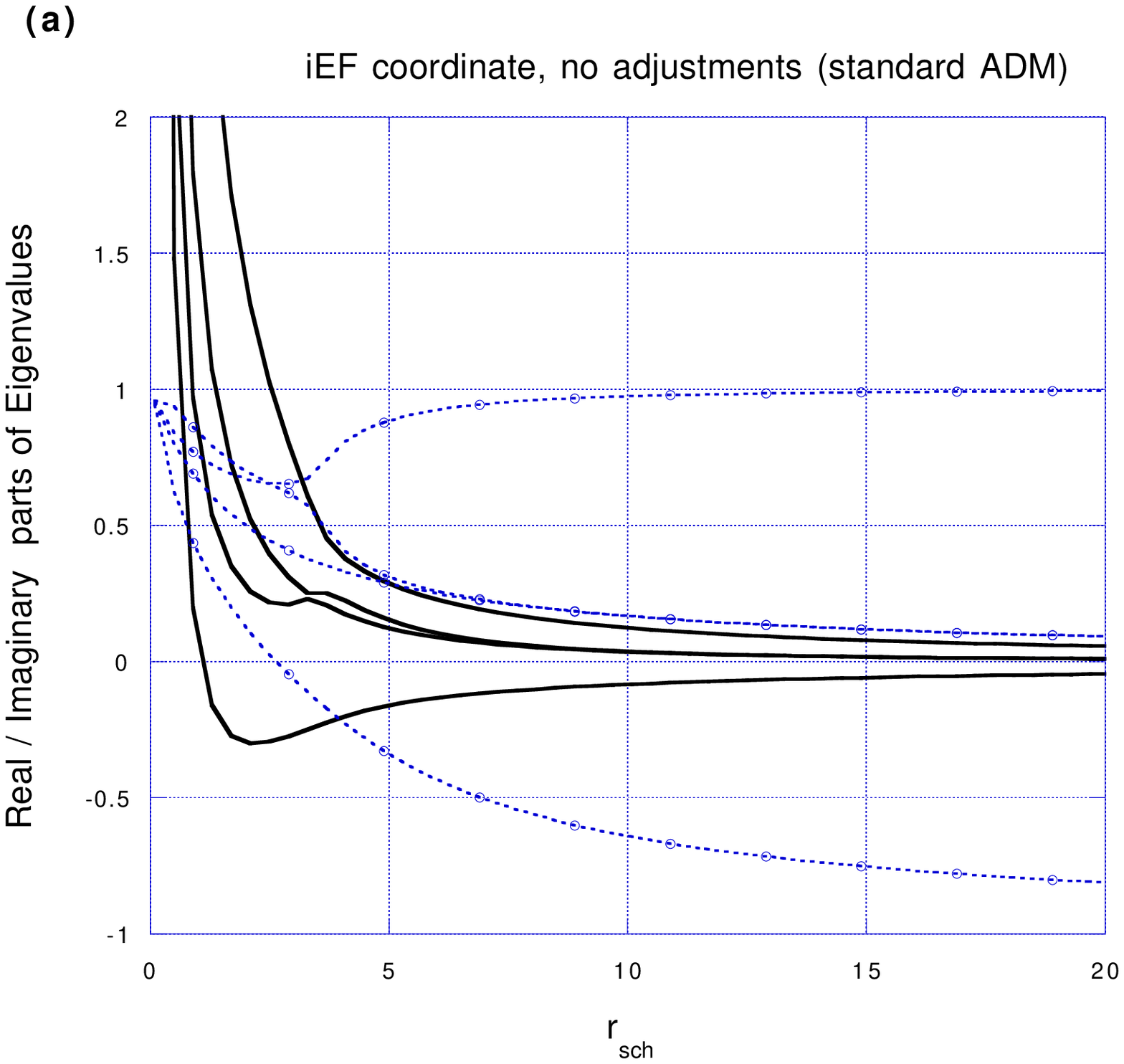} }
\put(8.5,0){\epsfxsize=6.5cm \epsffile{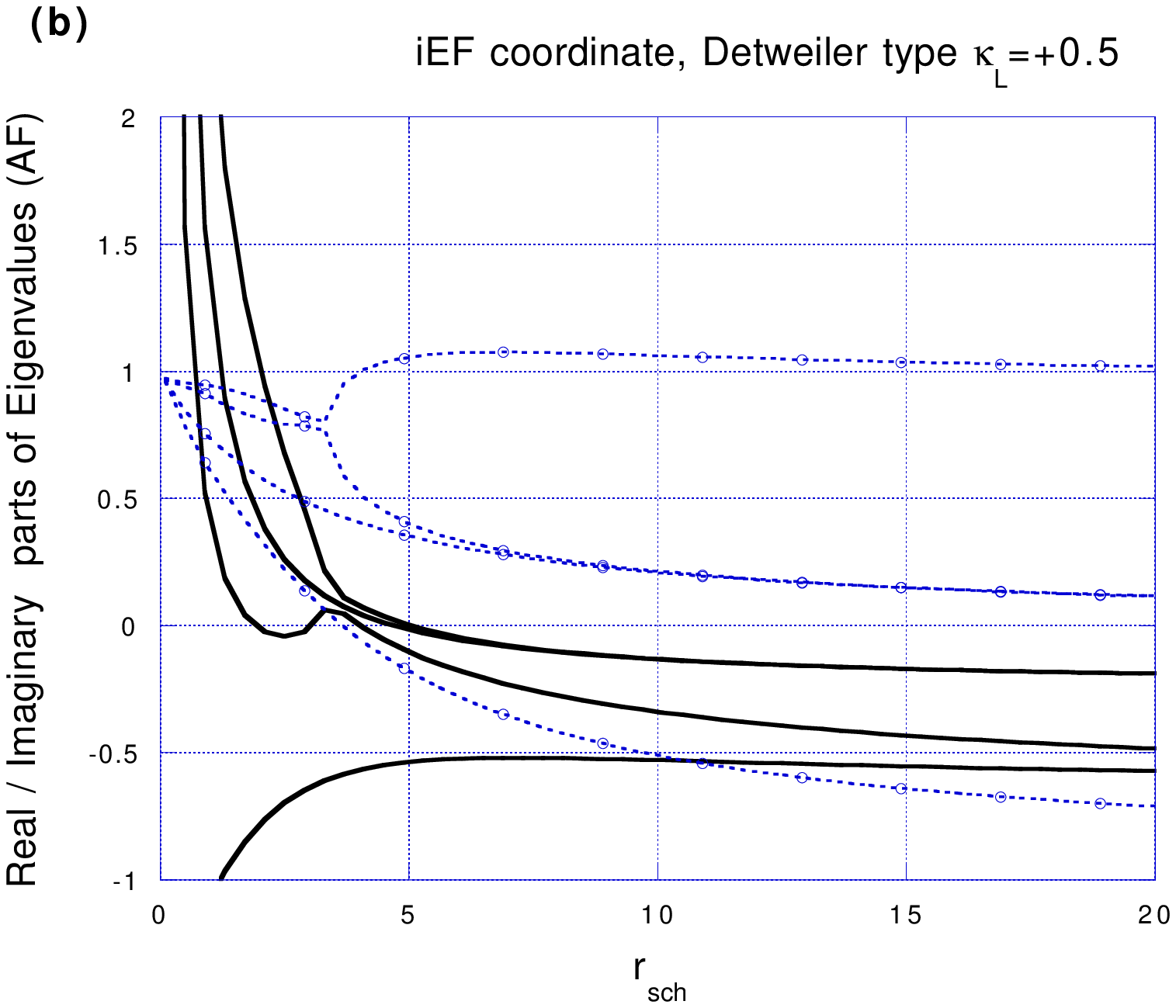} }
\end{picture}
\caption[quartic]{
CAFs in the iEF coordinate (\ref{iEFSch}) on $t=0$ slice 
for  the standard ADM formulation (i.e. no adjustments)
[See Fig.\ref{fig1schadm}(a) for the standard Schwarzschild coordinate.]
and for Detweiler adjustments with $\kappa_L=+1/2$  
[See Fig.\ref{fig2schdet}(a) for the standard Schwarzschild 
coordinate.]. 
The solid four lines and the dotted four lines with circles are 
real parts and imaginary parts, respectively.  
(Reprinted from \cite{adjADMsch}, \copyright APS 2002)
}
\label{fig4adm}
\end{figure}

We show CAFs in the following cases.  
More examples are available in \cite{adjADMsch}.  
\been
\item 
{\bf in the standard Schwarzschild metric expression (\ref{standardSch})}
\beit
\item 
{\bf For the adjustment (\ref{originalADMadjust})}, 
CAFs are obtained as 
\begin{eqnarray}
\Lambda^i &=&(0,0,\sqrt{a},-\sqrt{a}),  \label{eigenvalueADM} \\
a&=& -k^2 + { 4M k^2 r^2 (r-M)+ 2M(2r-M) + l(l+1) r (r-2M) + ikr(2r^2-3Mr-2M^2) 
\over r^4} \nonumber 
\end{eqnarray}
for the choice of $\kappa_1=0$ (the standard ADM), while they are
\begin{equation}
\Lambda^i =(0,0,\sqrt{b},-\sqrt{b}), \qquad
b= {M (2r-M) + irkM (2M-r) \over r^4} \label{eigenvalueOADM}
\end{equation}
for the choice of $\kappa_1=-1/4$ (the original ADM).  

These are plotted in Fig.\ref{fig1schadm}.  
The solid lines and dotted lines with circles are 
real parts and imaginary parts of CAFs, respectively.   
They are four lines each, but 
as we showed in (\ref{eigenvalueADM}), two of them are zero. 
The plotting range is $2 < r \le 20$ in Schwarzschild 
radial coordinate. The
CAFs at $r=2$ are $\pm \sqrt{3/8}$ and $0$. 
The existence of this positive real CAF near the horizon
is an important result. 
We show only the cases with $l=2$ and $k=1$, because we judged that the plots of 
$l=0$ and other $k$s are qualitatively the same. \\
The adjustment (\ref{originalADMadjust}) with 
$\kappa_F=-1/4$ returns the system 
back to the original ADM.  CAFs are (\ref{eigenvalueOADM}) 
and we plot them in 
Fig.\ref{fig1schadm}(b).  We can see that the imaginary parts are 
apparently different from those of 
the standard ADM [Fig.\ref{fig1schadm}(a)].  
This is the same feature 
as in the case of the flat background \cite{adjADM}.
According to our conjecture, 
the non-zero imaginary values are better than zeros, so 
we expect that the standard
ADM has a better evolution property than the original ADM system.  
Negative  $\kappa_F$ always makes the asymptotical 
real values finite. 
\item {\bf The Detweiler-type adjustment (\ref{Det1})-(\ref{Det4})}
makes the feature 
completely different.  
Fig.\ref{fig2schdet}(a) and (b) are the cases of $\kappa_L= \pm 1/2$.
A great improvement can be seen in the positive $\kappa_L$ case
where {\it all} real parts 
become negative in large $r$. 
Moreover all imaginary parts are apart from zero. 
These are 
the desired features according to our conjecture. 
Therefore we expect the Detweiler adjustment has good stability properties
{\it except} near the black hole region. 
The CAF near the horizon {\it has} a positive real component. 
This is not contradictory with the 
Detweiler's original idea.  
His idea came from 
suppressing the {\it total} L2 norm of constraints on the spatial slice, 
while our plot indicates 
the existence of a {\it local} violation mode.   
The change of signature of $\kappa_L$
can be understood just by changing the signature of CAFs, and 
this fact can also be seen  to the other plot.  
In \cite{adjADMsch}
we reported that a partial adjustment (apply only (\ref{Det1}) or (\ref{Det4}))
is also effective. 
\enit
\item 
{\bf in the iEF coordinates (\ref{iEFSch})}
\beit
\item 
{\bf For the adjustment (\ref{originalADMadjust})}, we plotted CAFs in  
Figure \ref{fig4adm}. 
We Figure \ref{fig4adm}(a) is qualitatively different 
{}from Fig.\ref{fig1schadm}(a).  
This is because the iEF expression is 
asymmetric to time, i.e. has non-zero extrinsic curvature. 
We notice that while some CAFs in iEF 
remain positive in large $r$ region, 
that their nature changes due to the adjustments. 
\item 
{\bf For the Detweiler-type  adjustment (\ref{Det1})-(\ref{Det4})}, 
CAFs are as in 
Figure \ref{fig4adm}(b). 
Interestingly, all plots indicate that 
 all real parts of CAFs are negative, and imaginary
parts are non-zero (again except near the  black hole region). 
By arranging the multiplier parameter, 
there is a chance to get
all negative real CAFs outside the black hole horizon. 
For example,  all the real-part goes negative
outside the black hole horizon if $\kappa_L > 3.1 $, while large
$\kappa_L$ may introduce another instability problem \cite{ronbun2}.   
\enit
\enen

Such kinds of test can be done with other combinations. 
In Table \ref{tableADMADJ}, we listed our results for more examples. 
We defined the adjustment terms so that their positive 
multiplier parameter, $\kappa > 0$, makes the system 
{\it better} in stability according to our conjecture.  
(Here {\it better} means in accordance with 
our conjecture in Box 3.2 and 3.3). 
The table includes the above results and is intended to extract the
contributions of each term in (\ref{adjADM1}) and (\ref{adjADM2}).
The effects of adjustments (of each $\kappa>0$ case) to CAFs are commented upon 
for each
coordinate system and for real/imaginary parts of CAFs, respectively. 
These judgements are made at
the $r \sim O(10M)$ region on their $t=0$ slice.  
Among them, No. R-2 in Table \ref{tableADMADJ} explains
why a particular adjustment by PennState group \cite{PennState_Sch1D}
gives better stability than before.

\begin{table}[t]
{\scriptsize
\begin{tabular}{l|ll|c||c|c|c|c|c}
\hline
No. & \multicolumn{2}{l|}{adjustment} &  1st? &
 \multicolumn{3}{c|}{Sch coord.} & \multicolumn{2}{c}{iEF  coord.} 
\\
    &  &    &   & TRS  & real. & imag. &  real. & imag.  
\\ \hline \hline 
0 &  -- & no adjustments  & yes & -- & -- & -- & -- & -- 
\\ \hline 
P-1 &
$P_{ij}$ & $-\kappa_L \alpha^3 \gamma_{ij}$ & no & no
& makes 2 Neg. & not app. & makes 2 Neg. & not app. 
\\ \hline 
P-2 & 
$P_{ij}$ & $-\kappa_L \alpha \gamma_{ij}$ & no & no  
& makes 2 Neg. & not app. & makes 2 Neg. & not app. 
\\ \hline 
P-3 & 
$P_{ij}$ & $P_{rr}=-\kappa$ 
& no &   no
& slightly enl.Neg.  & not app.  & slightly enl.Neg.  & not app. 
\\ \hline 
P-4 & 
$P_{ij}$ & $-\kappa \gamma_{ij}$ & no  &  no
& makes 2 Neg.  & not app.  & makes 2 Neg. &  not app.
\\ \hline 
P-5 & 
$P_{ij}$ & $-\kappa \gamma_{rr}$ & no &   no
& red. Pos./enl.Neg.  & not app.  &red.Pos./enl.Neg.  & not app.
\\ \hline 
Q-1  &
$Q^k{}_{ij}$ 
& $\kappa \alpha \beta^k \gamma_{ij}$   & no &no  & N/A & N/A & 
$\kappa\sim1.35$ min.vals. & not
app.
\\ \hline 
Q-2  &  
$Q^k{}_{ij}$ 
& $Q^r{}_{rr}=\kappa$   & no & yes & 
red. abs vals. & not app.  & red. abs vals.  & not app.
\\ \hline 
Q-3  & 
$Q^k{}_{ij}$ 
& $Q^r{}_{ij}=\kappa\gamma_{ij}$ 
& no & yes &  
red. abs vals. & not app. & enl.Neg. & enl. vals.
\\ \hline 
Q-4  &  
$Q^k{}_{ij}$ 
& $Q^r{}_{rr}=\kappa \gamma_{rr}$   & no & yes &  
red. abs vals. & not app.   & red. abs vals. & not app.
\\ \hline 
R-1 &  $R_{ij}$ & $\kappa_F \alpha \gamma_{ij}$ & yes & yes 
& \multicolumn{2}{c|}{$\kappa_F=-1/4$ min. abs vals.}  &
\multicolumn{2}{c}{$\kappa_F=-1/4$
min. vals.}
\\ \hline 
R-2 &  $R_{ij}$ & $R_{rr} = -\kappa_\mu \alpha $ 
& yes & no 
& not app. &  not app. & red.Pos./enl.Neg. & enl. vals.
\\ \hline 
R-3 &  $R_{ij}$ & $R_{rr} = -\kappa \gamma_{rr} $ & yes & no 
& enl. vals.  & not app.   & red.Pos./enl.Neg.  &  enl. vals.
\\ \hline 
S-1 & 
$S^k{}_{ij}$ & 
$S^k{}_{ij} =$ (\ref{Det3}) 
& yes & no  & not app. & not app. & 
not app. & not app.
\\ \hline 
S-2 &  $S^k{}_{ij}$ & 
$\kappa \alpha \gamma^{lk} (\partial_l \gamma_{ij}) $ & yes & no
& makes 2 Neg.  & not app.  & makes 2 Neg. & not app.
\\ \hline 
p-1 & 
$p^{k}{}_{ij}$ & $p^{r}{}_{ij}=-\kappa \alpha \gamma_{ij}$ & no  & no  
& red. Pos.   & red. vals.  & red. Pos.  & enl. vals.
\\ \hline 
p-2 & 
$p^{k}{}_{ij}$ & $p^{r}{}_{rr}=\kappa \alpha$ & no  &   no
& red. Pos.  & red. vals.  & red.Pos/enl.Neg. & enl. vals.
\\ \hline 
p-3 & 
$p^{k}{}_{ij}$ & $p^{r}{}_{rr}=\kappa \alpha \gamma_{rr}$ &  no &   no
& makes 2 Neg. & enl.vals. & red.Pos.vals. &  red.vals.
\\ \hline 
q-1 & 
$q^{kl}{}_{ij}$ & $q^{rr}{}_{ij}=\kappa \alpha \gamma_{ij}$ & no   & no  
& $\kappa=1/2$ min.vals.  & red.vals.  & not app. & enl.vals. 
\\ \hline 
q-2 & 
$q^{kl}{}_{ij}$ & $q^{rr}{}_{rr}=-\kappa \alpha \gamma_{rr}$ & no  &  yes
& red. abs vals.  & not app.  & not app. & not app.
\\ \hline 
r-1 &  
$r^{k}{}_{ij}$ & $r^{r}{}_{ij}=\kappa \alpha \gamma_{ij}$ & no  &  yes
& not app. & not app.  & not app. & enl. vals. 
\\ \hline 
r-2 & 
$r^{k}{}_{ij}$ & $r^{r}{}_{rr}=-\kappa \alpha$ &  no  &  yes  
& red. abs vals. & enl. vals.  & red. abs vals. &  enl. vals.
\\ \hline 
r-3 & 
$r^{k}{}_{ij}$ & $r^{r}{}_{rr}=-\kappa \alpha \gamma_{rr}$ & no  &  yes
& red. abs vals. & enl. vals. & red. abs vals. & enl. vals.
\\ \hline 
s-1 &  
$s^{kl}{}_{ij}$ & 
$s^{kl}{}_{ij} =$ (\ref{Det4}) 
& no & no 
& makes 4 Neg. & not app. & makes 4 Neg. & not app.
\\ \hline 
s-2 & 
$s^{kl}{}_{ij}$ & $s^{rr}{}_{ij}=-\kappa \alpha \gamma_{ij}$ & no  &  no
& makes 2 Neg.  & red. vals.  & makes 2 Neg. & red. vals.
\\ \hline 
s-3 & 
$s^{kl}{}_{ij}$ & $s^{rr}{}_{rr}= - \kappa \alpha \gamma_{rr}$ & no  &  no
& makes 2 Neg.  & red. vals.  & makes 2 Neg. & red. vals.
\\  \hline 
\end{tabular}
}
\caption{List of ADM adjustments we tested in the Schwarzschild spacetime.  
The column of adjustments are nonzero multipliers
in terms of (\ref{adjADM1}) and (\ref{adjADM2}).
The column `TRS' indicates whether each adjusting term satisfies the 
time reversal symmetry or not
on the standard Schwarzschild coordinate. 
(`No' means a candidate that makes asymmetric CAFs.)
 The column `1st?' indicates whether each adjusting term breaks the first-order feature of 
the standard constraint propagation equation, (\ref{CHproADM}) and (\ref{CMproADM}). 
(`Yes' keeps the system first-order, 
`No' may break hyperbolicity of 
constraint propagation depending a choice of $\kappa$.) 
The effects to CAFs (when $\kappa>0$) are commented 
for both coordinates
and both real/imaginary parts, respectively. 
The `N/A' means that there is no effect due to the coordinate properties;  
`not app.' (not apparent) means the adjustment does not change the CAFs
effectively according to our conjecture;
`enl./red./min.' means enlarge/reduce/minimize,  and 
`Pos./Neg.' means positive/negative, respectively. 
These judgements are made
at the $r \sim O(10M)$ region on their $t=0$ slice.  See more detail in \cite{adjADMsch}.
}
\label{tableADMADJ}
\end{table}
%

\subsubsection{Remarks}
\paragraph{Numerical demonstrations}

The above analyses are only predictions, and 
supporting numerical demonstrations are necessary for the next steps. 
Systematic numerical comparisons are progressing, and we show two sample
plots here. 

\begin{figure}[t]
\setlength{\unitlength}{1cm}
\begin{picture}(15,6)
\put(0.5,0){\epsfxsize=7.0cm \epsffile{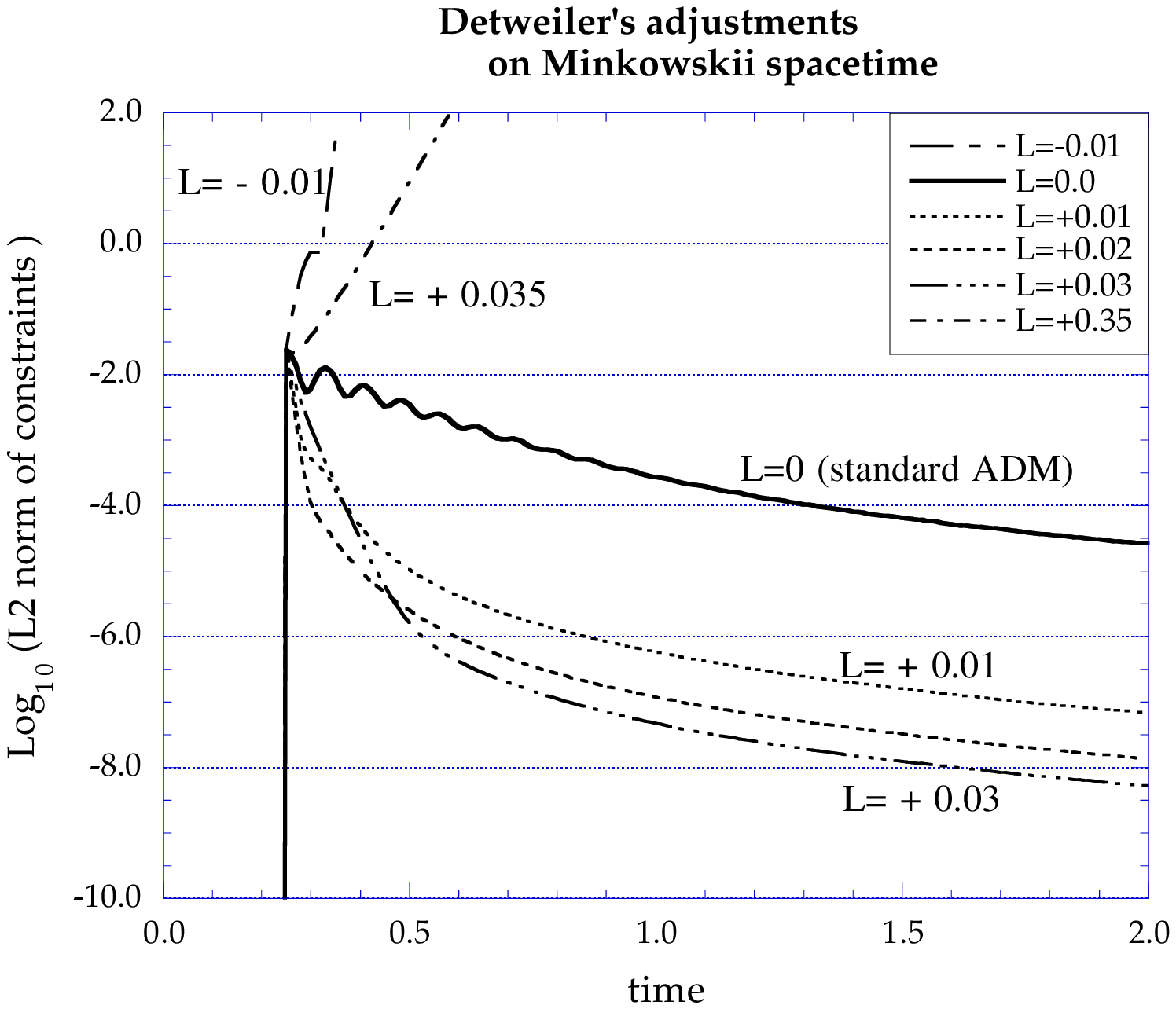}}
\put(9.0,0){\epsfxsize=7.0cm \epsffile{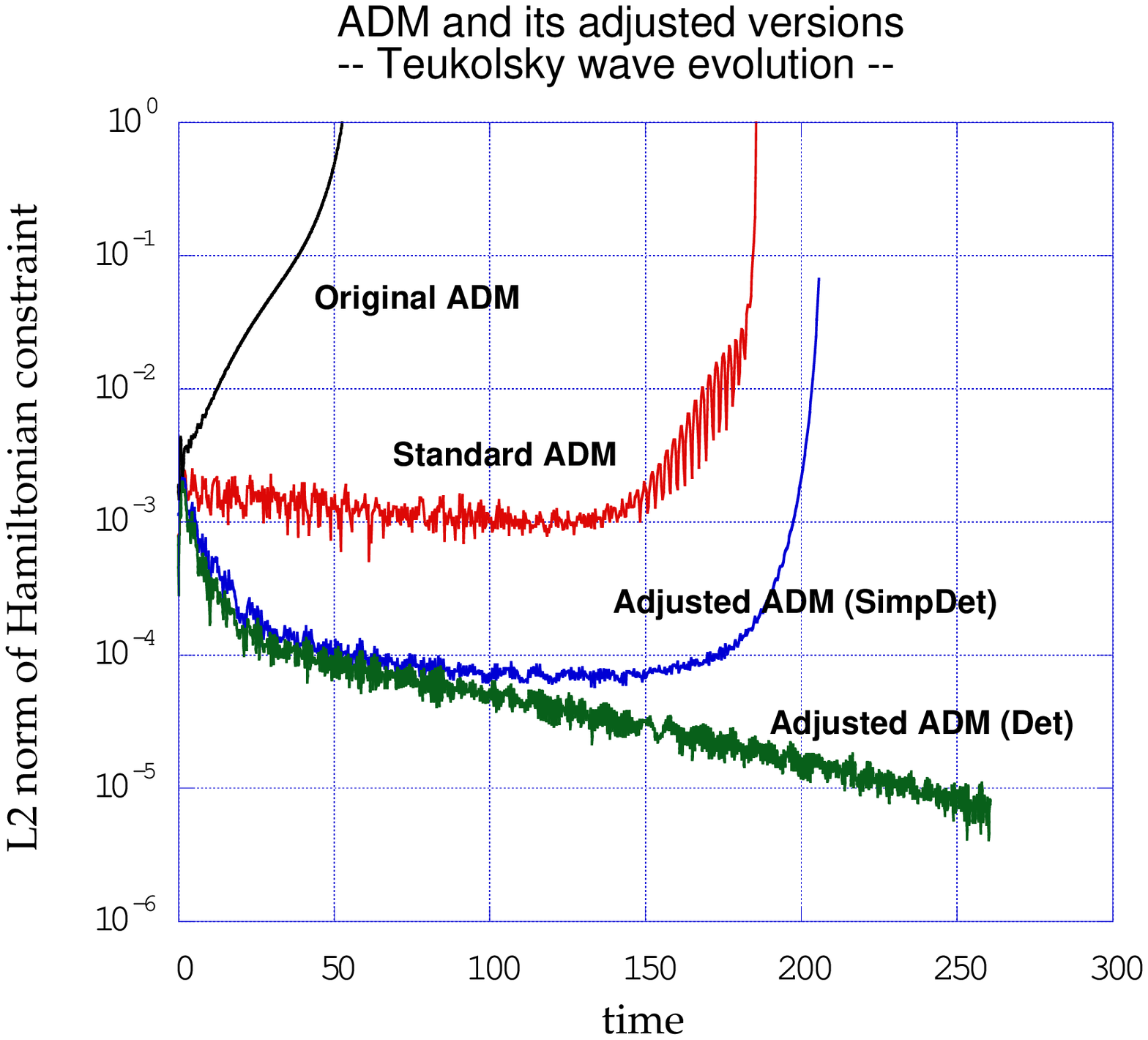} }
\end{picture}
\caption[quartic]{Comparisons of numerical evolution
between adjusted ADM systems. 
(a) Demonstration of the Detweiler's modified ADM system 
on Minkowskii background
spacetime, 1-dimensional simulation. 
The L2 norm of the constraints ${\cal H}^{ADM}$ and ${\cal M}^{ADM}$ 
is plotted in
the function of time.  Artificial error was
added at $t=0.25$.
$L$ is the parameter used in (\ref{Det1})-(\ref{Det3}).
We see the evolution is asymptotically constrained for small $\kappa>0$.
(Reprinted from \cite{adjADM}, \copyright APS 2001)
(b) L2 norm of the Hamiltonian constraint ${\cal H}^{ADM}$ 
of evolution using ADM/adjusted ADM formulations for the case
of Teukolsky wave, 3-dimensional simulation.  
Cactus-based original ADM
module (CactusGR) was used.  (Shinkai-Yoneda, in preparation). 
}
\label{figteukwaveADM}
\end{figure}

Figure \ref{figteukwaveADM} (a) is a test numerical evolution 
of Detweiler-type adjustment on the Minkowskii background. 
We see the adjusted version gives convergence on to the constraint surface
by arranging the magnitude of the adjusting parameter, $\kappa$. 

Figure \ref{figteukwaveADM} (b) is an example from the project of
``Comparison of formulations of the Einstein equations for numerical relativity"
\cite{mexico}. (We will describe the project in \S \ref{sec:outlook}).  
The plot was obtained by a 3-dimensional numerical evolution of weak 
gravitational wave, the so-called Teukolsky wave \cite{Teukolskywave}. 
The lines are of the original/standard ADM evolution equations, Detweiler-type
adjustment, and a part of Detweiler-type adjustment (actually used only
(\ref{Det1})).  For a particular choice of $\kappa$, we observe again the
L2 norm of constraint (violation of constraints) is reduced
than the standard ADM case, and can evolve longer than that. 

\paragraph{Notion of Time Reversal Symmetry}

During the comparisons of adjustments, we found that it is necessary to 
create time asymmetric structure of evolution equations in order to 
force the evolution on to the constraint surface.  
There are infinite ways of adjusting equations, but we found that if we 
follow the guideline Box 3.5, then such an adjustment will give us time
asymmetric evolution. 

\Largefbox{\boxwidth}{
{\bf Trick to obtain asymptotically constrained system}:
\hspace*{\fill} {\bf Box 3.5}\\
{\bf $=$ Break the time reversal symmetry (TRS) of the evolution equation. }
\been
\item Evaluate the parity of the evolution equation.\\
By reversing the time ($\partial_t \rightarrow -\partial_t$), there are
variables which change their signatures (parity $(-)$)
 [e.g. $K_{ij}, \partial_t \gamma_{ij}, {\cal M}_i, \cdots$], 
while not (parity $(+)$) [e.g. $g_{ij}, \partial_t K_{ij}, {\cal H}, \cdots$]. 
\item Add adjustments which have different parity of that equation.\\
For example,  for the parity $(-)$ equation $\partial_t \gamma_{ij}$, 
add a parity $(+)$ adjustment $\kappa {\cal H}$.  
\enen  
}
One of our criteria, the negative real CAFs, requires breaking
the time-symmetric features of the original evolution equations. 
Such CAFs are obtained by adjusting the terms which break the 
TRS of the evolution equations, and this is available even at the
standard ADM system. 
The TRS features (on the standard Schwarzschild metric, which is 
time-symmetric background) are also denoted 
in Table \ref{tableADMADJ}. 
The criteria of Box 3.5 will be applied also to the BSSN equation (see 
\S \ref{subsec:modifiedBSSN}).

\paragraph{Differences with Detweiler's requirement}
We comment on the differences between Detweiler's criteria 
\cite{detweiler} and ours.
Detweiler calculated the L2 norm of the constraints, $C_\rho$, 
over the 3-hypersurface
and imposed the negative definiteness of its evolution,
\begin{equation}
\mbox{Detweiler's criteria}
  \; \Leftrightarrow \;  \partial_t \int  C_\rho C^\rho \ dV <0,
\:  \ \forall \mbox{ non zero } C_\rho.
\end{equation}
where $C_\rho C^\rho =:G^{\rho\sigma} C_\rho C_\sigma$,  and
$G_{\rho\sigma}= diag [1, \gamma_{ij}]$ for the pair of 
$C_\rho=({\cal H}, {\cal M}_i)$.

Assuming the constraint propagation to be
$\partial_t \hat C_\rho=A_\rho{}^\sigma \hat C_\sigma$ 
in the Fourier components,
the time derivative of the L2 norm can be written as
\begin{equation}
\partial_t (\hat{C}_\rho \hat{C}^\rho)
=
(A^{\rho\sigma}+\bar{A}^{\sigma\rho}+\partial_t \bar{G}^{\rho\sigma})
\hat{C}_\rho \bar{\hat{C}}_\sigma.
\end{equation}
Together with the fact that the L2 norm is preserved by Fourier 
transform,
we can say, for the case of {\it static} background metric,
\bear
\mbox{Detweiler's criteria}
  & \Leftrightarrow &  \mbox{eigenvalues of } (A+A^\dagger)
\mbox{ are all negative} \ \forall k.
\\
\mbox{Our criteria}
  & \Leftrightarrow &  \mbox{eigenvalues of } A
\mbox{ are all negative} \ \forall k.
\enar
Therefore for the case of static background,
Detweiler's criterion is stronger than ours. For example, 
the matrix
\begin{equation}
A=
\left(\matrix{
-1&a \cr 0&-1
 }\right)
\mbox{~~where~} a \mbox{~is~constant,}
\end{equation}
for the evolution system $(\hat{C}_1, \hat{C}_2)$ 
satisfies our criterion
but not Detweiler's 
when $|a| \ge \sqrt{2}$.
This matrix however gives asymptotical decay for 
$(\hat{C}_1, \hat{C}_2)$.
Therefore we may say that
Detweiler requires the monotonic decay of the constraints,
while we assume only asymptotical decay.

We remark that Detweiler's truncations on higher order terms
in $C$-norm corresponds to our
perturbational analysis;  both are based on the idea that the deviations
from constraint surface
(the errors expressed non-zero constraint value) 
are initially small.

\paragraph{Ranges of effective $\kappa$}
We do not discuss the ranges of the effective
multiplier parameter, $\kappa$, since the range depends on the
characteristic speeds of the models and 
numerical integration schemes as we observed in \cite{ronbun2}.
At this moment, we can only estimate the maximum value for adjusting
parameter $\kappa$s from von Neumann's stability analysis, while 
we do not have a background theoretical explanation to predict the
optimized value of $\kappa$ for numerical evolution.  
We will comment
this point later again in \S \ref{sec:outlook}.

\subsection{Adjusted BSSN formulations} \label{secADJBSSN}

Next we apply the idea of adjusted system to 
the BSSN system, Box 2.3. 

\subsubsection{Constraint propagation analysis of the BSSN equations}
\label{subsec1:ADJBSSN}

The BSSN has 5 constraint equations, $({\cal H}, {\cal M}_i, 
{\cal G}^i,  {\cal A}, {\cal S})$, see
(\ref{BSSNconstraintH})-(\ref{BSSNconstraintS}). 
We begin identifying where the BSSN evolution equations  
were adjusted already
in its standard notation, (\ref{BSSNeqmPHI})-(\ref{BSSNeqmTG}). 

Taking careful account of these constraints, 
(\ref{BSSNconstraintH}) and (\ref{BSSNconstraintM}) can be expressed 
directly as 
\begin{eqnarray} 
{\cal H} &=& 
e^{-4\varphi}\tilde{R} 
-8e^{-4\varphi}\tilde{D}^j\tilde{D}_j\varphi 
-8e^{-4\varphi}(\tilde{D}^j\varphi)(\tilde{D}_j\varphi) 
+(2/3)K^2 
-\tilde{A}_{ij}\tilde{A}^{ij} 
-(2/3) {\cal A} K, 
\\ 
{\cal M}_i 
&=& 
6\tilde{A}^j{}_{i}(\tilde{D}_j \varphi) 
-2{\cal A}(\tilde{D}_i \varphi) 
-(2/3) (\tilde{D}_i K) 
+\tilde{\gamma}^{kj}(\tilde{D}_j\tilde{A}_{ki}). 
\end{eqnarray}

By a straightforward 
calculation,  we get: 
\begin{eqnarray} 
\partial_t^B \varphi&=& 
\partial_t^A \varphi 
+(1/6)\alpha{\cal A} 
-(1/12)\tilde{\gamma}^{-1}(\partial_j{\cal S})\beta^j, 
\label{BSSNeq1} 
\\ 
\partial_t^B \tilde{\gamma}_{ij}&=& 
\partial_t^A \tilde{\gamma}_{ij} 
-(2/3)\alpha \tilde{\gamma}_{ij}{\cal A} 
+(1/3)\tilde{\gamma}^{-1}(\partial_k{\cal S}) 
\beta^k\tilde{\gamma}_{ij}, 
\label{BSSNeq2} 
\\ 
\partial_t^B K 
&=& 
\partial_t^A K 
-(2/3)\alpha K {\cal A} 
-\alpha {\cal H} 
+\alpha e^{-4\varphi}(\tilde{D}_j{\cal G}^j), 
\label{BSSNeq3} 
\\ 
\partial_t^B \tilde{A}_{ij} 
&=& 
\partial_t^A \tilde{A}_{ij} 
+\big( 
(1/3)\alpha \tilde{\gamma}_{ij}K 
-(2/3)\alpha \tilde{A}_{ij}\big) 
{\cal A} 
\nonumber \\&& 
+\big( 
(1/2)\alpha e^{-4\varphi}(\partial_k\tilde{\gamma}_{ij}) 
-(1/6)\alpha e^{-4\varphi} 
\tilde{\gamma}_{ij}\tilde{\gamma}^{-1}(\partial_k{\cal S}) 
\big) 
{\cal G}^k 
\nonumber \\&& 
+\alpha e^{-4\varphi}\tilde{\gamma}_{k(i}(\partial_{j)}{\cal G}^k) 
-(1/3)\alpha e^{-4\varphi}\tilde{\gamma}_{ij}(\partial_k{\cal G}^k), 
\label{BSSNeq4} 
\\ 
\partial_t^B\tilde{\Gamma}^i 
&=& 
\partial_t^A\tilde{\Gamma}^i 
+\big( 
-(2/3)(\partial_j \alpha) \tilde{\gamma}^{ji} 
-(2/3)\alpha (\partial_j \tilde{\gamma}^{ji}) 
-(1/3)\alpha \tilde{\gamma}^{ji}\tilde{\gamma}^{-1} 
(\partial_j{\cal S}) 
+4\alpha\tilde{\gamma}^{ij}(\partial_j \varphi) 
\big){\cal A} 
\nonumber \\&& 
-(2/3)\alpha \tilde{\gamma}^{ji}(\partial_j {\cal A}) 
+2\alpha\tilde{\gamma}^{ij}{\cal M}_j 
-(1/2)(\partial_k\beta^i)\tilde{\gamma}^{kj} 
\tilde{\gamma}^{-1}(\partial_j{\cal S}) 
+(1/6)(\partial_j\beta^k)\tilde{\gamma}^{ij} 
\tilde{\gamma}^{-1}(\partial_k{\cal S}) 
\nonumber \\&& 
+(1/3)(\partial_k\beta^k)\tilde{\gamma}^{ij} 
\tilde{\gamma}^{-1}(\partial_j{\cal S}) 
+(5/6)\beta^k\tilde{\gamma}^{-2}\tilde{\gamma}^{ij} 
(\partial_k{\cal S})(\partial_j{\cal S}) 
+(1/2)\beta^k\tilde{\gamma}^{-1}(\partial_k 
\tilde{\gamma}^{ij})(\partial_j{\cal S}) 
\nonumber \\&& 
+(1/3)\beta^k\tilde{\gamma}^{-1}(\partial_j 
\tilde{\gamma}^{ji})(\partial_k{\cal S}). 
\label{BSSNeq5} 
\end{eqnarray} 
where $\partial_t^A $ denotes the part of no replacements, i.e. 
the terms only use the standard ADM evolution equations in its 
time derivatives.

{}From (\ref{BSSNeq1})-(\ref{BSSNeq5}), we understand that 
all BSSN evolution equations are {\it adjusted} using constraints. 
However, from the viewpoint of time reversal symmetry (Box 3.5), 
all the above adjustments in (\ref{BSSNeq1})-(\ref{BSSNeq5})
unfortunately keep the time reversal symmetry. 
Therefore we can not expect direct decays of constraint 
violation in the present form. 

The set of the constraint propagation equations, 
$\partial_t ({\cal H}, {\cal M}_i, 
{\cal G}^i,  {\cal A}, {\cal S})^T$, 
turns to be not a first-order hyperbolic 
and includes many non-linear terms, (see the Appendix in \cite{adjBSSN}). 
In order to understand the fundamental structure, we show an 
analysis on the flat spacetime background. 

For the flat background metric $g_{\mu\nu}=\eta_{\mu\nu}$, 
the first order perturbation equations of 
(\ref{BSSNeq1})-(\ref{BSSNeq5})  can be written as 
\begin{eqnarray} 
\partial_t{}^{\!(1)\!\!}\varphi &=& 
-(1/6){}^{\!(1)\!\!} K 
+(1/6)(\kappa_{\varphi}-1){}^{\!(1)\!\!}{\cal A}, 
\label{eqmMinkow1} 
\\ 
\partial_t{}^{\!(1)\!\!}\tilde \gamma_{ij}&=& 
-2{}^{\!(1)\!\!}\tilde A_{ij} 
-(2/3)(\kappa_{\tilde \gamma}-1)\delta_{ij}{}^{\!(1)\!\!}{\cal A}, 
\label{eqmMinkow2} 
\\ 
\partial_t {}^{\!(1)\!\!} K 
&=& 
-(\partial_j \partial_j {}^{\!(1)\!\!}\alpha) 
+(\kappa_{K1}-1)\partial_j{}^{\!(1)\!\!}{\cal G}^j 
-(\kappa_{K2}-1){}^{\!(1)\!\!}{\cal H}, 
\label{eqmMinkow3} 
\\ 
\partial_t {}^{\!(1)\!\!} \tilde A_{ij} 
&=& 
{}^{\!(1)\!\!}(R^{BSSN}_{ij}){}^{TF} 
-{}^{\!(1)\!\!}(\tilde D_i \tilde D_j \alpha){}^{TF} 
+(\kappa_{A1}-1)\delta_{k(i}(\partial_{j)}{}^{\!(1)\!\!}{\cal G}^k) 
-(1/3)(\kappa_{A2}-1)\delta_{ij}(\partial_k {}^{\!(1)\!\!} {\cal G}^k), 
\label{eqmMinkow4} 
\\ 
\partial_t {}^{\!(1)\!\!} \tilde\Gamma^i &=& 
-(4/3)(\partial_i {}^{\!(1)\!\!} K) 
-(2/3)(\kappa_{\tilde \Gamma 1}-1)(\partial_i{}^{\!(1)\!\!}{\cal A}) 
+2(\kappa_{\tilde\Gamma 2}-1){}^{\!(1)\!\!}{\cal M}_i, 
\label{eqmMinkow5} 
\end{eqnarray} 
where we introduced parameters $\kappa$s, 
all $\kappa=0$ reproduce no adjustment 
case from the standard ADM equations, 
and  all $\kappa=1$ correspond to the BSSN 
equations. 
We express them as 
\begin{eqnarray} 
\kappa_{adj}&:=&(\kappa_{\varphi},\kappa_{\tilde{\gamma}}, 
\kappa_{K1}, 
\kappa_{K2}, 
\kappa_{A1}, 
\kappa_{A2}, 
\kappa_{\tilde{\Gamma} 1}, 
\kappa_{\tilde{\Gamma} 2}). 
\end{eqnarray}

Constraint propagation equations at the first order 
in the flat spacetime, then, become: 
\begin{eqnarray} 
\partial_t{}^{\!(1)\!\!}{\cal H}&=& 
\left(\kappa_{\tilde{\gamma}}-(2/3)\kappa_{\tilde{\Gamma} 
1}-(4/3)\kappa_{\varphi}+2\right) 
\partial_j\partial_j{}^{\!(1)\!\!}{\cal A} 
+2(\kappa_{\tilde{\Gamma} 2}-1)(\partial_j{}^{\!(1)\!\!}{\cal M}_j), 
\label{CPMinkow1} 
\\ 
\partial_t {}^{\!(1)\!\!} {\cal M}_i&=& 
\left 
(-(2/3)\kappa_{K1}+(1/2)\kappa_{A1}-(1/3)\kappa_{A2}+(1/2)\right) 
\partial_i\partial_j{}^{\!(1)\!\!}{\cal G}^j 
\nonumber \\ && 
+(1/2)\kappa_{A1}\partial_j\partial_j{}^{\!(1)\!\!}{\cal G}^i 
+\left((2/3)\kappa_{K2}-(1/2)\right) 
\partial_i{}^{\!(1)\!\!}{\cal H}, 
\\ 
\partial_t {}^{\!(1)\!\!} {\cal G}^i 
&=& 
2\kappa_{\tilde{\Gamma} 2}{}^{\!(1)\!\!}{\cal M}_i 
+(-(2/3)\kappa_{\tilde{\Gamma} 
1}-(1/3)\kappa_{\tilde{\gamma}})(\partial_i{}^{\!(1)\!\!}{\cal A}), 
\\ 
\partial_t {}^{\!(1)\!\!} {\cal S}&=& 
-2\kappa_{\tilde{\gamma}}{}^{\!(1)\!\!}{\cal A}, 
\\ 
\partial_t {}^{\!(1)\!\!} {\cal A}&=& 
(\kappa_{A1}-\kappa_{A2})(\partial_{j}{}^{\!(1)\!\!}{\cal G}^j). 
\label{CPMinkow5} 
\end{eqnarray} 

\subsubsection{The origin of the advantages of the BSSN equations}
\label{subsec2:ADJBSSN}
We next discuss CAFs  of 
(\ref{CPMinkow1})-(\ref{CPMinkow5}). 
Hereafter we let 
$k^2=k_x^2+k_y^2+k_z^2$ for Fourier wave numbers. 
\begin{enumerate} 
\item The no-adjustment case, $\kappa_{adj}=$(all zeros). This is the 
starting point of the discussion.  In this case, 
$$ 
\mbox{CAFs}=(0 \,  (\times 7),\pm \sqrt{-k^2}), 
$$ 
i.e., $(0 \, (\times 7), \pm \mbox{pure imaginary} 
\, \mbox{(1 pair)})$. 
In the standard ADM formulation, which uses $(\gamma_{ij}, K_{ij})$, 
CAFs are $(0, 0, \pm \mbox{Pure Imaginary})$ \cite{adjADM}. 
Therefore if we do not apply adjustments in BSSN equations the
constraint propagation structure is quite similar to that of 
the standard ADM. 
\item For the BSSN equations,  $\kappa_{adj}=$(all 1s), 
$$ 
\mbox{CAFs}=(0 \,  (\times 3),\pm \sqrt{-k^2} \, \mbox{(3 pairs)}), 
$$ 
i.e., $(0 \, (\times 3), \pm \mbox{Pure Imaginary} 
\, \mbox{(3 pairs)})$. 
The number of pure imaginary CAFs is increased over that of No.1, 
and we 
conclude this is the advantage of adjustments used in BSSN equations. 
\item No ${\cal S}$-adjustment case. 
All the numerical experiments so far 
apply the scaling condition ${\cal S}$ 
for the conformal factor $\varphi$. 
The ${\cal S}$-originated terms appear many places in BSSN equations 
(\ref{BSSNeqmPHI})-(\ref{BSSNeqmTG}), so that we guess non-zero
${\cal S}$ is a kind of source of the constraint violation. 
However,
since all ${\cal S}$-originated terms do not appear in the 
flat spacetime background analysis, 
[no adjusted terms in (\ref{eqmMinkow1})-(\ref{eqmMinkow5})],
our case
can not say any effects due to S-constraint. 
\item No ${\cal A}$-adjustment case. 
The trace (or traceout) condition for 
the variables is also considered necessary (e.g. \cite{AB2001}). 
This can be checked with 
$\kappa_{adj}=(\kappa,\kappa, 1,1, 1,1, \kappa,1)$, and we get 
$$ 
\mbox{CAFs}=( 0 \, (\times 3), \pm \sqrt{-  k^2} \, \mbox{(3 pairs)} ), 
$$ 
independent of $\kappa$. 
Therefore the effect of ${\cal A}$-adjustment 
is not apparent from this analysis.
\item No ${\cal G}^i$-adjustment case. 
The introduction of $\Gamma^i$ is the 
key in the BSSN system.  
If we do not apply adjustments by ${\cal G}^i$,
($\kappa_{adj}=(1,1,0,1,0,0,1,1)$) then we get
$$ 
\mbox{CAFs}=( 0 (\times 7), \pm 
\sqrt{  - k^2  } ), 
$$ 
which is the same with No.1. 
That is, adjustments due to ${\cal G}^i$ terms 
are effective to make a progress from ADM. 
\item No ${\cal M}_i$-adjustment case.  This can be checked with 
$\kappa_{adj}=( 1,1,1,1,1,1,1,\kappa )$, and we get 
\begin{eqnarray*} 
\mbox{CAFs}&=&( 0, \pm \sqrt{-  \kappa k^2} \, \mbox{(2 pairs)}, 
\nonumber \\&& 
\pm \sqrt{ - k^2 ( -1 + 4 \kappa + |1-4\kappa|)/6}, 
\quad \pm \sqrt{ - k^2 ( -1 + 4 \kappa -  |1-4\kappa|)/6} 
). 
\end{eqnarray*} 
If $\kappa=0$, then 
$(0 (\times 7), \pm 
\sqrt{  k^2 /3 } )$, which is $( 0 (\times 7), \pm 
\mbox{real value} )$.  Interestingly, 
these real values indicate the existence of 
the error growing mode together with 
the decaying mode. 
Alcubierre et al. \cite{potsdam9908} found that the 
adjustment 
due to the  momentum constraint 
is crucial for obtaining stability.  We 
think  that they picked up this error growing mode. 
Fortunately at the BSSN 
limit ($\kappa=1$), this error growing mode disappears and turns into 
a propagation mode. 
\item No ${\cal H}$-adjustment case.  The set 
$\kappa_{adj}=( 1,1,1,\kappa,1,1,1,1) $ gives 
$$ 
\mbox{CAFs}=( 0 \, (\times 3), \pm \sqrt{-  k^2} \, \mbox{(3 pairs)} ), 
$$ 
independently to $\kappa$. 
Therefore the effect of ${\cal H}$-adjustment is not apparent from this analysis.
\end{enumerate} 
These tests are on the effects of adjustments that are already 
in the BSSN equations.
We will consider whether much 
better adjustments are possible in the next section.

We list the above results in Table \ref{table1ADJBSSN}. 
The most characteristic points of the above are No. 5 and No.6 
that 
denote the contribution of the momentum constraint adjustment and 
the importance of the new variable $\tilde{\Gamma}^i$. 
It is quite interesting that the unadjusted BSSN equations (case 2)
does not have apparent advantages from the ADM system.
As we showed in the case 5 and 6, if we missed a particular
adjustment,  then the expected stability behaviour occationally 
gets worse
than the  starting ADM system.  
Therefore we conclude that the better stability of the 
BSSN formulation is obtained by their adjustments in the equations, 
and the combination of the adjustments is in a good balance. 

\begin{table}[t] 
{\scriptsize
\begin{tabular}{ll|ccccc||c|l} 
\hline 
No.  & & 
\multicolumn{5}{l||}{Constraints (number of components)}  
& diag? & CAFs
\\ 
 & & ${\cal H}$ (1) &  ${\cal M}_i$ (3) 
&  ${\cal G}^i$ (3) & 
${\cal A} $ (1) &  ${\cal S} 
$ (1) & &
in Minkowskii background 
\\ 
\hline  &&&&&&&& \\ 
0.& standard ADM & use & use & - & -& -& yes & $(0,0,\Im,\Im)$ 
\\ 
1.& BSSN no adjustment & use & use & use & use & use & yes & 
$(0,0,0,0,0,0,0,\Im,\Im)$ 
\\ 
2. &the BSSN   & use+adj & use+adj & use+adj & use+adj & use+adj & no & 
$(0,0,0,\Im,\Im,\Im,\Im,\Im,\Im)$ 
\\ 
&&&&&&& \\ 
\hline 
&&&&&&& \\ 
3.& no ${\cal S} $ adjustment   & use+adj & use+adj 
& use+adj & use+adj & 
use  & no & 
no difference in flat background 
\\ 
4.& no ${\cal A} $ adjustment    & use+adj & use+adj 
& use+adj & use & use+adj & no 
& $(0,0,0,\Im,\Im,\Im,\Im,\Im,\Im)$ 
\\ 
5. & no ${\cal G}^i $ adjustment    & use+adj & use+adj 
& use & use+adj & 
use+adj & no & $(0,0,0,0,0,0,0,\Im,\Im)$ 
\\ 
6. &no ${\cal M}_i $ adjustment   & use+adj & use  & use+adj 
& use+adj & 
use+adj 
& no & $(0,0,0,0,0,0,0,\Re,\Re)$ 
\\ 
7. &no ${\cal H} $ adjustment   & use & use+adj  & use+adj 
& use+adj & 
use+adj & no
& $(0,0,0,\Im,\Im,\Im,\Im,\Im,\Im)$ 
\\ 
&&&&&&& \\ 
\hline 
\end{tabular} 
}
\caption{ 
Contributions of adjustments terms and 
effects of introductions of new constraints 
in the BSSN system. 
The center column indicates whether each constraints are taken as 
a component of constraints in each constraint propagation analysis 
(`use'), and whether each adjustments are on (`adj'). 
The column `diag?' indicates
diagonalizability of the constraint propagation matrix. 
The right column shows amplification factors, 
where 
$\Im$ and $\Re$ means pure imaginary and real eigenvalue, 
respectively. 
}\label{table1ADJBSSN}
\end{table} 

\subsubsection{Proposals of the modified BSSN equations}
\label{subsec:modifiedBSSN}

We next consider the possibility whether we can obtain a 
system which has much better properties; 
whether more pure imaginary CAFs 
or negative real CAFs. 

\paragraph{Heuristic examples} 
\beit
\item[(A)] 
A system which has 8 pure imaginary CAFs: \\ 
One direction is to seek a set of equations which 
make fewer zero CAFs than the standard BSSN case. 
Using the same set of adjustments in 
(\ref{eqmMinkow1})-(\ref{eqmMinkow5}), 
CAFs are written as
\begin{eqnarray*} 
\mbox{CAFs} &=& \Big(0,\pm\sqrt{-k^2 \kappa_{A1} 
\kappa_{\tilde{\Gamma} 2}} \, \mbox{(2 pairs)}, 
\pm \mbox{complicated expression}, 
\pm \mbox{complicated expression} \Big). 
\end{eqnarray*} 
The terms in the first line certainly give 
four pure imaginary CAFs (two 
positive and negative real pairs) if 
$\kappa_{A1}\kappa_{\tilde{\Gamma} 2}>0 \, (<0)$. 
Keeping this in  mind, by choosing 
$\kappa_{adj}=(1,1,1, 1,1,\kappa,1,1)$, we find 
\begin{eqnarray*} 
\mbox{CAFs}&=&\Big(0,\pm\sqrt{-k^2 } \, \mbox{(2 pairs)}, 
\pm\sqrt{-k^2 (2+\kappa+|\kappa-4|)/6}, 
\pm\sqrt{-k^2 (2+\kappa-|\kappa-4|)/6}, 
\Big). 
\end{eqnarray*} 
Therefore the adjustment 
$\kappa_{adj}=( 1,1,1, 1,1,4,1,1 )$ gives 
$ 
\mbox{CAFs}=\Big(0,\pm\sqrt{-k^2 } \, \mbox{(4 pairs)} \Big), 
$  
which is one step advanced from BSSN according our guidelines. 

We note that 
such a system can be obtained in many ways, e.g. 
$\kappa_{adj}=(0,0,1,0,2,1,0,1/2)$ also gives four pairs of pure 
imaginary CAFs. 
\item[(B)] A system which has negative real CAF: \\ 
One criterion to obtain a decaying constraint mode 
(i.e. an asymptotically 
constrained system) is to adjust an evolution equation as it breaks 
time reversal symmetry (Box 3.5). 
For example, we consider an additional adjustment 
to the BSSN equation as 
\begin{equation} 
\partial_t \tilde{\gamma}_{ij} = \partial_t^B 
\tilde{\gamma}_{ij}+\kappa_{SD} 
\alpha \tilde{\gamma}_{ij} {\cal H}, 
\end{equation} 
which is a similar adjustment of the simplified 
Detweiler-type \cite{detweiler}. 
The constaint amplification factors become 
\begin{eqnarray*} 
\mbox{CAFs} &=&(0 \, (\times2), 
\pm\sqrt{-k^2} \mbox{(3 pairs)}, \, (3/2)k^2 \kappa_{SD}), 
\end{eqnarray*} 
in which the last one becomes negative real if 
$\kappa_{SD} <0$.  
\item[(C)]
 Combination of above (A) and (B) \\ 
Naturally we next consider both  adjustments: 
\begin{eqnarray} 
\partial_t \tilde{\gamma}_{ij} &=& \partial_t^B 
\tilde{\gamma}_{ij}+\kappa_{SD} 
\alpha \tilde{\gamma}_{ij} {\cal H} 
\\ 
\partial_t \tilde{A}_{ij} 
&=& 
\partial_t^B \tilde{A}_{ij} 
-\kappa_{8}\alpha e^{-4\varphi}\tilde{\gamma}_{ij} \partial_k{\cal G}^k 
\end{eqnarray} 
where the second one produces the 8 pure imaginary CAFs. 
We then obtain 
\begin{eqnarray*} 
\mbox{CAFs}&=&\Big( 
0,\pm\sqrt{-k^2} \, \mbox{(3 pairs)}, 
(3/4)k^2 \kappa_{SD} 
\pm \sqrt{k^2(-\kappa_{8}+(9/16) k^2 \kappa_{SD})} 
\Big) 
\end{eqnarray*} 
which reproduces case (A) when $\kappa_{SD}=0,\kappa_{8}=1$, 
and case (B) when $\kappa_{8}=0$. 
These CAFs can become 
(0, pure imaginary (3 pairs), complex numbers 
with a negative real part (1 pair)), 
with an appropriate combination of $\kappa_{8}$ and $\kappa_{SD}$. 
\enit
\paragraph{Possible adjustments} 
In order to break time reversal symmetry of the evolution equations 
(Box 3.5), 
the possible simple adjustments are 
(1) to add ${\cal H}$, ${\cal S}$ or ${\cal G}^i$ terms 
to the equations of 
$\partial_t \phi$, $\partial_t \tilde{\gamma}_{ij}$, or
$\partial_t \tilde{\Gamma}^i$, and/or (2) to add ${\cal M}_{i}$ 
or ${\cal A}$ terms to 
$\partial_t  K$ or $\partial_t {\tilde A}_{ij}$. 
We write them generally, including the above proposal (B),  as 
\begin{eqnarray} 
\partial_t \phi &=& \partial_t^B \phi 
+ \kappa_{\phi{\cal H}} \, \alpha{\cal H} 
+ \kappa_{\phi{\cal G}}\,  \alpha{\tilde D}_k{\cal G}^k,  
\label{negadj1} 
\\ 
\partial_t \tilde{\gamma}_{ij} &=& \partial_t^B \tilde{\gamma}_{ij} 
+ \kappa_{SD} \, \alpha\tilde{\gamma}_{ij}{\cal H} 
+ \kappa_{\tilde{\gamma}{\cal G} 1} \, 
\alpha\tilde{\gamma}_{ij}{\tilde D}_k{\cal G}^k 
+ \kappa_{\tilde{\gamma}{\cal G} 2} \, 
\alpha\tilde{\gamma}_{k(i}{\tilde D}_{j)}{\cal G}^k 
\nonumber \\&& \qquad ~~
+ \kappa_{\tilde{\gamma} {\cal S} 1}\, 
\alpha\tilde{\gamma}_{ij}{\cal S} 
+ \kappa_{\tilde{\gamma}{\cal S} 2} \, 
\alpha{\tilde D}_i{\tilde D}_j{\cal S}, 
\\ 
\partial_t K &=&\partial_t^B K 
+ \kappa_{K {\cal M}} \, 
\alpha \tilde{\gamma}^{jk}({\tilde D}_j{\cal M}_k), 
\\ 
\partial_t {\tilde A}_{ij}&=&\partial_t^B {\tilde A}_{ij} 
+ \kappa_{A {\cal M} 1} \, 
\alpha \tilde{\gamma}_{ij} ({\tilde D}^k{\cal M}_k) 
+ \kappa_{A {\cal M} 2} \, 
\alpha ({\tilde D}_{(i}{\cal M}_{j)}) 
\nonumber \\&& \qquad ~~
+ \kappa_{A {\cal A}1} \, 
\alpha \tilde{\gamma}_{ij} {\cal A} 
+ \kappa_{A {\cal A}2} \, 
\alpha {\tilde D}_i{\tilde D}_j{\cal A}, 
\\ 
\partial_t \tilde{\Gamma}^i&=&\partial_t^B \tilde{\Gamma}^i 
+ \kappa_{\tilde{\Gamma} {\cal H}} \, 
\alpha {\tilde D}^i{\cal H} 
+ \kappa_{\tilde{\Gamma} {\cal G} 1} \, 
\alpha {\cal G}^i 
+ \kappa_{\tilde{\Gamma} {\cal G} 2} \, 
\alpha {\tilde D}^j{\tilde D}_j{\cal G}^i 
+ \kappa_{\tilde{\Gamma} {\cal G} 3} \, 
\alpha {\tilde D}^i{\tilde D}_j{\cal G}^j, 
\label{negadj5} 
\end{eqnarray} 
where $\kappa$s are possible multipliers  (all $\kappa=0$ reduce the 
system the standard BSSN system). 

\begin{table}[thb] 
{\scriptsize
\begin{tabular}{ll|l|c|ll} 
\hline 
\multicolumn{2}{c|}{adjustment} & 
\multicolumn{1}{c|}{CAFs} & diag? & 
\multicolumn{2}{|c}{effect of the adjustment} 
\\ 
\hline 
$\partial_t \phi$ & 
$\kappa_{\phi{\cal H}} \, \alpha {\cal H}$ & 
$(0,0,\pm\sqrt{-k^2}(*3),8\kappa_{\phi{\cal H}}k^2)$ & no & 
$\kappa_{\phi{\cal H}}<0$ makes 1 Neg. & 
\\ 
$\partial_t \phi$ & 
$\kappa_{\phi{\cal G}} \, \alpha{\tilde D}_k{\cal G}^k$& 
$(0,0,\pm\sqrt{-k^2}(*2)$, long expressions) & yes &
$\kappa_{\phi{\cal G}}<0$ makes 2 Neg. 1 Pos. & 
\\ 
$\partial_t \tilde\gamma_{ij}$ & 
$\kappa_{SD} \, \alpha\tilde{\gamma}_{ij}{\cal H} $& 
$(0,0,\pm\sqrt{-k^2}(*3),(3/2)\kappa_{SD}k^2)$ & yes &
$\kappa_{SD}<0$ makes 1 Neg. & Case (B) 
\\ 
$\partial_t \tilde\gamma_{ij}$ & 
$\kappa_{\tilde{\gamma}{\cal G} 1} \, 
\alpha\tilde{\gamma}_{ij}{\tilde D}_k{\cal G}^k$& 
$(0,0,\pm\sqrt{-k^2}(*2)$, long expressions)& yes &
$\kappa_{\tilde{\gamma}{\cal G} 1}>0$ makes 1 Neg. 
\\ 
$\partial_t \tilde\gamma_{ij}$ & 
$\kappa_{\tilde{\gamma}{\cal G} 2} \, 
\alpha\tilde{\gamma}_{k(i}{\tilde D}_{j)}{\cal G}^k$& 
(0,0, 
long expressions)
& yes &
$\kappa_{\tilde{\gamma}{\cal G} 2}<0$ makes 6 Neg. 1 Pos. & Case (E1) 
\\ 
$\partial_t \tilde\gamma_{ij}$ & 
$\kappa_{\tilde{\gamma} {\cal S} 1} \, 
\alpha\tilde{\gamma}_{ij}{\cal S}$& 
$(0, 0,\pm\sqrt{-k^2}(*3),3\kappa_{\tilde{\gamma}{\cal S} 1})$& no &
$\kappa_{\tilde{\gamma}{\cal S} 1}<0$ makes 1 Neg.   & 
\\ 
$\partial_t \tilde\gamma_{ij}$ & 
$\kappa_{\tilde{\gamma}{\cal S} 2} \, 
\alpha{\tilde D}_i{\tilde D}_j{\cal S}$& 
$(0, 0,\pm\sqrt{-k^2}(*3),-\kappa_{\tilde{\gamma}{\cal S} 2} k^2) $& no &
$\kappa_{\tilde{\gamma}{\cal S} 2}>0$ makes 1 Neg.  & 
\\ 
$\partial_t K$ & 
$\kappa_{K {\cal M}} \,  \alpha 
\tilde{\gamma}^{jk}({\tilde D}_j{\cal M}_k)$& 
$(0,0,0,\pm\sqrt{-k^2}(*2), $ long expressions)  
& no & $\kappa_{K {\cal M}}<0$ makes 2 Neg. & 
\\ 
$\partial_t \tilde{A}_{ij}$ & 
$\kappa_{A {\cal M} 1} \, 
\alpha \tilde{\gamma}_{ij} ({\tilde D}^k{\cal M}_k)$& 
$(0,0,\pm\sqrt{-k^2}(*3),-\kappa_{A {\cal M} 1}k^2)$ & yes &
$\kappa_{A {\cal M} 1}>0$ makes 1 Neg. & 
\\ 
$\partial_t \tilde{A}_{ij}$ & 
$ \kappa_{A {\cal M} 2} \,  \alpha ({\tilde D}_{(i}{\cal M}_{j)})$& 
(0,0, 
 long expressions)
& yes &
$\kappa_{A {\cal M} 2}>0$ makes 7 Neg  & Case (D) 
\\ 
$\partial_t \tilde{A}_{ij}$ & 
$\kappa_{A {\cal A}1} \,  \alpha \tilde{\gamma}_{ij} {\cal A}$& 
$(0,0,\pm\sqrt{-k^2}(*3),3\kappa_{A {\cal A}1})$ & yes &
$\kappa_{A {\cal A}1}<0$ makes 1 Neg. & 
\\ 
$\partial_t \tilde{A}_{ij}$ & 
$\kappa_{A {\cal A}2} \,  \alpha {\tilde D}_i{\tilde D}_j{\cal A}$& 
$(0,0,\pm\sqrt{-k^2}(*3),-\kappa_{A {\cal A}2}k^2)$ & yes &
$\kappa_{A {\cal A}2}>0$ makes 1 Neg. & 
\\ 
$\partial_t \tilde{\Gamma}^i$ & 
$\kappa_{\tilde{\Gamma} {\cal H}} \,  \alpha {\tilde D}^i{\cal H}$& 
$(0,0,\pm\sqrt{-k^2}(*3),-\kappa_{A {\cal A}2}k^2)$ & no &
$\kappa_{\tilde{\Gamma} {\cal H}} >0$ makes 1 Neg. & 
\\ 
$\partial_t \tilde{\Gamma}^i$ & 
$ \kappa_{\tilde{\Gamma} {\cal G} 1} \,  \alpha {\cal G}^i$& 
(0,0, 
long expressions)& yes &
$\kappa_{\tilde{\Gamma} {\cal G} 1}<0$ makes 6 Neg. 1 Pos. & Case (E2) 
\\ 
$\partial_t \tilde{\Gamma}^i$ & 
$ \kappa_{\tilde{\Gamma} {\cal G} 2} \, 
\alpha {\tilde D}^j{\tilde D}_j{\cal G}^i$& 
(0,0, 
long expressions)& yes &
$\kappa_{\tilde{\Gamma} {\cal G} 2}>0$ makes 2 Neg. 1 Pos. & 
\\ 
$\partial_t \tilde{\Gamma}^i$ & 
$ \kappa_{\tilde{\Gamma} {\cal G} 3} \, 
\alpha {\tilde D}^i{\tilde D}_j{\cal G}^j$& 
(0,0, 
long expressions)& yes &
$\kappa_{\tilde{\Gamma} {\cal G} 3}>0$ makes 2 Neg. 1 Pos. & 
\\ 
\hline 
\end{tabular} 
}
\caption{ 
Possible adjustements which make a real-part 
constraint amplification
factors negative. 
The column of adjustments are nonzero multipliers in terms of 
(\ref{negadj1})-(\ref{negadj5}), which all violate 
time reversal symmetry of the equation. 
The column `diag?' indicates
diagonalizability of the constraint propagation matrix. 
Neg./Pos. means negative/positive respectively. 
}\label{table2ADJBSSN} 
\end{table} 


We show the effects of each terms in Table \ref{table2ADJBSSN}. 
The CAFs in the table are on the 
flat space background.  We see several terms make 
negative CAFs, which might improve the stability than 
the previous system.  For the readers convenience, we list 
up several best candidates here. 

\beit
\item[(D)] 
{ A system which has 7 negative CAFs}\\ 
Simply adding $\tilde{D}_{(i}{\cal M}_{j)}$ term to 
$\partial_t \tilde{A}_{ij}$ equation, say 
\begin{equation} 
\partial_t \tilde{A}_{ij}=\partial^{BSSN}_t \tilde{A}_{ij} 
+\kappa_{A {\cal M} 2} \alpha (\tilde{D}_{(i}{\cal M}_{j)}) 
\end{equation} 
with 
$\kappa_{A {\cal M} 2}>0$, CAFs on the flat background are 7 negative 
real CAFs. \\ 

\item[(E)] 
 { A system which has 6 negative and 1 positive CAFs}\\ 
The below two adjustments will make 
6 negative real CAFs, while they also produce one positive real CAF 
(a constraint violating mode).  The effectiveness is not clear 
at this moment, but we think they are worth to be tested in numerical 
experiments. 
\begin{eqnarray} (E1) &&\qquad 
\partial_t \tilde{\gamma}_{ij} =\partial^{BSSN}_t \tilde{\gamma}_{ij} 
+ \kappa_{\tilde{\gamma}{\cal G} 2} \, 
\alpha\tilde{\gamma}_{k(i}{\tilde D}_{j)}{\cal G}^k,  
\qquad \mbox{with} \qquad \kappa_{\tilde{\gamma} {\cal G} 2} <0. 
\\
(E2) && \qquad 
\partial_t \tilde{\Gamma}^i = \partial^{BSSN}_t \tilde{\Gamma}^i 
+ \kappa_{\tilde{\Gamma} {\cal G} 2} \, 
\alpha {\tilde D}^j{\tilde D}_j{\cal G}^i, \quad  
\qquad \mbox{with} \qquad \kappa_{\tilde{\Gamma} {\cal G} 2} <0. 
\end{eqnarray} 
\enit

\subsubsection{Remarks}
\label{subsec:modifiedBSSNremark}
We have studied step-by-step where the replacements in the equations 
affect and/or newly added constraints work, by checking whether the 
error of constraints (if it exists) will decay or propagate away. 
Alcubierre et al \cite{potsdam9908} pointed out  
the importance of the 
replacement (adjustment) of terms in the evolution equation 
due to the momentum constraint, 
and our analysis clearly explain why they concluded 
this is the key. 
Not only this adjustment, we found, 
but also other adjustments and other 
introductions of new constraints also contribute 
to making the evolution system more stable. 
We found that if we missed a particular adjustment, 
then the expected stability behaviour occationally gets worse than the 
ADM system. 
We further propose other adjustments of 
the set of equations which may have better features 
for numerical treatments. 

The discussion was only in the flat 
background spacetime, and further analysis is in progress. 
However, we rather 
believe that the general fundamental aspects of constraint
propagation analysis are already 
revealed here.  This is because, 
for the ADM and its adjusted formulation cases, we found that 
the better formulations 
in the flat background are also better 
in the Schwarzschild spacetime, while 
there are differences on the effective adjusting multipliers or the 
effective coordinate ranges \cite{adjADMsch,adjADM}. 
Actually, recently Yo, Baumgarte and Shapiro \cite{YBS} reported 
their simulations of stationary rotating black hole, and mentioned that
the above proposal (B) was contributed to maintain their evolution of 
Kerr black hole ($J/M$ up to $0.9M$) for long time ($t \sim 6000M$).  Their 
results also indicates that the evolved solution is closed to the exact one,
that is, the constrained surface.

\newpage
\section{Outlook}\label{sec:outlook}
\setcounter{equation}{0}

\subsection{What we have achieved}

Let us summarize our story first.  The beginnings of the study are:
\beit
\item General relativistic numerical simulations are quite important in 
astrophysical studies, but
we do not have a definite recipe to integrate the Einstein equations
in a long-term stable and accurate manner.  
\item The most standard approach
is to decompose the space-time into 3-space and time 
($3+1$ decomposition), to solve the constraints for obtaining the 
initial data, and to evolve the space-time applying 
a {\it free evolution} scheme. 
\item 
Over a period of decades, the community has been observing the violation 
of constraints in their simulations, and their consensus is that 
we have to find a better formulation of the Einstein evolution equations. 
\enit
We reviewed recent efforts on this problem 
by categorizing them into   
\begin{indention}{1cm}
\noindent
(0)~The standard ADM formulation (\S \ref{secADM}, Box 2.1), 
\\
(1)~The modified ADM (so-called BSSN) formulation (\S \ref{secBSSN}, Box 2.3), 
\\
(2)~Hyperbolic formulations (\S \ref{secHYP}, Box 2.5), and 
\\
(3)~Asymptotically constrained formulations (\S \ref{secASYMPT}). 
\end{indention}
\noindent
Among them, the approach (2) is perhaps best justified on mathematical
grounds. 
However, as we critically reviewed in \S \ref{sec:hypRemark}, the practical
advantages may not be available unless we kill the lower-order terms in its
hyperbolized equation, as in KST's formulation (Box 2.6). 

We therefore proceeded in the direction (3). 
Our approach, which we term  {\it adjusted system}, is to 
construct a system that has its constraint surface as an attractor. 
Our unified view is to understand the evolution system by evaluating its constraint
propagation.  Especially we proposed to analyze the constraint 
amplification factors (Box 3.1) which are the eigenvalues of the homogenized constraint
propagation equations. 
We analyzed the system based on our conjecture (Box 3.2/3.3) whether the constraint
amplification factors suggest the constraint to decay/propagate or not.  
We concluded that
\beit
\item The constraint propagation features become different by simply adding 
constraint terms to the original evolution equations  
(we call this the {\it adjustment} of the evolution equations). 
\item There {\it is} a constraint-violating mode in the standard ADM evolution system
when we apply it to a single non-rotating black hole space-time, and its 
growth rate is larger near the black-hole horizon. 
\item Such a constraint-violating mode can be killed if we adjust the
evolution equations with a particular modification using constraint terms (Box 2.7). 
An effective guideline is to adjust terms as they break the time-reversal symmetry
of the equations (Box 3.5). 
\item Our expectations are borne out in simple 
numerical experiments using the Maxwell, Ashtekar, and ADM systems.  
However the modifications
are not yet perfect to prevent non-linear growth of the constraint violation. 
\item 
We understand why the BSSN formulation works better than the ADM one
in the limited case (perturbative analysis in the flat background), 
 and further we 
proposed modified evolution equations along the lines of our previous procedure. 
\enit

The common key to the problem is how to adjust the evolution equations with 
constraints.  
Any adjusted systems are mathematically equivalent if the constraints are
completely satisfied, but this is not the case for numerical simulations. 
Replacing terms with constraints is one of the normal steps 
when people hyperbolize equations. 
Our approach is to employ the evaluation process of constraint 
amplification factors for an alternative guideline to 
hyperbolization of the system. 

\subsection{Next steps}
Here are some directions for future researches in this 
formulation problem of the Einstein equations. 

\paragraph{Generalize the adjusted systems:}
We have tried to unify all the efforts of
the community applying the idea of ``adjusted systems".  
Our newly modified equations are
working as desired up to the current numerical tests, but we see also that 
the effect is not yet perfect to remove non-linear error growth 
that terminates numerical simulations.  
We suspect that this is due to the current eigenvalue analysis 
based on a perturbative analysis with fixing adjusting multiplier, $\kappa$. 
One remedy is to generalize the determination process of $\kappa$, say, 
dynamically and automatically under a suitable principle. 
We are now working on a method to control the violation of
{\it each} constraint independently, or an additional supporting 
mathematical criteria to realize more robust stabilizations. 


\paragraph{More on hyperbolic formulations:}
We have already pointed out several
directions on hyperbolic efforts in \S \ref{sec:hypRemark}.  
We here only point out the links to the 
initial-boundary value problem (IBVP). 
In order to avoid unphysical incoming information from the boundary
of computation, a proper treatment of the boundary is quite important
problem in numerical simulations. 
If we can treat boundary condition as a part of mathematical framework, 
that would be a great step to promote researches. 
The IBVP approach to the Einstein equations is quite new and has only 
been studied in a restricted cases.  
The current proposals are based on a particular symmetric
hyperbolic formulation or under a certain assumption to the symmetry
of space-time.  More general treatments on IBVP are expected.  


\paragraph{Alternative new ideas?:}
We have to be open to develop an alternative approach to this formulation 
problem.  If our goal is to obtain a stable system
on its constrained surface, then there may be 
at least three fundamental approaches: 
(1) construct an improved evolution system (as we discussed most), 
(2) develop a ``maintenance method" for the system, or 
(3) redefine or classify the stability of the system. 
An idea of automatic adjustment of multipliers in adjusted system is 
a sort of ``maintenance".  
We guess there might be a key in existing theories such as 
control theories, optimization methods (convex functional theories), 
mathematical programming methods, or others.  

\paragraph{Numerical comparisons of formulations:}
Fortunately, an effort toward systematic numerical comparisons of different 
formulations of the Einstein equations has been organized and started
recently.  
The workshop ``Comparisons of Formulations of Einstein's
equations for Numerical Relativity" was held at Mexico City in May 2002
\footnote{ Follow-up information is 
available at {\tt http://www.nuclecu.unam.mx/$\tilde{~}$gravit/main\_rn.html}.},  
and more than 20 people
attended this 2-week workshop.   This project intends to provide a format for
comparisons in  a common numerical framework. 
That is,  comparing different formulations
using 
the same initial data, 
the same resolution, 
the same integration scheme, 
the same boundary treatment, and 
the same output. 
Comparisons are now in progress for vacuum and regular space-time, and
we hope to extend them to black-hole space-time next.  
The first reports are in preparation \cite{mexico}.

\subsection{Final remarks}

If we say the final goal of this project is to find a robust algorithm
to obtain long-term accurate and stable time-evolution method, 
then the recipe should be a combination of 
(a) formulations of the evolution equations, 
(b) choice of gauge conditions, 
(c) treatment of boundary conditions, and 
(d) numerical  integration methods. 
We are in the stages of solving this mixed puzzle. 
The ideal almighty algorithm may not exit, but we believe our accumulating
experience will make the ones we do have more 
robust and automatic. 

We have written this review from the viewpoint that the
general relativity is a constrained dynamical system.  
This is not only a proper problem 
in general relativity, but also in many physical systems such as 
electrodynamics, magnetohydrodynamics, 
molecular dynamics, mechanical dynamics, and so on. 
Therefore sharing the thoughts between different field will 
definitely accelerate the progress.  

When we discuss asymptotically constrained manifolds, we implicitly 
assume that  the dynamics could be expressed on a wider manifold.  
Recently we found that such a proposal is quite similar with 
techniques in e.g. molecular dynamics.  
For example, people assume an extended environment when simulating 
molecular dynamics under constant pressure (with a potential piston) or 
a constant temperature (with a potential thermostat) (see e.g.
\cite{NosePTP}).   
We have also noticed that a dynamical adjusting method of Lagrange
multipliers  has been developed in multi-body 
mechanical dynamics (see e.g. \cite{Nagata}).  
We are now trying to apply these ideas back into numerical relativity. 

In such a way, 
communication and interaction between different fields is encouraged. 
Cooperation between numerical and mathematical scientists is
necessary.  By interchanging ideas, we hope we will reach our goal in 
next few years, and obtain interesting physical results and predictions. 
It is our personal view that
exciting revolutions in numerical relativity 
are coming soon.


\newpage
\appendix

\section{General expressions of ADM constraint propagation equations}
\label{appADMconpro}
For the reader's convenience, we express here the
constraint propagation equations generally, considering the adjustments
to the evolution equations.

\subsection{The standard ADM equations and  constraint propagations}
We start by analyzing the standard ADM system, that is,
with evolution equations (\ref{adm_evo1}) and (\ref{adm_evo2})
and constraint equations (\ref{admCH}) and (\ref{admCM}).

The constraint propagation equations,
which are the time evolution equations
of the Hamiltonian constraint (\ref{admCH}) and
the momentum constraints (\ref{admCM}).

\paragraph{Expression using ${\cal H}$ and ${\cal M}_i$}
The constraint propagation equations
can be written as [these are the same with (\ref{CHproADM}) and
(\ref{CMproADM}) ]
\begin{eqnarray}
\partial_t {\cal H}&=&
\beta^j (\partial_j {\cal H})
+2\alpha K{\cal H}
-2\alpha \gamma^{ij}(\partial_i {\cal M}_j)
\nonumber \\&&
+\alpha(\partial_l \gamma_{mk})
   (2\gamma^{ml}\gamma^{kj}-\gamma^{mk}\gamma^{lj}){\cal M}_j
-4\gamma^{ij} (\partial_j\alpha){\cal M}_i,
\label{ACHproADM}
\\
\partial_t {\cal M}_i&=&
-(1/2)\alpha (\partial_i {\cal H})
-(\partial_i\alpha){\cal H}
+\beta^j (\partial_j {\cal M}_i)
\nonumber \\&&
+\alpha K {\cal M}_i
-\beta^k\gamma^{jl}(\partial_i\gamma_{lk}){\cal M}_j
+(\partial_i\beta_k)\gamma^{kj}{\cal M}_j.
\label{ACMproADM}
\end{eqnarray}
This is a suitable form to discuss hyperbolicity of the system.
The simplest derivation of (\ref{ACHproADM}) and (\ref{ACMproADM})
is by using the Bianchi identity, which
can be seen in Frittelli \cite{Fri-con}.

A shorter expression is available, e.g.
\begin{eqnarray}
\partial_t {\cal H}
&=&
  \beta^l \partial_l {\cal H} +2\alpha K{\cal H}
-2\alpha \gamma^{-1/2}\partial_l(\sqrt\gamma {\cal M}^l)
-4(\partial_l\alpha){\cal M}^l
\nonumber
\\
&=&
   \beta^l \nabla_l {\cal H} +2\alpha K{\cal H}
-2\alpha (\nabla_l {\cal M}^l)
-4(\nabla_l\alpha){\cal M}^l,
\label{YS35b}
\\
  \partial_t   {\cal M}_i &=&
-(1/2)\alpha(\partial_i {\cal H})
-(\partial_i \alpha) {\cal H}
+ \beta^l \nabla_l {\cal M}_i
+\alpha K {\cal M}_i
+ (\nabla_i \beta_l ) {\cal M}^l
\nonumber \\
&=&
-(1/2)\alpha(\nabla_i {\cal H})
-(\nabla_i \alpha) {\cal H}
+ \beta^l \nabla_l {\cal M}_i
+\alpha K {\cal M}_i
+ (\nabla_i \beta_l ) {\cal M}^l,
\label{YS36b}
\end{eqnarray}
or by using Lie derivatives along $\alpha n^\mu$, 
\begin{eqnarray}
\pounds_{\alpha n^\mu} {\cal H}
&=&
2\alpha K{\cal H}
-2\alpha \gamma^{-1/2}\partial_l(\sqrt\gamma {\cal M}^l)
-4(\partial_l\alpha){\cal M}^l,
\label{YS35alie}
\\
\pounds_{\alpha n^\mu}  {\cal M}_i &=&
-(1/2)\alpha(\partial_i {\cal H})
-(\partial_i \alpha) {\cal H}
+\alpha K {\cal M}_i.
\label{YS35blie}
\end{eqnarray}

\paragraph{Expression using $\gamma_{ij}$ and $K_{ij}$}
In order to check the effects of the adjustments in (\ref{adm_evo1})
and (\ref{adm_evo2}) to constraint propagation,
it is useful to re-express (\ref{ACHproADM}) and (\ref{ACMproADM})
using $\gamma_{ij}$ and $K_{ij}$.
By a straightforward calculation, we obtain an expression as
\begin{eqnarray}
\partial_t {\cal H} &=&
H_1^{mn}(\partial_t\gamma_{mn})
+H_2^{imn}\partial_i(\partial_t \gamma_{mn})
+H_3^{ijmn}\partial_i\partial_j (\partial_t \gamma_{mn})
+H_4^{mn}(\partial_t K_{mn}), \label{CHpro_new}
\\
\partial_t {\cal M}_i
&=&
M_1{}_i{}^{mn}(\partial_t \gamma_{mn})
+M_2{}_i{}^{jmn}\partial_j(\partial_t \gamma_{mn})
+M_3{}_i{}^{mn}(\partial_t K_{mn})
+M_4{}_i{}^{jmn} \partial_j(\partial_t K_{mn}), \label{CMpro_new}
\end{eqnarray}
where
\begin{eqnarray}
H_1^{mn}&:=&
-2R^{(3)}{}^{mn}
-\Gamma^p_{kj}\Gamma^k_{pi}\gamma^{mi}\gamma^{nj}
+\Gamma^m\Gamma^n
\nonumber \\ &&
+\gamma^{ij}\gamma^{np}(\partial_i\gamma^{mk})(\partial_j\gamma_{kp})
-\gamma^{mp}\gamma^{ni}(\partial_i\gamma^{kj})(\partial_j\gamma_{kp})
-2K K^{mn}+2K^n{}_jK^{mj},
\\
H_2^{imn}&:=&
-2\gamma^{mi}\Gamma^n
-(3/2)\gamma^{ij}(\partial_j\gamma^{mn})
+\gamma^{mj}(\partial_j\gamma^{in})
+\gamma^{mn}\Gamma^i,
\\
H_3^{ijmn}&:=&
-\gamma^{ij}\gamma^{mn}+\gamma^{in}\gamma^{mj},
\\
H_4^{mn}&:=&
2(K\gamma^{mn}-K^{mn}),
\\
M_1{}_i{}^{mn}&:=&
   \gamma^{nj}(\partial_i K^m{}_j)
-\gamma^{mj}(\partial_j K^n{}_{i})
+(1/2)(\partial_j\gamma^{mn})K^j{}_i
+\Gamma^n K^m{}_{i},
\\
M_2{}_i{}^{jmn}&:=&
-\gamma^{mj}K^n{}_i
+ (1/2) \gamma^{mn}K^j{}_i
+(1/2)  K^{mn}\delta^j_i,
\\
M_3{}_i{}^{mn}&:=&
-\delta^n_i \Gamma^m
-(1/2)(\partial_i\gamma^{mn}),
\\
M_4{}_i{}^{jmn}&:=&
\gamma^{mj}\delta^n_i -\gamma^{mn}\delta^j_i,
\end{eqnarray}
where we expressed $\Gamma^m=\Gamma^m_{ij}\gamma^{ij}$. 

\subsection{Constraint propagations for the adjusted ADM systems}

Generally, we here write the adjustment terms to
(\ref{adm_evo1}) and (\ref{adm_evo2})
using (\ref{admCH}) and (\ref{admCM}) by the following combinations,
[these are the same with (\ref{adjADM1}) and (\ref{adjADM2})]
\begin{eqnarray}
&\mbox{adjustment term of }\quad
\partial_t \gamma_{ij}:& \quad
+P_{ij} {\cal H}
+Q^k{}_{ij}{\cal M}_k 
+p^k{}_{ij}(\nabla_k {\cal H})
+q^{kl}{}_{ij}(\nabla_k {\cal M}_l),
\label{AadjADM1}
\\
&\mbox{adjustment term of}\quad
\partial_t K_{ij}:& \quad
+R_{ij} {\cal H}
+S^k{}_{ij}{\cal M}_k 
+r^k{}_{ij} (\nabla_k{\cal H})
+s^{kl}{}_{ij}(\nabla_k {\cal M}_l),
\label{AadjADM2}
\end{eqnarray}
where $P, Q, R, S$ and $p, q, r, s$
are multipliers  (please do not
confuse $R_{ij}$ with
three Ricci curvature  that we write as $R^{(3)}_{ij}$).
We adjust them only using up to the first derivatives in order to make 
the discussion simple. 

By substituting the above adjustments into (\ref{CHpro_new}) and (\ref{CMpro_new}),
we can write the adjusted constraint propagation equations as
\begin{eqnarray}
\partial_t {\cal H} &=&
\mbox{(original terms)}
\nonumber \\&&
  + H_1^{mn}[P_{mn} {\cal H}+Q^k_{mn}{\cal M}_k
            +p^k{}_{mn}(\nabla_k {\cal H})
            +q^{kl}{}_{mn}(\nabla_k {\cal M}_l)]
\nonumber \\&&
+H_2^{imn}\partial_i [P_{mn} {\cal H}+Q^k_{mn}{\cal M}_k
            +p^k{}_{mn}(\nabla_k {\cal H})
            +q^{kl}{}_{mn}(\nabla_k {\cal M}_l)]
\nonumber \\&&
+H_3^{ijmn}\partial_i\partial_j [P_{mn} {\cal H}+Q^k_{mn}{\cal M}_k
            +p^k{}_{mn}(\nabla_k {\cal H})
            +q^{kl}{}_{mn}(\nabla_k {\cal M}_l)]
\nonumber \\&&
+H_4^{mn} [R_{mn} {\cal H}+S^k_{mn}{\cal M}_k
            +r^k{}_{mn} (\nabla_k{\cal H})
            +s^{kl}{}_{mn}(\nabla_k {\cal M}_l)], \label{CHpro_newhosei}
\\
\partial_t {\cal M}_i &=&
\mbox{(original terms)}
\nonumber \\&&
+ M_1{}_i{}^{mn} [P_{mn} {\cal H}+Q^k_{mn}{\cal M}_k
            +p^k{}_{mn}(\nabla_k {\cal H})
            +q^{kl}{}_{mn}(\nabla_k {\cal M}_l)]
\nonumber \\&&
+M_2{}_i{}^{jmn} \partial_j [P_{mn} {\cal H}+Q^k_{mn}{\cal M}_k
            +p^k{}_{mn}(\nabla_k {\cal H})
            +q^{kl}{}_{mn}(\nabla_k {\cal M}_l)]
\nonumber \\&&
+M_3{}_i{}^{mn}  [R_{mn} {\cal H}+S^k_{mn}{\cal M}_k
            +r^k{}_{mn} (\nabla_k{\cal H})
            +s^{kl}{}_{mn}(\nabla_k {\cal M}_l)]
\nonumber \\&&
+M_4{}_i{}^{jmn} \partial_j [R_{mn} {\cal H}+S^k_{mn}{\cal M}_k
            +r^k{}_{mn} (\nabla_k{\cal H})
            +s^{kl}{}_{mn}(\nabla_k {\cal M}_l)]. \label{CMpro_newhosei}
\end{eqnarray}
Here the ``original terms" can be understood either as 
(\ref{ACHproADM}) and (\ref{ACMproADM}), or as 
(\ref{CHpro_new}) and (\ref{CMpro_new}).
Therefore, for example, we can see that adjustments to 
$\partial_t \gamma_{ij}$ 
do not always keep 
the constraint propagation equations in the first order form,
due to their contribution in the third adjusted term in (\ref{CHpro_newhosei}).

We note that these expressions of constraint propagation equations are
equivalent when we include the cosmological constant and/or matter terms.

\newpage
\section{Numerical demonstrations using 
 the Ashtekar formulation}\label{App_Ashtekar}

This appendix is devoted to introduce our numerical comparisons between 
three levels of hyperbolicity using Ashtekar variables. 
Details are available in \cite{ronbun1,ronbun2}.

\subsection{The Ashtekar formulation}

The key feature of  Ashtekar's formulation of general relativity
\cite{Ashtekar} is the introduction of a self-dual connection 
  as one of the basic dynamical variables.
Let us write
the metric $g_{\mu\nu}$ using the tetrad
$E^I_\mu$ as $g_{\mu\nu}=E^I_\mu E^J_\nu \eta_{IJ}$
\footnote{We use
$I,J=(0),\cdots,(3)$ and
$a,b=(1),\cdots,(3)$ are $SO(1,3)$, $SO(3)$ indices respectively.
We raise and lower
$\mu,\nu,\cdots$ by $g^{\mu\nu}$ and $g_{\mu\nu}$
(the Lorentzian metric);
$I,J,\cdots$ by $\eta^{IJ}={\rm diag}(-1,1,1,1)$ and $\eta_{IJ}$;
$i,j,\cdots$ by $\gamma^{ij}$ and $\gamma_{ij}$ (the three-metric);
$a,b,\cdots$ by $\delta^{ab}$ and $\delta_{ab}$.
We also use volume forms $\epsilon_{abc}$:
$\epsilon_{abc} \epsilon^{abc}=3!$.}.
Define its inverse, $E^\mu_I$, by
$E^\mu_I:=E^J_\nu g^{\mu\nu}\eta_{IJ}$ and we impose
$E^0_a=0$ as the gauge condition.
We define SO(3,C) 
connections
$
{}^{\pm\!}{\cal A}^a_{\mu}
:= \omega^{0a}_\mu \mp ({i / 2}) \epsilon^a{}_{bc} \, \omega^{bc}_\mu,
$
where $\omega^{IJ}_{\mu}$ is a spin connection 1-form (Ricci
connection), $\omega^{IJ}_{\mu}:=E^{I\nu} \nabla_\mu E^J_\nu.$
Ashtekar's plan is to use  only the self-dual part of
the connection
$^{+\!}{\cal A}^a_\mu$
and to use its spatial part $^{+\!}{\cal A}^a_i$
as a dynamical variable.
Hereafter,
we simply denote $^{+\!}{\cal A}^a_\mu$ as ${\cal A}^a_\mu$.

The lapse function, $N$, and shift vector, $N^i$, both of which we
treat as real-valued functions\footnote{$N$ and $N^i$ are 
the same with $\alpha$ and $\beta^i$
in the previous text.  We follow this conventional notation in this
appendix.},
are expressed as $E^\mu_0=(1/N, -N^i/N$).
This allows us to think of
$E^\mu_0$ as a normal vector field to $\Sigma$
spanned by the condition $t=x^0=$const.,
which plays the same role as that of
the ADM formulation.
Ashtekar  treated the set  ($\tilde{E}^i_{a}$, ${\cal A}^a_{i}$)
as basic dynamical variables, where
$\tilde{E}^i_{a}$ is an inverse of the densitized triad
defined by
$
\tilde{E}^i_{a}:=e E^i_{a},
$
where $e:=\det E^a_i$ is a density
\footnote{For later convenience, 
$e^2=\det\tilde{E}^i_a
=(\det E^a_i)^2=
(1/6)\epsilon^{abc}
\null\!\mathop{\vphantom {\epsilon}\smash \epsilon}
\limits ^{}_{^\sim}\!\null_{ijk}\tilde{E}^i_a \tilde{E}^j_b
\tilde{E}^k_c$, where
$\epsilon_{ijk}:=\epsilon_{abc}E^a_i E^b_j E^c_k$
  and $\null\!\mathop{\vphantom {\epsilon}\smash \epsilon}
\limits ^{}_{^\sim}\!\null_{ijk}:=e^{-1}\epsilon_{ijk}$. 
When $(i,j,k)=(1,2,3)$,
we have
$\epsilon_{ijk}=e$,
$\null\!\mathop{\vphantom {\epsilon}\smash \epsilon}
\limits ^{}_{^\sim}\!\null_{ijk}=1$,
$\epsilon^{ijk}=e^{-1}$, and
$\tilde{\epsilon}^{ijk}=1$.} 
.
This pair forms the canonical set.
In the case of pure gravitational spacetime,
the Hilbert action takes the form
\begin{eqnarray}
S&=&\int {\rm d}^4 x
[ (\partial_t{\cal A}^a_{i}) \tilde{E}^i_{a}
+(i/2) \null \! \mathop {\vphantom {N}\smash N}\limits ^{}_{^\sim}\!\null
\tilde{E}^i_a \tilde{E}^j_b F_{ij}^{c} \epsilon^{ab}{}_{c}
-N^i F^a_{ij} \tilde{E}^j_a
+{\cal A}^a_{0} \, {\cal D}_i \tilde{E}^i_{a} ],
  \label{AshtekarAction}
\end{eqnarray}
where
$\null \! \mathop {\vphantom {N}\smash N}\limits ^{}_{^\sim}\!\null
:= e^{-1}N$,
${F}^a_{\mu\nu}
:=
2 \partial_{[\mu} {\cal A}^a_{\nu]}
  - i \epsilon^{a}{}_{bc} \, {\cal A}^b_\mu{\cal A}^c_\nu
$
is the curvature 2-form,
${\cal D}_i \tilde{E}^j_{a}
     :=\partial_i \tilde{E}^j_{a}
-i \epsilon_{ab}{}^c  \, {\cal A}^b_{i}\tilde{E}^j_{c}$.

The action (\ref{AshtekarAction}) gives us the following evolution 
equations and constraints:
\Largefbox{\boxwidth}{
{\bf The Ashtekar formulation \cite{Ashtekar}:}
\hspace*{\fill} {\bf Box B.1}\\
The dynamical variables are ($\tilde{E}^i_a, {\cal A}^a_i$).
\\The evolution equations for a set of
$(\tilde{E}^i_a, {\cal A}^a_i)$ are
\begin{eqnarray}
\partial_t {\tilde{E}^i_a}
&=&-i{\cal D}_j( \epsilon^{cb}{}_a  \, \null \!
\mathop {\vphantom {N}\smash N}\limits ^{}_{^\sim}\!\null
\tilde{E}^j_{c}
\tilde{E}^i_{b})
+2{\cal D}_j(N^{[j}\tilde{E}^{i]}_{a})
+i{\cal A}^b_{0} \epsilon_{ab}{}^c  \, \tilde{E}^i_c,  \label{eq-E}
\\
\partial_t {\cal A}^a_{i} &=&
-i \epsilon^{ab}{}_c  \,
\null \! \mathop {\vphantom {N}\smash N}\limits ^{}_{^\sim}\!\null
\tilde{E}^j_{b} F_{ij}^{c}
+N^j F^a_{ji} +{\cal D}_i{\cal A}^a_{0},
\label{eq-A}
\end{eqnarray}
where
${\cal D}_jX^{ji}_a:=\partial_jX^{ji}_a-i
  \epsilon_{ab}{}^c {\cal A}^b_{j}X^{ji}_c,$
and 
${F}^a_{ij}
:=
2 \partial_{[i} {\cal A}^a_{j]}
  - i \epsilon^{a}{}_{bc} \, {\cal A}^b_i{\cal A}^c_j
$.
\\Constraint equations: (Hamiltonian, momentum and Gauss constraints)
\begin{eqnarray}
{\cal C}^{\rm ASH}_{H} &:=&
  (i/2)\epsilon^{ab}{}_c \,
\tilde{E}^i_{a} \tilde{E}^j_{b} F_{ij}^{c}
    \approx 0, \label{const-ham} \\
{\cal C}^{\rm ASH}_{M i} &:=&
   -F^a_{ij} \tilde{E}^j_{a} \approx 0, \label{const-mom}\\
{\cal C}^{\rm ASH}_{Ga} &:=&  {\cal D}_i \tilde{E}^i_{a}
  \approx 0,  \label{const-g}
\end{eqnarray}
}

We have to consider the reality conditions when we use this
formalism to describe the classical Lorentzian spacetime.
Fortunately, 
the metric will remain on
its real-valued constraint surface during time evolution
  automatically if we prepare initial data which satisfies the
reality condition. 
More practically, we further require that triad is
real-valued.  But again this reality condition appears as a gauge
restriction on ${\cal A}^a_0$\cite{ys-con}, 
which can be imposed at every time step. In our actual simulation, we
prepare our initial data using the standard ADM approach, so that
we have no difficulties in maintaining  these reality conditions.

\subsection{Reformulate the Ashtekar evolution equations}
\subsubsection{Strongly and Symmetric Hyperbolic systems}

The authors' recent studies showed the following:
\begin{itemize}
\item[(a)] 
The original set of dynamical equations (\ref{eq-E}) and (\ref{eq-A})
 [the {\it original} equations]  already forms 
a weakly hyperbolic system \cite{ysIJMPD}. So that we regard  
the mathematical structure of the original equations as one step
advanced from the standard ADM.  
\item[(b)]  
Further, we can construct higher levels of
hyperbolic systems by restricting the gauge condition and/or
by adding constraint terms,
${\cal C}^{\rm ASH}_{H}, {\cal C}^{\rm ASH}_{Mi}$ and
${\cal C}^{\rm ASH}_{Ga}$, to the original equations.
\beit
\item 
by requiring additional gauge conditions {\it or} adding constraints to
the dynamical equations, we can obtain a
strongly hyperbolic system \cite{ysIJMPD},
\item  by requiring additional gauge conditions {\it and} adding constraints to
the dynamical equations,  we can obtain a symmetric hyperbolic system
\cite{ysPRL,ysIJMPD}.
\enit
\item[(c)]  Based on the above symmetric hyperbolic system,
we can construct an Ashtekar version of the 
$\lambda$-system \cite{BFHR}
which is robust against perturbative
errors for both constraints and
reality conditions \cite{SY-asympAsh}.
\end{itemize}
In order to obtain a
symmetric hyperbolic system
\footnote{
Iriondo et al \cite{Iriondo} presented 
a symmetric hyperbolic expression 
in a different form.
The differences between ours and theirs are discussed in
\cite{ysPRL,ysIJMPD}.}, we add constraint terms to
the right-hand-side of
(\ref{eq-E}) and (\ref{eq-A}).  The adjusted dynamical equations, 
\begin{eqnarray}
\partial_t {\tilde{E}^i_a}
&=&-i{\cal D}_j( \epsilon^{cb}{}_a  \, \null \!
\mathop {\vphantom {N}\smash N}\limits ^{}_{^\sim}\!\null
\tilde{E}^j_{c}
\tilde{E}^i_{b})
+2{\cal D}_j(N^{[j}\tilde{E}^{i]}_{a})
+i{\cal A}^b_{0} \epsilon_{ab}{}^c  \, \tilde{E}^i_c
+\kappa_1 P^i{}_{ab}  \, {\cal C}^{\rm ASH}_G{}^b,  \label{eqE2} \\
&~& {\rm where} \qquad P^i{}_{ab} \equiv
N^i \delta_{ab}+i\null \! \mathop {\vphantom {N}\smash N}
\limits ^{}_{^\sim}\!\null  \epsilon_{ab}{}^{c}\tilde{E}^i_c,
\nonumber
\\
\partial_t {\cal A}^a_{i} &=&
-i \epsilon^{ab}{}_c  \,
\null \! \mathop {\vphantom {N}\smash N}\limits ^{}_{^\sim}\!\null
\tilde{E}^j_{b} F_{ij}^{c}
+N^j F^a_{ji} +{\cal D}_i{\cal A}^a_{0}
+\kappa_2Q^a_i {\cal C}^{\rm ASH}_H
+\kappa_3R_i{}^{ja}  \, {\cal C}^{\rm ASH}_{Mj}, \label{eqA2}
\\
&~& {\rm where} \qquad Q^a_{i} \equiv
e^{-2}
\null \! \mathop {\vphantom {N}\smash N}\limits ^{}_{^\sim}\!\null
\tilde{E}^a_i, \qquad
R_i{}^{ja} \equiv
ie^{-2}
\null \! \mathop {\vphantom {N}\smash N}\limits ^{}_{^\sim}\!\null
\epsilon^{ac}{}_b \tilde{E}^b_i \tilde{E}^j_c
\nonumber
\end{eqnarray}
form a symmetric hyperbolicity if we further 
require $\kappa_1=\kappa_2=\kappa_3=1$ and the gauge conditions,
\begin{equation}
{\cal A}^a_0={\cal A}^a_i N^i, \qquad \partial_i N =0.
\label{symhypgauge}
\end{equation}
We remark that the adjusted coefficients,
$P^i{}_{ab}, Q^a_i, R_i{}^{ja}$, for
constructing the symmetric  hyperbolic system are uniquely determined,
and
there are no other additional terms (say, no ${\cal C}^{\rm ASH}_H,
{\cal C}^{\rm ASH}_M$ for $\partial_t \tilde{E}^i_a$,
no ${\cal C}^{\rm ASH}_G$ for $\partial_t {\cal A}^a_i$)
\cite{ysIJMPD}.
The gauge conditions, (\ref{symhypgauge}), are
consequences of the consistency with (triad) reality conditions.

We can also construct a strongly (or diagonalizable) hyperbolic system
by restricting to a gauge
$N^l \neq 0, \pm N \sqrt{\gamma^{ll}}$
(where $\gamma^{ll}$ is the three-metric and we do not sum indices here)
for the original equations (\ref{eq-E}), (\ref{eq-A}).
Or we can also construct from the adjusted equations,
(\ref{eqE2}) and (\ref{eqA2}),
  together with the gauge condition
\begin{equation}
{\cal A}^a_0={\cal A}^a_i N^i.  \label{diagohypgauge}
\end{equation}
As for the strongly hyperbolic system, we hereafter take the latter
expression.  

In Table \ref{eqmtable}, we summarized the equations to
be used for our comparisons.
\begin{table}[t]
\begin{center}
\begin{tabular}{cl||c|l|c}
\hline
& system & variables & Eqs of motion & remark
\\
\hline \hline
I & Ashtekar (weakly hyp.) & ($\tilde{E}^i_a, {\cal A}^a_i$) &
(\ref{eq-E}), (\ref{eq-A}) (original) & ``original" eqs.
\\ \hline
II & Ashtekar (strongly hyp.) & ($\tilde{E}^i_a, {\cal A}^a_i$) &
(\ref{eqE2}), (\ref{eqA2}) (with $\kappa=1$)&
(\ref{diagohypgauge}) required
\\ \hline
III & Ashtekar (symmetric hyp.) & ($\tilde{E}^i_a, {\cal A}^a_i$) &
(\ref{eqE2}), (\ref{eqA2}) (with $\kappa=1$)&
(\ref{symhypgauge}) required
\\ \hline
adj & Ashtekar (adjusted) &
($\tilde{E}^i_a, {\cal A}^a_i$) &
(\ref{eqE2}), (\ref{eqA2}) (with $\kappa\neq 1$)&
\\ \hline
$\lambda$ & Ashtekar-$\lambda$-system &
\begin{tabular}{c}
$(\tilde{E}^i_a, {\cal A}^a_i, $ \\
$\lambda, \lambda_i, \lambda_a)$
\end{tabular} &
(\ref{DClambda-system}) &
controls ${\cal C}_H, {\cal C}_{Mi}, {\cal C}_{Ga}$
\\
\hline
\end{tabular}
\caption{List of systems that we compare in this appendix.  }
\label{eqmtable}
\end{center}
\end{table}

\subsubsection{Ashtekar-$\lambda$-system} \label{AshtekarLambdasystem}
In \S \ref{sec:lambdasystem}, we introduced an idea to construct
a robust evolution system against a perturbative error, named 
``$\lambda$-system" \cite{BFHR}.
Based on the above symmetric hyperbolic equations, we constructed the
Ashtekar version of the ``$\lambda$-system" \cite{SY-asympAsh}.  
Here we present 
our system which evolves the
spacetime to the constraint surface, ${\cal C}_H \approx {\cal C}_{Mi}
\approx {\cal C}_{Ga} \approx 0$ as the attractor.
In \cite{SY-asympAsh}, we also presented a system which controls the 
perturbative violation of the reality condition. 

We introduce new variables ($\lambda, \lambda_i, \lambda_a$),
as they obey the dissipative evolution equations
\begin{eqnarray}
\partial_t\lambda &=&
\alpha_1 \,{\cal C}_H
-\beta_1 \, \lambda, \label{lambdaC1}
\\
\partial_t\lambda_i &=&
 \alpha_2 \,\tilde{{\cal C}}_{Mi}
 -\beta_2 \,\lambda_i, \label{lambdaC2}
\\
\partial_t\lambda_a &=&
\alpha_3 \, {\cal C}_{Ga}
-\beta_3 \, \lambda_a, \label{lambdaC3}
\end{eqnarray}
where $\alpha_i \neq 0$ (allowed to be complex numbers) and $\beta_i > 0$
(real numbers) are constants.

If we take
${u}^{(DL)}_\alpha=(\tilde{E}^i_a, {\cal A}^a_i, \lambda, \lambda_i, \lambda_a)$
as a set of dynamical variables, then the
principal part of (\ref{lambdaC1})-(\ref{lambdaC3})
can be written as
\begin{eqnarray}
\partial_t\lambda &\cong&
 -i\alpha_1\epsilon^{bcd} \tilde{E}^j_c \tilde{E}^l_d  (\partial_l{\cal A}^b_j),
\\
\partial_t\lambda_i&\cong&
\alpha_2
[-e \delta^l_i \tilde{E}^j_b
+e \delta^j_i \tilde{E}^l_b
](\partial_l{\cal A}^b_j),
\\
\partial_t\lambda_a&\cong&\alpha_3
\partial_l\tilde{E}^l_a.
\end{eqnarray}

The characteristic matrix of the system ${u}^{(DL)}_\alpha$ does not
form a Hermitian matrix.  However,
if we modify the right-hand-side of
the evolution equation of ($\tilde{E}^i_a, {\cal A}^a_i$), 
then the set becomes
a symmetric hyperbolic system.
This is done by adding
$\bar{\alpha}{}_3 \gamma^{il}(\partial_l \lambda_a)$
to the equation of $\partial_t \tilde{E}^i_a$,
and by adding
$i\bar{\alpha}{}_1\epsilon^a{}_c{}^d \tilde{E}^c_i \tilde{E}^l_d
(\partial_l \lambda)
+
\bar{\alpha}{}_2
(-e \gamma^{lm} \tilde{E}^a_i
+e \delta^m_i \tilde{E}^{la}  )
(\partial_l \lambda_m)
$ to the equation of $\partial_t{\cal A}^a_i$.
The final principal part, then, is written as
{\small
\begin{equation}
\partial_t \left(
\matrix{\tilde{E}^i_a \cr {\cal A}^a_i \cr \lambda
 \cr \lambda_i \cr \lambda_a}
\right)
\cong
\left(
\matrix{
A^l {}_a {}^{bi}{}_m & 0 & 0 & 0&
 \bar{\alpha}{}_3 \gamma^{il}\delta_a{}^b  
\cr 
0&  D^l{}^a{}_i{}_b{}^m&
i\bar{\alpha}{}_1\epsilon^a{}_c{}^d \tilde{E}^c_i \tilde{E}^l_d
&
\bar{\alpha}{}_2 e 
(
 \delta^m_i \tilde{E}^{la} - \gamma^{lm} \tilde{E}^a_i )
 & 0 
\cr 
0 & 
-i\alpha_1\epsilon_b{}^{cd} \tilde{E}^m_c \tilde{E}^l_d
& 0 & 0 & 0 
\cr 
0 &
\alpha_2 e 
(\delta^m_i \tilde{E}^l_b -\delta^l_i \tilde{E}^m_b)
& 0 & 0 & 0  
\cr 
\alpha_3\delta_a{}^b \delta^l{}_m& 0 & 0 & 0& 0
}
\right)
\partial_l \left(
\matrix{\tilde{E}^m_b \cr {\cal A}^b_m \cr \lambda
 \cr \lambda_m \cr \lambda_b}
\right).  
\label{DClambda-system}
\end{equation}
}
Clearly, the solution
$(\tilde{E}^i_a, {\cal A}^a_i, \lambda, \lambda_i, \lambda_a)
=(\tilde{E}^i_a, {\cal A}^a_i, 0, 0, 0)$ represents the original solution
of the Ashtekar system.  If the $\lambda$s decay to zero
after the evolution, then the solution also describes the original
solution of the Ashtekar system in that stage.
Since the dynamical system of ${u}^{(DL)}_\alpha$, 
(\ref{DClambda-system}),
 constitutes a symmetric
hyperbolic form, the solutions to the $\lambda$-system are unique.
Therefore, the dynamical system, (\ref{DClambda-system}), is useful
for stabilizing numerical simulations from the point that it recovers
the constraint surface automatically.

\subsubsection{Adjusted Ashtekar system}
We also try to compare a set of evolution system, which we proposed as
``adjusted-system".
The fundamental equations that we will demonstrate 
are the same with (\ref{eqE2}) and (\ref{eqA2}), but here the
real-valued constant multipliers $\kappa$s are not necessary equals to
unity.  We set $\kappa\equiv\kappa_1=\kappa_2=\kappa_3$ for simplicity. 
Apparently 
the set of (\ref{eqE2}) and (\ref{eqA2})
becomes the original weakly hyperbolic system if $\kappa=0$,
becomes the symmetric hyperbolic system if $\kappa=1$ and
$N=const.$. The set remains strongly hyperbolic systems
for other choices of $\kappa$ except $\kappa=1/2$ which only forms
 weakly hyperbolic system.
\subsection{Comparing numerical performance}
\subsubsection{Model and Numerical method}
The model we present here is gravitational wave propagation in a
planar spacetime under periodic boundary condition. 
We performed a full numerical simulation using
Ashtekar's variables.  
We prepare two $+$-mode strong pulse waves initially
by solving the ADM Hamiltonian constraint equation, 
using York-O'Murchadha's conformal approach. Then we
transform the initial Cauchy data (3-metric and extrinsic curvature)
into the connection variables, $(\tilde{E}^i_a, {\cal A}^a_i)$,
and evolve them using the dynamical equations. 
For the presentation in this article, we apply the geodesic slicing condition
(ADM lapse $N=1$ or densitized lapse $\ut N=1$, 
with zero shift and zero triad
lapse).  We have used both the Brailovskaya integration scheme, which is
a second order predictor-corrector method, and the so-called iterative 
Crank-Nicholson integration scheme for numerical time evolution.    
The details of the numerical method are
described in \cite{ronbun1}. 

 More specifically, we set our initial
guess 3-metric as
\begin{equation}
\hat{\gamma}_{ij}=
\left(\matrix{
1&0&0 \cr
sym.& 1+ K (e^{ -  (x-L)^2}+  e^{ -  (x+L)^2})&0 \cr
sym.& sym.& 1- K (e^{ - (x-L)^2} + e^{ - (x+L)^2})
}\right), 
\label{plusmetric}
\end{equation}
in the periodically bounded region $x=[-5, +5]$.
 Here $K$ and $L$ are constants
and we set $K=0.3$ and $L=2.5$ for the plots. 

In order to show the expected ``stabilization behavior" clearly,
we artificially add an error in the
middle of the time evolution. We added an artificial
inconsistent rescaling once at time $t=6$ for the ${\cal A}^2_y$ component as
${\cal A}^2_y \rightarrow {\cal A}^2_y (1+ {\rm error})$.

\begin{figure}[ht]
\setlength{\unitlength}{1in}
\begin{picture}(6.0,2.2)
\put(0.0,0.0){\epsfxsize=2.8in \epsfysize=1.8in \epsffile{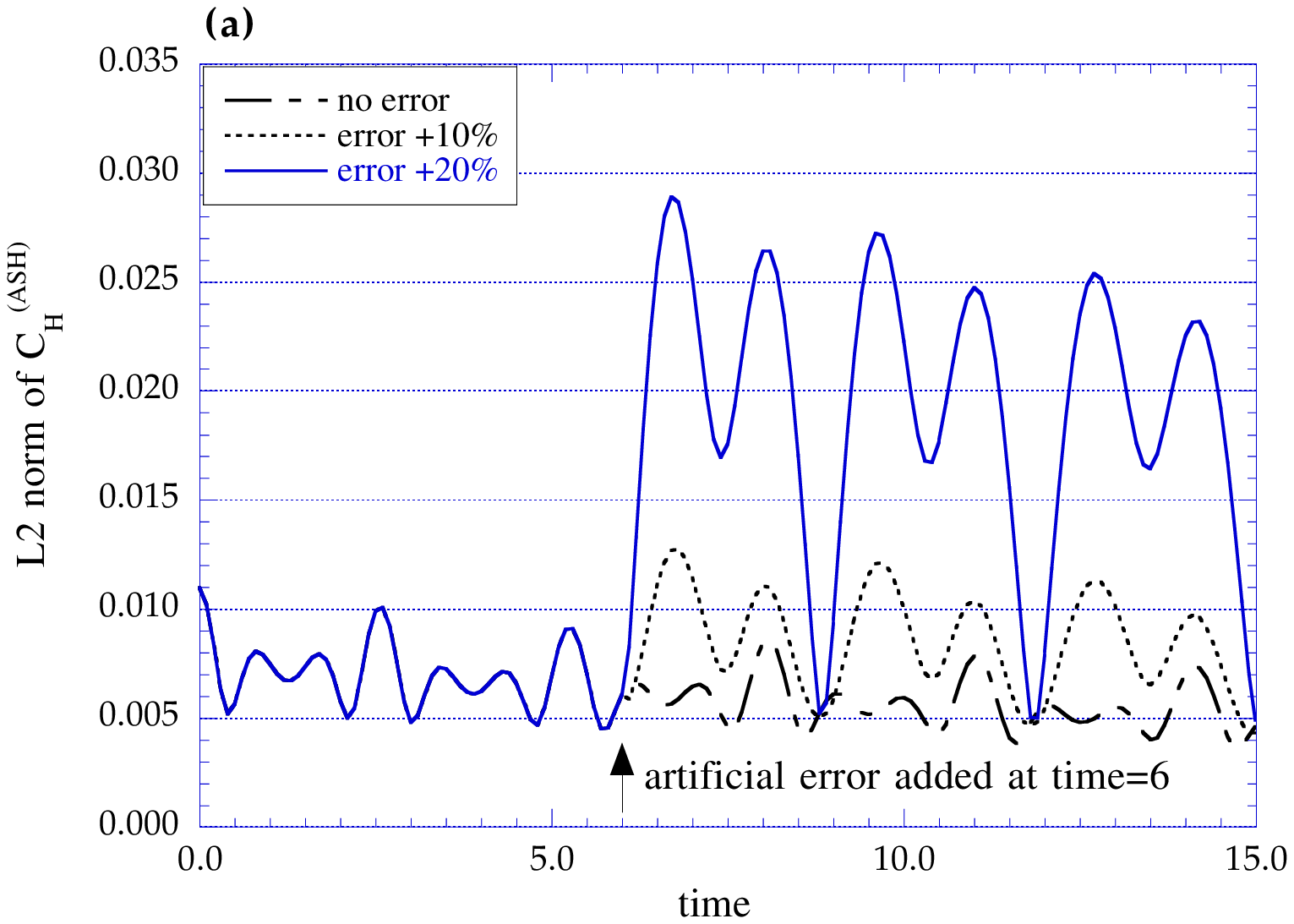} }
\put(3.2,0.0){\epsfxsize=2.8in \epsfysize=1.8in \epsffile{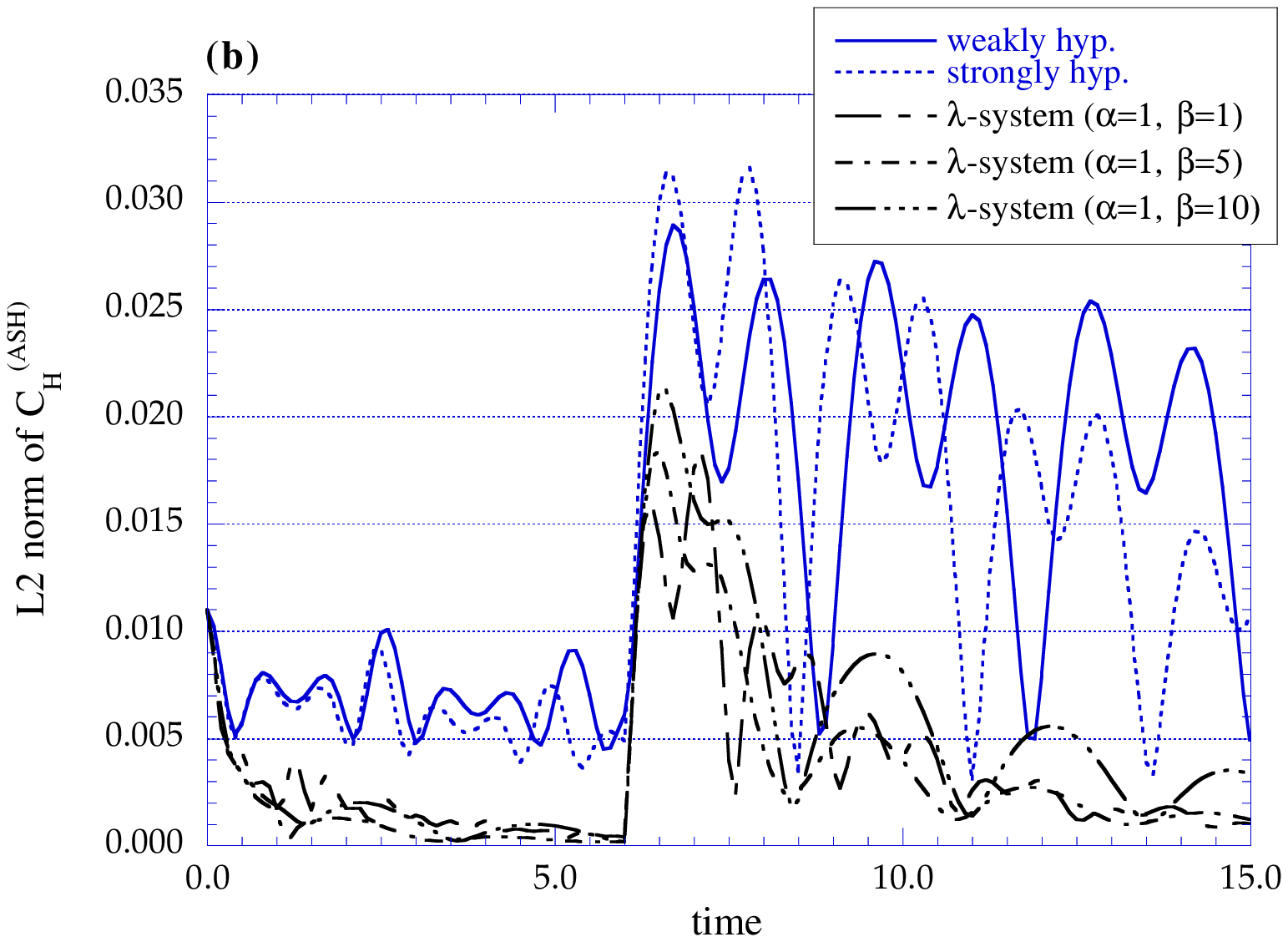} }
\end{picture}
\caption[fig-acc]{\small
Demonstration of the Ashtekar-$\lambda$-system
for the cases of plane wave propagation under the
periodic boundary.  
We plot the 
L2 norm of the Hamiltonian
constraint equation, ${\cal C}_H$.  
Fig. (a) shows how the system goes bad depending on the amplitude of
artificial error at $t=6$.  
All the lines are of the
evolution of Ashtekar's original equation (no $\lambda$-system).
Fig. (b) shows the effect of $\lambda$-system.
All the lines are 20\% error amplitude, but shows the difference of
evolution equations. The solid line is for Ashtekar's original equation
(the same as in Fig.(a)), the dotted line is for the strongly hyperbolic
Ashtekar's equation.  Other lines are of $\lambda$-systems, which produces
better performance than that of the strongly hyperbolic system. 
(Reprinted from \cite{ronbun2}, \copyright IOP 2001)
}
\label{errwave}
\end{figure}


\subsubsection{Differences between three levels of hyperbolicity}
We have performed comparisons of stability and/or accuracy between weakly and
strongly hyperbolic systems, and between weakly and symmetric hyperbolic
systems\cite{ronbun1}. (We can not compare strongly and symmetric hyperbolic
systems directly, because these two requires different gauge conditions.)

\begin{figure}[htb]
\setlength{\unitlength}{1in}
\begin{picture}(6.0,2.4)
\put(0.0,0.0){\epsfxsize=2.8in \epsfysize=1.8in \epsffile{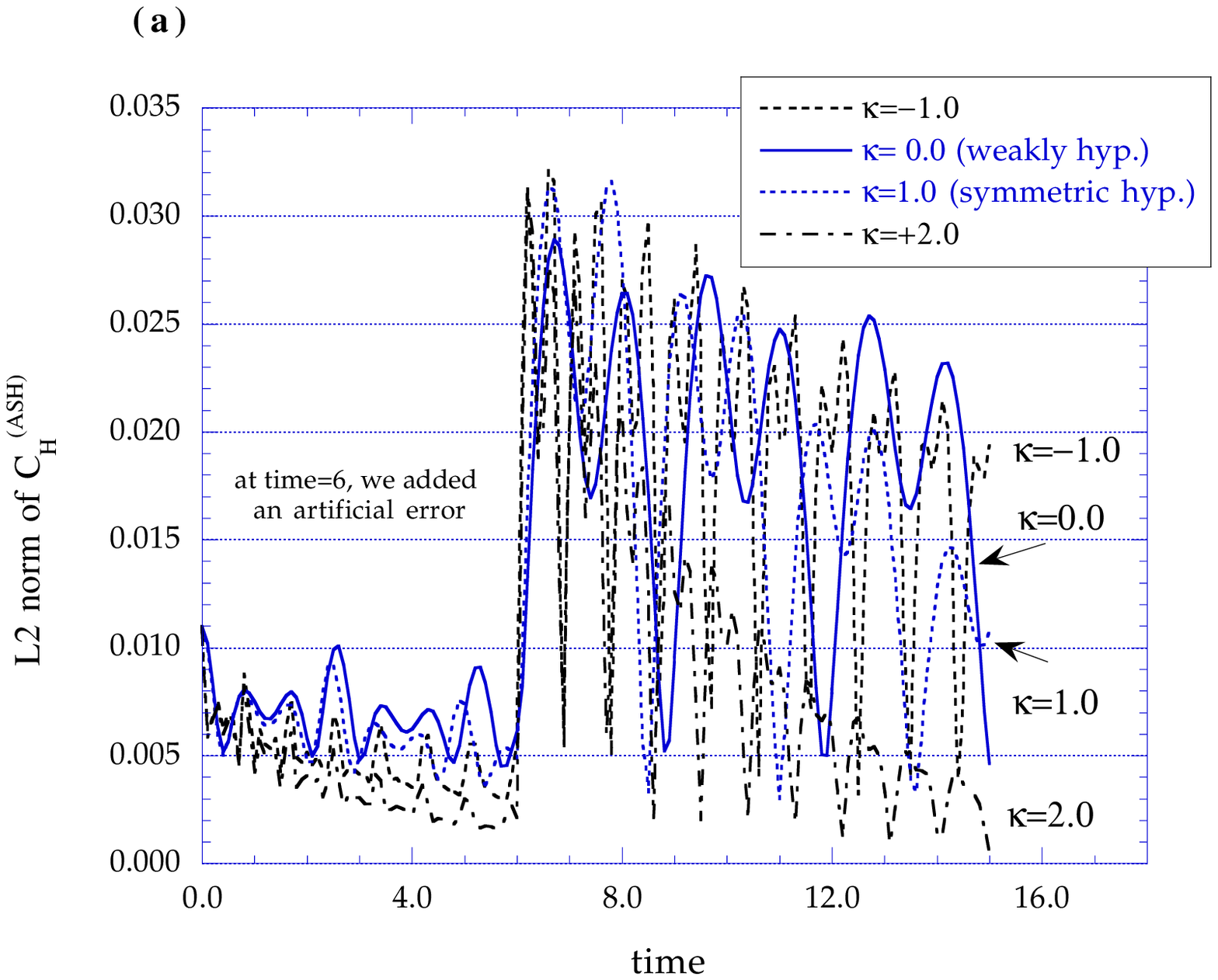} }
\put(3.2,0.0){\epsfxsize=2.8in \epsfysize=1.8in \epsffile{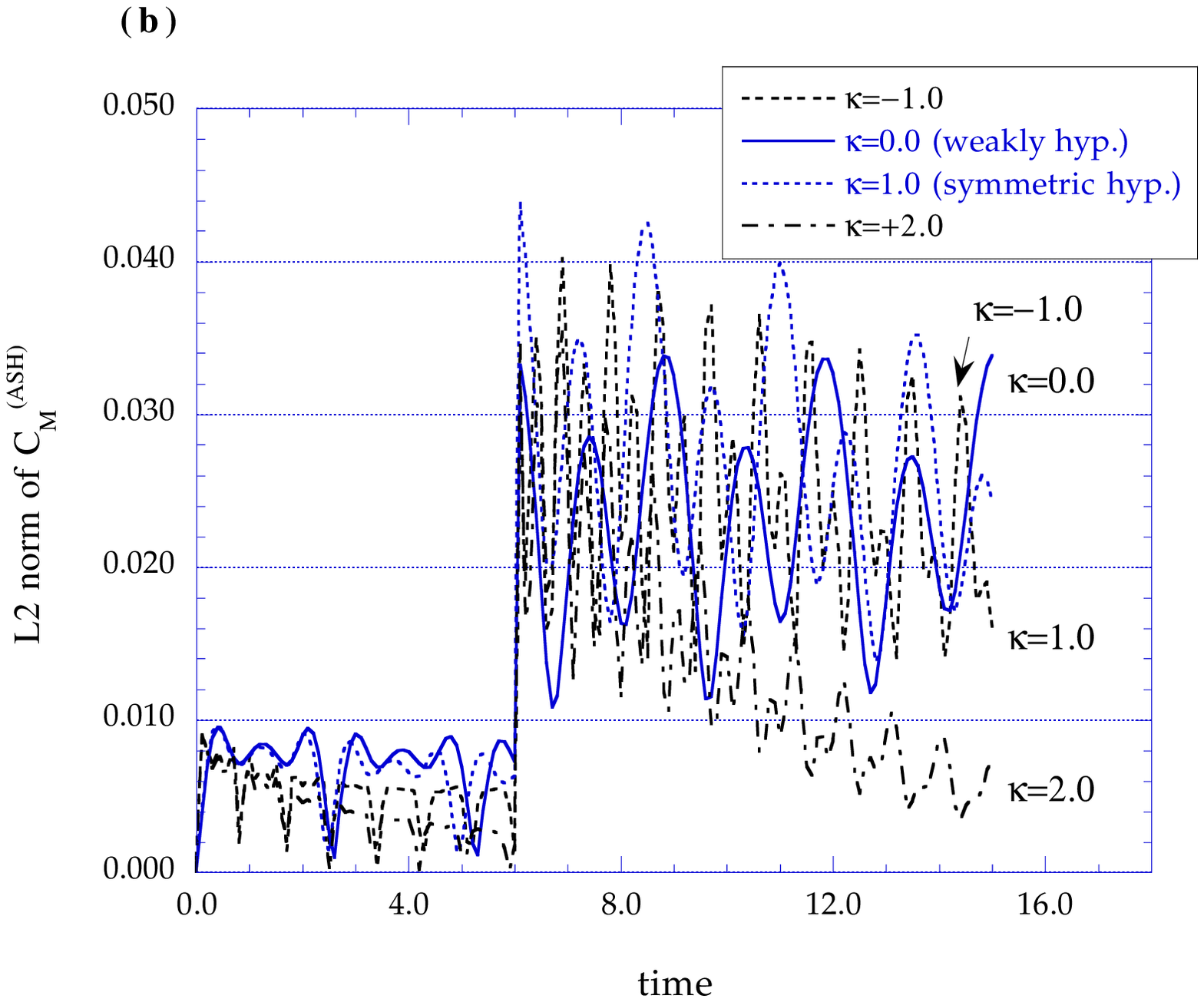} }
\end{picture}
\caption[fig-acc]{\small 
Demonstration of the adjusted-Ashtekar systems. 
The experiments as in Fig.\ref{errwave}(b).
Fig. (a) and (b) are L2 norm of the Hamiltonian
constraint equation, ${\cal C}_H$, and momentum constraint equation,
${\cal C}_{Mx}$, respectively.
The solid line is the case of $\kappa=0$, that is the case of ``no adjusted"
original Ashtekar equation (weakly hyperbolic system).
The dotted line is for $\kappa=1$, equivalent to the symmetric hyperbolic
system.  We see other line  ($\kappa= 2.0$) shows better performance
than the symmetric hyperbolic case.
(Reprinted from \cite{ronbun2}, \copyright IOP 2001)
}
\label{errwave2}
\end{figure}

We omit figures in this report, but one can see a part of results in 
Fig.\ref{errwave} and Fig.\ref{errwave2}. 
We may conclude that 
higher level hyperbolic system
gives us slightly accurate evolution. 
However, if we evaluate the magnitude of L2 norms, then 
we also conclude that there is no measurable differences between 
strongly and symmetric hyperbolicities.  
This last fact will be supported more affirmatively in the next
experiments.

\subsubsection{Demonstrating ``$\lambda$-system"}\label{AshtekarLambdaNum}
Next, we show a result of the ``$\lambda$-system" \cite{ronbun2}.
Fig.\ref{errwave} (a) shows how the violation of the Hamiltonian
constraint equation, ${\cal C}_H$, become  worse depending on the
term ${\rm error}$.
The oscillation of the L2 norm ${\cal C}_H$ in the figure due to the
pulse waves collide periodically in the numerical region.
We, then, fix the error term as a 20\% spike, and try to evolve the
same data in different equations of motion, i.e., the original Ashtekar's
equation [solid line in Fig.\ref{errwave} (b)], strongly hyperbolic
version of Ashtekar's equation (dotted line) and the above $\lambda$-system
equation (other lines) with different $\beta$s but the same $\alpha$.
As we expected, all the $\lambda$-system cases result in reducing the 
Hamiltonian constraint errors.


\subsubsection{Demonstrating ``adjusted system"}\label{AshtekarAdjNum}

We here 
examine how the adjusted multipliers contribute to the system's stability
\cite{ronbun2}.
In Fig.\ref{errwave2}, we show the results of this experiment.
We plot the violation of the constraint equations both ${\cal C}_H$
and ${\cal C}_{Mx}$.
An artificial error term was added in the same way as above.
The solid line is the case of $\kappa=0$, that is the case of ``no adjusted"
original Ashtekar equation (weakly hyperbolic system).
The dotted line is for $\kappa=1$, equivalent to the symmetric hyperbolic
system.  We see other line ($\kappa=2.0$) shows better performance
than the symmetric hyperbolic case.

\vfill

\section*{Acknowledgment}
HS thanks 
M. Alcubierre, A. Ashtekar, T. Baumgarte, C. Bona, G. Calabrese, 
S. Detweiler, N. Dorband, S. Husa,  P. Laguna, C. Lechner, L.Lehner, 
L. Lindblom, V. Moncrief, T. Nakamura, 
D. Pollney, O. Sarbach, M. Scheel, E. Seidel, M. Shibata, D. Shoemaker, 
and J. Winicour for scientific discussion in the past year.   
He also thanks Albert Einstein Institute Cactus terms 
for technical helps at the Mexico numerical relativity workshop.
He also thanks T. Ebisuzaki for continuous encouragement.  
HS is supported by the special postdoctoral researchers
program at RIKEN, and this work was supported partially 
by the Grant-in-Aid for
Scientific Research Fund of Japan Society of the Promotion of Science,
No.~14740179. 
We thank J. Overduin for careful reading a part of the manuscript. 
GY thanks H. Shinkai for writing all the manuscript. 

\vfill


\newpage
\addcontentsline{toc}{section}{\protect\numberline{}{References}}

\end{document}